\documentclass[12pt]{article}
\input epsfig.sty
\usepackage{graphicx} 
\def\laq{\raise 0.4ex\hbox{$<$}\kern -0.8em\lower 0.62
ex\hbox{$\sim$}}
\def\gaq{\raise 0.4ex\hbox{$>$}\kern -0.7em\lower 0.62
ex\hbox{$\sim$}}

\begin{document}

\begin{titlepage}
\begin{flushright}
CERN-PH-TH/2007-260
\end{flushright}
\vspace*{1cm}

\begin{center}

{\large{\bf Magnetized CMB observables: a dedicated numerical approach}}
\vskip1.cm

Massimo Giovannini$^{a,c}$ and Kerstin E. Kunze$^{b,c}$

\vskip2.cm

{\sl $^a$Centro ``Enrico Fermi", Compendio del Viminale, Via 
Panisperna 89/A, 00184 Rome, Italy}
\vskip 0.2cm 
{\sl $^b$ Departamento de F\'\i sica Fundamental, \\
 Universidad de Salamanca,Plaza de la Merced s/n, E-37008 Salamanca, Spain}
\vskip 0.2cm 
{\sl $^c$  Department of Physics, Theory Division, CERN, 1211 Geneva 23, Switzerland}

\vspace*{1cm}

\begin{abstract}
Large-scale magnetic fields affect the scalar modes of the geometry whose ultimate effect is to determine the anisotropies of the Cosmic Microwave Background (CMB in what follows).
For the first time, a  consistent numerical approach to the magnetized CMB anisotropies is pursued 
with the aim of assessing the angular power spectra of temperature and polarization
when the scalar 
modes of the geometry and a stochastic background of inhomogeneous magnetic fields are simultaneously 
present in the plasma.
The effects related to the magnetized nature of the plasma are taken into account both at the level of the dynamical equations and at the level of the initial conditions of the Einstein-Boltzmann hierarchy. 
The temperature and polarization observables are exploited to infer the peculiar signatures 
of a pre-equality magnetic field. Using the extrapolated best fit to the three year WMAP data the increase and distortions of the first seven peaks in the TT autocorrelations are monitored for different values 
of the regularized magnetic field intensity and for the physical range of spectral indices.
Similar analyses are also conducted for the first few anticorrelation (and corrrelation) peaks of the TE 
power spectra.  Possible interesting degeneracies and stimulating perspectives are pointed out and explored. 
\end{abstract}

\end{center}
\end{titlepage}

\newpage
\renewcommand{\theequation}{1.\arabic{equation}}
\setcounter{equation}{0}
\section{The general framework}
\label{sec1}
The origin of large-scale magnetic fields is still a poorly understood 
subject as much as the origin of the primordial spectrum of the density contrast 
was unclear almost two score years ago when the first attempts of understanding   the origin of large-scale structure actually begun. The first observational evidence 
of the existence of large-scale magnetic fields in our galaxy goes back to the work of Hiltner \cite{hiltner} and 
Hall \cite{hall} (see also \cite{davis}). 
The problem, in itself, has many facets which can be summarized as follows:
\begin{itemize}
\item{} observational evidence of the existence of large-scale magnetic fields is indisputable 
from a number of different observations ranging from Faraday rotation to synchrotron emission;
\item{} objects of different sizes (and different evolutionary histories) possess 
sizable  magnetic fields in the $\mu$ G range: a striking example of this statement are, for instance, spiral galaxies, elliptical galaxies and a class of x-ray bright Abell clusters;
\item{} theoretical evidence of the mechanism responsible of the degree of magnetization of the 
present  Universe is still under active discussion.
\end{itemize}
We have a rather plausible
control of the dynamics of electromagnetic fields in plasmas: since the pioneering work of 
Alfv\'en \cite{alfven} the various descriptions of weakly coupled plasmas have been tested both in astrophysical 
systems and, most importantly, laboratory experiments \cite{spitzer,krall,biskamp, boyd}. 
Various plasma descriptions (covering different branches of the spectrum 
of plasma excitations) allow to predict instabilities in terrestrial tokamaks. An example 
of successful (but not unique) framework is the one-fluid plasma theory which, under 
some circumstances, reduces to the well known magnetohrdrodynamics (MHD in what follows).
All the astrophysical attempts for the justification of large scale magnetic fields rest 
on the assumption that, at some time prior to galaxy formation, appropriate initial conditions for the MHD evolution
should be present. While this point of view is shared by the whole community, opinions 
vary as far as the real primordial nature of magnetic fields is concerned. For informative 
reviews on this broad subject the interested reader may usefully consult the publications 
reported in \cite{review1} (see also \cite{review2} for more observational aspects 
of large-scale magnetization).

Theoretical diatribes cannot decide wether or not large-scale magnetic fields 
are a cosmic relic.
It is more rewarding, in our opinion, to apply Occam's razor and pose 
a more modest but yet experimentally answerable question: were large-scale 
magnetic fields present prior to matter-radiation equality?
The latter question is answerable since, as it will be specifically shown, 
the effects of large-scale magnetic fields can be read-off from the properties 
of CMB observables. 

In recent years diverse data sets seem to conspire towards a sort of paradigm 
which is customarily employed in the interpretation of cosmological data. This paradigm is 
often dubbed as $\Lambda$CDM lore where $\Lambda$ stands for the dark-energy component 
(parametrized as a cosmological constant) and CDM stands for the cold dark matter component. The data sets supporting this general 
view range from the CMB data\footnote{ Among them, the WMAP data \cite{WMAP1,WMAP2,WMAP3,WMAP4,WMAP5}, the data of other 
balloon borne experiments as well as the data of various terrestrial arrays \cite{AC1}.} to the large-scale structure 
data \cite{LSS1,LSS2} and, finally, to the type Ia supernova data in their various incarnations \cite{SN1,SN2} (see 
also \cite{lens} for weak lensing data). 
The combined analysis of these three  classes of data support the $\Lambda$CDM paradigm.
A primordial spectrum 
of adiabatic perturbations, present for typical wavelengths larger than the Hubble 
radius prior to Hydrogen recombination and after matter-radiation equality, is the main 
responsible for the normalization of the CMB temperature autocorrelations (TT correlations 
in what follows). The cross-correlations between temperature and polarization (TE correlations in what follows) 
lead to a typical anticorrelation peak \cite{WMAP4} which is often quoted as the 
golden signature of the predominant adiabaticity of the CMB initial conditions.
The polarization autocorrelations (EE correlations in what follows) have been 
partially observed by the WMAP experiment itself \cite{WMAP2} as well as by other terrestrial 
arrays. 
The viability of the $\Lambda$CDM paradigm is customarily completed by the addition 
of various parameters whose explicit determination can either confirm or improve
the pure $\Lambda$CDM scenario. Among 
these models we can mention, just as an example, the addition of a tensor component, the inclusion of a 
specific form of the barotropic index for the dark energy component (different from the one of a pure cosmological constant), 
the search for specific corrections in the scalar power spectrum motivated, with various 
degrees of theoretical accuracy, by the (yet unknown) physics of the Planck energy scale (see, for instance,
some of the analyses reported in \cite{WMAP1}). There is the hope that the unprecedented 
accuracy of the data of the Planck explorer mission \cite{Planck}  will allow to rule in (or rule out) 
some of these theoretical possibilities \footnote{Various models going beyond the pure $\Lambda$CDM scenario
can be confronted with the foreseeable sensitivities of the high and low frequency instruments embarked on the 
Planck spacecraft. See, for instance, \cite{zalsel,pav4}.}.

The aim of this paper is to complement  the $\Lambda$CDM paradigm with the presence 
of a pre-decoupling magnetic field (see also \cite{giokunz1}). Such a completion is not only motivated 
theoretically but also observationally: since we do observe large-scale 
magnetic fields later on, it is plausible to posit, as falsifiable hypothesis, their existence also
prior to equality. Definite answers to such a question will come, in this context, by confronting the 
completion of the $\Lambda$CDM paradigm with all the available cosmological data 
in the same way as, for instance, a pristine tensor contribution to the CMB anisotropies 
can be constrained by adding, in the parameter estimation, a stochastic background of long-wavelength gravitons.
In this investigation one of the first goals of the program will be reached: a dedicated 
numerical approach for the magnetized CMB  observables will be constructed. Without 
this step sound strategies of parameter estimation will be forlorn. This step 
is often rather straightforward in different cases since all the available 
codes contemplate, for instance, the inclusion of tensor modes or the inclusion 
of a peculiar barotropic index for the dark energy component. However 
the latter statement does not apply to the case of large-scale magnetic fields.

The analysis of the interplay between 
large-scale magnetic fields and CMB observables might be traced back 
to the pioneering works of Zeldovich \cite{zeldovich1} and Harrison \cite{harrison}.
In recent years it has been understood that large-scale magnetic fields 
may affect the vector and tensor modes of the geometry and may also 
affect, indirectly, the CMB polarization \cite{vt1,vt2} (see, for instance, \cite{maxrev1} for a 
topical review on this subject). The main obstacle to a systematic analysis of the current data 
in the light of a magnetized component is represented by our lack 
of understanding of the close relationship between the  large-scale magnetic fields 
and the {\em scalar} modes of the geometry. Indeed, as explicitly suggested 
by observations, the inhomogeneities in the CMB temperature and polarization 
can be attributed to curvature perturbations. 

An impasse then seems to arise. The vector and tensor modes 
induced by large-scale magnetic fields are very small at large length scales (i.e. small 
multipoles). It is thus rather hard to imagine the possibility of including the magnetic field 
contribution as an appropriate fit parameter in an extended version of the $\Lambda$CDM 
paradigm if the effects of magnetic fields on the scalar 
modes of the geometry are unknown. 

A theoretical framework 
for the analysis of scalar modes of the geometry and large-scale magnetic fields 
has been recently developed in a series of papers ranging from the accurate determinations 
of adiabatic initial conditions in the presence of a scalar mode of the geometry \cite{mg1}, to the 
extension of the tight-coupling expansion \cite{mg2} and to the  semi-analytical calculation 
of temperature autocorrelations \cite{mg3}.
One of the aims of the present paper is to translate into a dedicated numerical approach 
all the theoretical understanding of the interplay between the scalar modes 
of the geometry and the large scale magnetic fields that has been 
pursued, through various steps, in \cite{mg1,mg2,mg3}.
The CMB effects related to the scalar modes of the geometry are the 
most difficult ones already in the absence of large-scale magnetic fields. The 
complication comes from the physical observation that the density 
contrasts of the various species do not couple to the tensor and the vector modes but they do couple 
to the scalar modes of the geometry. Conversely 
the curvature perturbations are the source of the evolution of the density 
contrasts for photons, for baryons and for the CDM species. In the magnetized 
case this occurrence is even more acute since large-scale 
magnetic fields and plasma effects propagate both at the level of the Boltzmann 
hierarchy and at the level of the perturbed Einstein equations. 

The approach studied in the present investigation is based on a faithful 
MHD description of the pre-decoupling plasma. More refined 
descriptions of the high frequency branch of the spectrum of plasma excitations (valid for
frequencies comparable with the electron plasma frequency) must reproduce anyway 
the approach described here when the typical length-scales and dynamical times 
are, respectively, much larger than the Debye length and much larger than the inverse 
of the plasma frequency \cite{spitzer,krall}.

The plasma will not only be populated by an electromagnetic component but also by 
fluctuations of the geometry which should be treated relativistically since 
their typical wavelengths, at the onset of the numerical calculation, will be much larger 
than the Hubble radius at the corresponding epoch. The description of the scalar modes of the geometry
will be formulated in the peculiar language of the synchronous coordinate system.
One of the first codes developed for the analysis of CMB anisotropies is 
 COSMICS \cite{cosmics1,cosmics2}. The CMBFAST \cite{cmbfast1,cmbfast2} code is based 
 originally on COSMICS and in many respects it can be said that COSMICS 
 is the ancestor of CMBFAST.  As in COSMICS, also in CMBFAST the dynamical evolution 
 across equality and decoupling is followed in the synchronous coordinate system.
 The synchronous approach carries necessarily a possible ambiguity on the
 complete removal of the gauge freedom. It is actually known since the early 
 eighties \cite{PV1} that, in the synchronous gauge, the coordinate 
 system is only fixed up to a pair of space-dependent integration constants. The
 remaining gauge freedom must be removed from the initial conditions of the 
 Boltzmann hierarchy to avoid the dangerous presence of spurious (i.e. gauge) 
 modes. The way to handle this potentially annoying problem resides 
 in the ability of treating the same problem also in contiguous gauges where 
 the gauge parameters are completely fixed. An example of this technique 
 is the longitudinal gauge \cite{harrison,cosmics2,bardeen} (see also \cite{weinberg})
 which has been 
 also exploited for this purpose in the absence of magnetic fields. This 
 will also be the path followed in the present analysis (see also \cite{mg1,mg2}) by including 
 consistently the effects due to the global magnetization of the plasma.
Finally, in the present study  the scalar vector and tensor modes 
are separated on the basis of their transformation properties under three-dimensional rotations 
\cite{bardeen} (see also \cite{lif1}).
There exist approaches which are fully covariant \cite{EB} and which have been 
also applied to the case of large-scale magnetic fields \cite{cov} without leading, however, to any explicit estimate 
either of the Sachs-Wolfe plateau or of the temperature autocorrelations in the Doppler region as reported 
in \cite{mg1}.

 With these specifications in mind, our code is based on CMBFAST with integration along the line of sight. The main difference stems from the consistent 
introduction of a magnetized component both al the level of the 
initial conditions and at the level of the dynamical equations. 
This choice 
has been also dictated by the fact that the WMAP collaboration 
 used also CMBFAST for the analysis of the observational data. 
It is plausible to think that this numerical approach can be also effective throughout the 
next decade for the analysis of forthcoming data like the ones of the Planck explorer \cite{Planck}.

The plan of our paper is therefore the following.  In Section \ref{sec2} we will review 
the main evolution equations to be integrated.
Particular attention will be given to the way large-scale magnetic fields are 
 included in the pre-equality and pre-decoupling physics.
In Section \ref{sec3} a theory of the magnetized 
initial conditions will be formulated  when the  dominant source of curvature 
inhomogeneities is the standard adiabatic mode.  Section \ref{sec4} is devoted to the calculation 
of temperature autocorrelations. The  results on the polarization 
observables are collected in Section \ref{sec5}.
The distinctive features 
introduced in the angular power spectra by the presence of large-scale magnetic fields will be further 
scrutinized in Section \ref{sec6}. Section \ref{sec7} contains our concluding remarks.
In the Appendix we collected auxiliary material on the longitudinal 
gauge description sticking, however, only to those themes that are germane to our
calculations.
\renewcommand{\theequation}{2.\arabic{equation}}
\setcounter{equation}{0}
\section{The full content of the magnetized plasma}
\label{sec2}
Below the temperature of neutrino decoupling, the content of the plasma is formed 
both by neutral species and charged species. The neutral species are 
cold dark matter (CDM in what follows), neutrinos (which will be taken to be massless) 
and photons. The charged species are baryons and electrons. 
The three observational data sets
(i.e. CMB anisotropies \cite{WMAP1,WMAP2}, large-scale structure \cite{LSS1,LSS2} and type Ia supernovae \cite{SN1,SN2}) suggest, that a cosmological constant term should also be added.
The pivotal $\Lambda$CDM lore is then delicately improved by positing, already prior to equality,
that the primeval plasma is effectively magnetized.

Neutral species are indirectly affected by the presence 
of large-scale magnetic fields. In fact, magnetic fields gravitate and 
contribute both to the  Hamiltonian and momentum constraints 
as well as to the dynamical evolution of the gravitational inhomogeneities.
The evolution of the charged species will be followed using MHD  \cite{spitzer,krall,boyd}
generalized to the situation where 
the geometry is dynamical and where gravitational inhomogeneities are also 
simultaneously present. The adoption of this scheme is dictated 
by the typical hierarchy of the Coulomb and Thompson scatterings.

The position of the first Doppler peak implies, in a $\Lambda$CDM framework,
that the background geometry is spatially flat. The inclusion of spatial curvature amounts to an overall shift 
of the TT power spectra and hence to a change in the position of the first peak. The  line element can then be written, in the conformal 
time coordinate $\tau$, as:
\begin{equation}
ds^2 = a^2(\tau) [ d\tau^2 - d\vec{x}^2],\qquad a(\tau) d\tau = dt, 
\qquad {\mathcal H} = \frac{a'}{a} = a H.
\label{line}
\end{equation}
In Eq. (\ref{line}), $t$ denotes the cosmic time coordinate and $H$ the Hubble rate;
the prime will denote, throughout the paper, a derivation with respect to the conformal 
time coordinate $\tau$.
The evolution of the scale factor $a(\tau)$ is determined by the appropriate 
Friedmann-Lema\^itre equations:
\begin{eqnarray}
&& {\mathcal H}^2 = \frac{8\pi G}{3} a^2 \rho_{\mathrm{t}},
\label{FL1}\\
&& {\mathcal H}^2 - {\mathcal H}' = 4 \pi G a^2 ( p_{\mathrm{t}} + \rho_{\mathrm{t}}),
\label{FL2}\\
&& \rho_{\mathrm{t}}' + 3 {\mathcal H} (\rho_{\mathrm{t}} + p_{\mathrm{t}}) =0,
\label{FL3}
\end{eqnarray}
where $\rho_{\mathrm{t}}$ and $p_{\mathrm{t}}$ denote, respectively, 
the total energy density and pressure of the plasma, i.e. 
\begin{equation}
\rho_{\mathrm{t}} = \rho_{\mathrm{c}} + \rho_{\nu} + \rho_{\gamma} + \rho_{\mathrm{e}} + 
\rho_{\mathrm{b}} + \rho_{\Lambda},\qquad 
p_{\mathrm{t}} = \frac{\rho_{\nu}}{3} + \frac{\rho_{\gamma}}{3} - \rho_{\Lambda}.
\label{FL4}
\end{equation}
The subscripts in Eq. (\ref{FL4}) refer to the various components 
of the plasma mentioned in the first paragraph of the current section. 

The time scales involved in the present study are the ones encountered in CMB physics: 
the equality time (at which the radiation and the matter 
component have equal weight in Eq. (\ref{FL1})), the recombination time (at which 
the ionization fraction drops because neutral Hydrogen is formed), the decoupling time 
(at which the mean free path of the photons becomes 
comparable with the Hubble radius). The exact solution of Eqs. (\ref{FL1}), (\ref{FL2}) and (\ref{FL3})
in the absence of dark energy component (which can be neglected around decoupling)
stipulates that the scale factor interpolates, in the conformal time coordinate $\tau$, between a linear 
evolution (typical of the pre-equality plasma) and a quadratic evolution 
(typical of the plasma around recombination and decoupling):
\begin{equation}
a(\tau) = a_{\mathrm{eq}} \biggl[ \biggl(\frac{\tau}{\tau_{1}}\biggr)^2 + 2 \biggl(\frac{\tau}{\tau_{1}}\biggr)\biggr], \qquad 
\tau_{1} = \frac{2}{H_{0}} \sqrt{\frac{a_{\mathrm{eq}}}{\Omega_{\mathrm{M}  0}}}  \simeq 288.15 \,\, \biggl(\frac{ \omega_{\mathrm{M}}}{0.134}\biggr)^{-1}\, \mathrm{Mpc}.
\label{SCF}
\end{equation}
In Eq. (\ref{SCF}) $\omega_{\mathrm{M}} = h_{0}^2 \Omega_{\mathrm{M}0}$ where 
$h_{0}$ (of the order of $0.7$ in the standard $\Lambda$CDM framework) is the 
current indetermination on the Hubble rate. Given a generic species 
$X$, $\omega_{X} = h_{0}^2 \Omega_{X0}$: while 
$\Omega_{X0}$ is proportional to $h_{0}^{-2}$, $\omega_{X}$ is, by its definition, independent 
of $h_{0}$ (this is the reason why it is sometimes called Hubble-free critical fraction). To estimate
$\tau_{1}$ in Eq. (\ref{SCF}) it has been used that $\omega_{\gamma} = 2.47 \times 10^{-5}$, 
$\omega_{\nu} = 1.68 \times 10^{-5}$ and $\omega_{\mathrm{R}} = \omega_{\nu} + \omega_{\gamma} = 4.15 
\times 10^{-5}$. In the code Eqs. (\ref{FL1}), (\ref{FL2}) and (\ref{FL3}) are 
integrated numerically. Equation (\ref{SCF}) can be anyway used for 
semi-analyitical estimates \cite{mg2} and for the important problem of correctly setting the initial conditions 
of the Einstein-Boltzmann hierarchy (see Section \ref{sec3}).  According to Eq. (\ref{SCF}) 
$\tau_{\mathrm{eq}} = (\sqrt{2} -1) \tau_{1} \simeq \tau_{1}/2$.
The redshift is defined as $1 + z = a_{0}/a$ by fixing 
$a_{0}=1$. The recombination redshift (taken for instance between 
$1050$ and $1100$) will determine, via Eq. (\ref{SCF}) $\tau_{\mathrm{rec}}$ (note that $\tau_{\mathrm{rec}} > \tau_{\mathrm{eq}}$). 

The coupled evolution of the fluctuations of the geometry (\ref{line}) with the 
fluctuations of the plasma quantities will determine, thanks 
to the interaction with the magnetic fields, the peculiar features of the magnetized 
CMB anisotropies. 
The synchronous coordinate system is most easily presented already 
in Fourier space and it can be written as\footnote{Our conventions on the perturbations are 
summarized by Eq. (\ref{pert1}). It should be borne in mind that the signature  
of the metric is mostly minus (see Eq. (\ref{line})). Various treatments of this problem 
adopt the opposite convention (i.e. the signature $(-,+,+,+)$).}
\begin{equation}
\delta_{\rm s} g_{i j}(k,\tau) = 
a^2(\tau)\biggl[ \hat{k}_{i} \hat{k}_{j} h(k,\tau) + 6 \xi(k,\tau)\biggl(  
\hat{k}_{i} \hat{k}_{j} - \frac{1}{3} \delta_{ij}\biggr)\biggr],
\label{pert1}
\end{equation}
where $\hat{k}_{i}= k_{i}/|\vec{k}|$ denotes the direction of the Fourier 
wave-vector and $\delta_{\mathrm{s}}$ reminds that we are 
considering here only the effects of the scalar modes of the geometry 
which are, as already mentioned, the most difficult (but also the most relevant) 
when the plasma is effectively magnetized.  
 \subsection{CDM particles and neutrinos}
\label{sec2a}
Defining as $\delta_{\mathrm{c}}$ and $\theta_{\mathrm{c}}$ the density contrast 
of CDM particles and the three-divergence of the corresponding 
peculiar velocity, the evolution equations of the CDM sector are:
\begin{equation}
\delta_{\mathrm{c}}' = - \theta_{\mathrm{c}} + \frac{h'}{2},\qquad 
\theta_{\mathrm{c}}' + {\mathcal H} \theta_{\mathrm{c}} =0.
\label{CDMS1}
\end{equation}
In spite of the fact that the CDM velocity field will be used, in Section \ref{sec3} to remove partially 
the remaining gauge freedom, it is relevant to appreciate that Eq. (\ref{CDMS1}) 
can be also written as 
\begin{equation}
\biggl(\delta_{\mathrm{c}} - \frac{h}{2} \biggr)'' + {\mathcal H} \biggl(\delta_{\mathrm{c}} - \frac{h}{2} \biggr)'=0.
\label{CDMS2}
\end{equation}
Defining, in analog terms, $\delta_{\nu}$ and  $\theta_{\nu}$ as the neutrino 
density contrast and as the three-divergence of the neutrino peculiar velocity, the corresponding evolution equations 
are:
\begin{eqnarray}
&& \delta_{\nu}' = -\frac{4}{3} \theta_{\nu} + \frac{2}{3} h',
\label{NUS1}\\
&& \theta_{\nu}' = - k^2 \sigma_{\nu}  + \frac{k^2}{4} \delta_{\nu},
\label{NUS2}\\
&& \sigma_{\nu}' = \frac{4}{15} \theta_{\nu} - \frac{3}{10} k {\mathcal F}_{\nu 3} - \frac{2}{15} h' - \frac{4}{5} \xi',
\label{NUS3}
\end{eqnarray}
where $\sigma_{\nu}$ is the neutrino anisotropic stress (also appearing in the perturbed Einstein
equations) which is related to the quadrupole of the (perturbed) phase space distribution 
as $\sigma_{\nu} = {\mathcal F}_{\nu 2}/2$; ${\mathcal F}_{\nu 3}$ is the octupole 
of the (perturbed) phase space distribution. The presence of the quadrupole 
and octupole reflects the occurrence that neutrinos are collisionless below temperatures 
of the order of the MeV and should therefore be treated in the framework of the 
appropriate Boltzmann hierarchy. Equations (\ref{NUS1}), (\ref{NUS2}) and (\ref{NUS3}) 
couple together the lowest multipoles and will be the ones used, in Section \ref{sec3},
to set initial conditions of the CMB anisotropies in the pre-equality 
regime. At later time, in the code, the neutrinos 
will be integrated using the perturbed form of the collisionless
Boltzmann equation written in the synchronous coordinate 
system:
\begin{equation}
{\mathcal F}_{\nu}' + i k \mu {\mathcal F}_{\nu} = 4 \biggl[ - \xi' + \frac{\mu^2}{2} (h' + 6\xi')\biggr],
\label{NUS4}
\end{equation}
where $\mu = \hat{k}\cdot\hat{n}$. Equations (\ref{NUS1}), (\ref{NUS2}) and (\ref{NUS3})
can be derived from Eq. (\ref{NUS4}) by expanding ${\mathcal F}_{\nu}$ in series of Legendre 
polynomials according to the convention:
\begin{equation}
{\mathcal F}_{\nu}(\vec{k},\hat{n},\tau)= \sum_{\ell} (-i)^{\ell}( 2\ell + 1) {\mathcal F}_{\nu\ell}(k,\tau)P_{\ell}(\mu).
\label{NUS5}
\end{equation}
For larger multipoles Eqs. (\ref{NUS1}), (\ref{NUS2}) and (\ref{NUS3}) 
can be written as 
\begin{equation}
{\mathcal F}_{\nu\ell}'= \frac{k}{2\ell + 1} [ \ell {\mathcal F}_{\nu(\ell -1)} - (\ell + 1) {\mathcal F}_{\nu(\ell+1)}],\qquad \ell\geq 3.
\label{NUS6}
\end{equation}
Equation (\ref{NUS6}) can be derived from Eq. (\ref{NUS4}) if we multiply  both sides by $P_{\ell}(\mu)$, i.e. 
by a generic Legendre Polynomial. By then integrating over $\mu$ (between $-1$ and $1$) and by using 
the convention established in Eq. (\ref{NUS5}), Eq. (\ref{NUS6}) follows by appropriate use 
of the recurrence relation of the Legendre polynomials stipulating that 
$(\ell + 1) P_{\ell + 1}(\mu) = ( 2 \ell + 1) \mu P_{\ell}(\mu) - \ell P_{\ell -1}(\mu)$ \cite{abr,grad}.
The numerical integration will demand to cut the hierarchy at an appropriately (large) multipole.
The neutrino fraction in the pre-equality plasma will be denoted by $R_{\nu}$ and it is defined 
as:
\begin{equation}
R_{\nu} = \frac{r}{1 + r},\qquad 
r = \frac{7}{8} N_{\nu} \biggl(\frac{4}{11}\biggr)^{4/3} \equiv 0.681 \biggl(\frac{N_{\nu}}{3}\biggr).
\label{rnu}
\end{equation}
By definition, the photon fraction at the corresponding time will be given by $R_{\gamma} = 1 - R_{\nu}$.
\subsection{Pre-decoupling plasma}
Overall, the plasma 
will obey the Gauss constraint whose explicit form can  be expressed, in the 
present case, as 
\begin{equation}
\vec{\nabla}\cdot \vec{E} =  4\pi e (n_{\mathrm{i}} - n_{\mathrm{e}}), \qquad 
\vec{E} = a^2 \vec{{\mathcal E}},
\label{gauss}
\end{equation}
where $n_{\mathrm{e}}$ is the electron concentration and  $n_{\mathrm{i}}$ is 
the concentration of the ions (to be soon identified with protons); $\vec{E}$ is the 
electric field rescaled through the second power of the scale factor \footnote{In the 
present paper we will denote with calligraphic style the electric and magnetic fields 
which are not rescaled through $a^2$, i.e. $\vec{E} = a^2 \vec{{\mathcal E}}$ and 
$\vec{B} = a^2 \vec{{\mathcal B}}$.}. Electrons and ions are approximately 
in thermal equilibrium for temperatures smaller than the MeV. The electric field 
appearing in the Gauss constraint will be screened for length scales $L > \lambda_{\mathrm{D}}$:
\begin{equation}
\lambda_{\mathrm{D}}(\tau_{\mathrm{eq}}) = \sqrt{\frac{T}{8\pi e^2 n_{0}} }= 2.8 \biggl(\frac{\omega_{\mathrm{M}}}{0.134}\biggr)^{-1} 
\biggl(\frac{\omega_{\mathrm{b}}}{0.023}\biggr)^{-1/2} \biggl(\frac{T}{\mathrm{eV}}\biggr)^{-1} \mathrm{m},
\label{BL1}
\end{equation}
where $T\simeq T_{\mathrm{e}} \simeq T_{\mathrm{i}} \simeq T_{\gamma}$. In Eq. (\ref{BL1}) 
the common value of the electron and ion concentrations is simply given by 
\begin{equation}
n_{0} = n_{\mathrm{e}} = n_{\mathrm{i}} = \eta_{\mathrm{b}} n_{\gamma}, \qquad 
\eta_{\mathrm{b}} = 6.27 \times 10^{-10} \biggl(\frac{\omega_{\mathrm{b}}}{0.023}\biggr),
\label{BL2}
\end{equation}
where the ionization fraction $x_{\mathrm{e}}$ is set to $1$ as it is consistent 
for pre-recombination temperatures. Since the Hubble rate at equality is given by 
$H_{\mathrm{eq}} = 1.65 \times 10^{-56} (\omega_{\mathrm{M}}/0.134)^{2} M_{\mathrm{P}}$ we have that 
\begin{equation}
H_{\mathrm{eq}} \lambda_{\mathrm{D}}(\tau_{\mathrm{eq}})\simeq 2.85 \times 10^{-21} \biggl(\frac{\omega_{\mathrm{M}}}{0.134}\biggr) \biggl(\frac{\omega_{\mathrm{b}}}{0.023}\biggr)^{-1/2} \biggl(\frac{T}{\mathrm{eV}}
\biggr)^{-1}.
\label{BL3}
\end{equation}
The hierarchy between the Debye length $\lambda_{\mathrm{D}}$ 
and the other length-scales of the problem 
persists if we move from the equality time to the decoupling time. 
According to Eq. (\ref{BL1}) $\lambda_{\mathrm{D}}$ scales \footnote{In principle the electron and baryon temperature (in the absence of energy exchange and in the absence of photons) should scale as $a^{-2}$. However, as it will be discussed in a moment, the Coulomb and Thompson rates of interactions 
are both larger than the Hubble rate and $T_{\mathrm{e}} \simeq T_{\mathrm{i}} \simeq T_{\gamma} = T$ 
(see Eqs. (\ref{Tem1})--(\ref{Tem6}) and discussions therein).} as $a(\tau)$ 
 (since $T\simeq a^{-1}$ and $n_{0} \simeq a^{-3}$).
Thus $\lambda(\tau_{\mathrm{dec}}) \simeq 3 (\omega_{\mathrm{M}}/0.134)\lambda_{\mathrm{eq}}$ 
where the factor $3  (\omega_{\mathrm{M}}/0.134)$ arises because $ (a_{\mathrm{dec}}/a_{\mathrm{eq}}) \simeq 
2.393 (\omega_{\mathrm{M}}/0.134)$ (taking, as an example, $1 + z_{\mathrm{dec}}
\simeq 1100$ and fixing the ionization fraction as $x_{\mathrm{e}} \simeq 1$). 

Recombination entails a sudden drop in the 
ionization fraction. From the usual considerations involving Saha's equation, around decoupling, $x_{\mathrm{e}} \simeq 10^{-5}$. The Debye length 
increases then (see Eq. (\ref{BL1})) by a factor $10^{2.5}$ which is still minute in comparison with 
all the other lengths of the problem. Note that $\lambda_{\mathrm{D}}(\tau_{\mathrm{dec}})$ is not only parametrically smaller 
than the Hubble radius, but it is also negligible in comparison with the sound horizon 
at the corresponding epoch, i.e. 
\begin{equation}
r_{\mathrm{s}}(\tau_{\mathrm{dec}})= \int_{0}^{\tau_{\mathrm{dec}}} d\tau c_{\mathrm{sb}}(\tau) = \int_{0}^{\tau_{\mathrm{dec}}} \frac{d\tau}{\sqrt{3[1 + R_{\mathrm{b}}(\tau)]}},
\label{SEE3}
\end{equation}
where $c_{\mathrm{sb}}(\tau)$ (the characteristic sound speed of the baryon-photon system 
in the tight coupling approximation) is defined in terms of the baryon to photon ratio $R_{\mathrm{b}}$ (see 
also Eqs. (\ref{BR4}) and (\ref{INC1})):
\begin{equation}
R_{\mathrm{b}}(z) = \frac{3}{4} \frac{\rho_{\mathrm{b}}}{\rho_{\gamma}}  = 0.664 \biggl(\frac{ \omega_{\mathrm{b}}}{0.023}\biggr) \biggl(\frac{1051}{z + 1}\biggr).
\label{Rbdef}
\end{equation}
Since the dark energy component is negligible around decoupling 
the integral appearing in Eq. (\ref{SEE3}) can be estimated analytically and the overall result can be 
expressed as:
\begin{equation}
\frac{r_{\mathrm{s}}(\tau_{\mathrm{dec}})}{\mathrm{Mpc}} = \frac{2998}{\sqrt{1 + z_{\mathrm{dec}}}}
\frac{2}{\sqrt{3 \, \omega_{\mathrm{M}} c_1}} \ln{\biggl[ \frac{\sqrt{1 + c_1} + \sqrt{c_1 + c_1 c_{2}}}{1 + \sqrt{c_1 c_2}}\biggr]},
\label{SEE7}
\end{equation}
where
\begin{equation}
c_{1} = 27.6\,\,  \omega_{\mathrm{b}}\,\,\biggl(\frac{1100}{1 + z_\mathrm{dec}}\biggr),\hspace{2cm}
c_{2} =\frac{0.045}{h_{0}^2 \omega_{\mathrm{M}}} \biggl(\frac{1 + z_{\mathrm{dec}}}{1100}\biggr).
\end{equation}
 With our fiducial values of the parameters, $r_{\mathrm{s}}(\tau_{\mathrm{dec}})$ lies between $150$ 
 and $200$ Mpc. But now this figure should be compared with the Debye length 
 $\lambda(\tau_{\mathrm{dec}}) \simeq 2.5 \times 10^{3}$ m (having taken into account 
 the drop in the ionization fraction). Thus, as anticipated, $\lambda_{\mathrm{D}}(\tau_{\mathrm{dec}})/r_{\mathrm{s}}(\tau_{\mathrm{dec}}) \simeq 10^{-21}$. So the plasma is, to a very good approximation globally neutral.

As already mentioned, for temperatures smaller than the 
temperature of neutrino decoupling baryons and electrons interact strongly through 
Coulomb scattering. The corresponding rate, for $T> \mathrm{eV}$, is 
\footnote{In the case of a proton (or of an electron) impinging 
on an electron (or on a proton) the Rutherford cross section is logarithmically 
divergent at large impact parameters when the particles are free. In the plasma around decoupling the logarithmic 
divergence is avoided  because of the Debye screening length: the cross section is then known as Coulomb cross section and the 
logarithmic divergence is replaced by the so-called Coulomb logarithm.}:
\begin{equation}
\frac{\Gamma_{\mathrm{Coul}}}{H} \simeq 4 \times 10^{11} \, x_{\mathrm{e}} 
\biggl(\frac{T}{\mathrm{eV}}\biggr)^{-1/2} \biggl(\frac{\omega_{\mathrm{b}}}{0.023}\biggr).
\label{BL4}
\end{equation}
Since after equality $H\propto T^{3/2}$ ( and ignoring for the 
moment the drop in the ionization fraction) the ratio of Eq. (\ref{BL4}) gets frozen. 
Equation (\ref{BL4}) justifies to consider 
a unique baryon-lepton fluid as a single dynamical entity.  In Eq. (\ref{BL4}) the Coulomb rate 
has been computed by recalling that 
\begin{eqnarray}
&&\Gamma_{\mathrm{Coul}} = n_{\mathrm{e}} \, x_{\mathrm{e}}\,v_{\mathrm{th}} \sigma_{\mathrm{Coul}},\qquad 
v_{\mathrm{th}} \simeq \sqrt{\frac{T}{m_{\mathrm{e}}}},
\nonumber\\
&& \sigma_{\mathrm{Coul}} = \frac{\alpha_{\mathrm{em}}^2}{T^2} \ln{\Lambda_{\mathrm{C}}},
\qquad \Lambda_{\mathrm{C}} = \frac{3}{2} \biggl(\frac{T^3}{\pi n_{\mathrm{e}}}\biggr)^{1/2} \frac{1}{e^3}.
\label{BL5}
\end{eqnarray}
where $\ln{\Lambda_{\mathrm{C}}} \simeq 14.71$ for typical values of $\omega_{\mathrm{b}}$. 
The Coulomb cross section is the main responsible for the conductivity of the plasma 
which can be estimated as the ratio between the square of the plasma frequency and the Coulomb 
rate (which is also, by definition, the collision frequency), namely 
\begin{equation}
\sigma_{\mathrm{c}}(T) = \frac{\omega_{\mathrm{pe}}^2}{4\pi \Gamma_{\mathrm{Coul}}} = 
\frac{4\pi}{\alpha_{\mathrm{em}}\ln{\Lambda_{\mathrm{C}}}} T \biggl(\frac{T}{m_{\mathrm{e}}}\biggr)^{1/2}\simeq 
0.16 \biggl(\frac{T}{\mathrm{eV}}\biggr)^{3/2} \,\, \mathrm{eV},
\label{BL6}
\end{equation}
where, as it should, the electron concentration effectively simplifies in the final expression.
It is useful also to estimate, at this point, the plasma frequency of the electrons, i.e.
\begin{equation}
\omega_{\mathrm{pe}} = \sqrt{\frac{4\pi e^2 n_{\mathrm{e}}}{m_{\mathrm{e}}}}= 28.05 \sqrt{x_{\mathrm{e}}}
\biggl(\frac{T}{\mathrm{eV}}\biggr)^{3/2} \biggl(\frac{\omega_{\mathrm{b}}}{0.023}\biggr)^{1/2}\,\, \mathrm{MHz}.
\label{B7}
\end{equation}
Thus the typical length-scales are much larger than the Debye scale.
The typical time-scales greatly exceed  $\omega_{\mathrm{pe}}^{-1}$. This is 
the realm of MHD.

Prior to equality electrons and protons 
interact also with photons via Thompson cross-section. Protons can be neglected  in the Thompson mean free path (determined by electron-photon interactions); the Thompson rate in 
units of the Hubble rate is,  for $T> \mathrm{eV}$,
\begin{equation}
\frac{\Gamma_{\mathrm{Th}}}{H} \simeq 5.9 \times 10^{4} x_{\mathrm{e}}\biggl(\frac{\omega_{\mathrm{b}}}{0.023}\biggr)
\biggl(\frac{T}{\mathrm{eV}}\biggr).
\label{B8}
\end{equation}
After equality $\Gamma_{\mathrm{Th}}/H$ is proportional to $T^{3/2}$ and it becomes 
eventually much smaller than one as $x_{\mathrm{e}}$ drops at recombination. 
Deep in the radiation epoch, i.e. when the initial conditions of CMB anisotropies are 
set numerically, the Coulomb rate of Eq. (\ref{BL4}) is larger than the Thompson rate 
but while the Thompson rate increases with the temperature (see Eq. (\ref{B8})) the 
Coulomb rate decreases. The meeting point of the two rates occurs 
close to the MeV. Initial conditions 
will then be set in the radiation epoch when both Coulomb and Thompson scattering are large.
Also photons are strongly coupled to the baryon-lepton 
fluid. So, a unique physical entity emerges, i.e. the so-called baryon-lepton-photon
fluid.  This fluid is often dubbed as the baryon-photon fluid by implicitly 
including the electrons in the baryonic component thanks to the strength 
of Coulomb coupling. 

The various species have all a 
putative common temperature $T$, i.e. $T_{\mathrm{e}} \simeq T_{\mathrm{p}} \simeq T_{\gamma} = T$. 
This statement will now be justified.
Electrons and protons, being massive, have energy densities and pressures 
which can be written, respectively, as
\begin{eqnarray}
&& \rho_{\mathrm{e}} = n_{\mathrm{e}}\biggl[ m_{\mathrm{e}} + \frac{3}{2} T_{\mathrm{e}}\biggr],\qquad 
p_{\mathrm{e}} = n_{\mathrm{e}} T_{\mathrm{e}},
\label{Tem1}\\
&& \rho_{\mathrm{p}} = n_{\mathrm{p}}\biggl[ m_{\mathrm{p}} + \frac{3}{2} T_{\mathrm{p}}\biggr],\qquad 
p_{\mathrm{p}} = n_{\mathrm{p}} T_{\mathrm{p}}.
\label{Tem2}
\end{eqnarray}
The photon energy density and pressure will be instead, as it is well known, 
$\rho_{\gamma} = (\pi^2/15) T_{\gamma}^4$ and $ p_{\gamma} = \rho_{\gamma}/3$.
Ignoring, for the moment, the other species of the plasma,
 the total conservation equation assumes the form $d(a^3 \rho) + p d (a^3) =0$ where 
$\rho = (\rho_{\mathrm{e}} + \rho_{\mathrm{p}} + \rho_{\gamma})$ and $ p = 
(p_{\mathrm{e}} + p_{\mathrm{p}} + p_{\gamma})$. Since $n_{\mathrm{e}} = n_{\mathrm{p}} = n_{0}$ (and both scale as $a^{-3}$) the total conservation equation implies the following differential relation: 
\begin{equation}
a \lambda d (a T_{\gamma}) + d [ a^2 ( T_{\mathrm{e}} + T_{\mathrm{p}})] =0, \qquad 
\lambda = 2.8 \times 10^{9} \biggl(\frac{\omega_{\mathrm{b}}}{0.023}\biggr)^{-1}.
\label{Tem3}
\end{equation}
Up to numerical factors $\lambda$ is the ratio between the entropy density of the photons and 
$n_{0}$ which is, in turn, roughly $10$ orders of magnitude smaller than the photon concentration. 
The electron and proton temperatures vary adiabatically as $a^{-2}$ (in the absence of photons) while the radiation temperature varies as $a^{-1}$ (in the absence of protons and electrons). But we do know that electron-photon and electron-proton interactions tie the temperatures 
close together. Equation (\ref{Tem3}) can then be solved assuming, to lowest order,  $T_{\mathrm{e}} \simeq T_{\mathrm{p}} \simeq T_{\gamma} = T$. The differences in the various temperatures can be 
estimated: if the differences  are small  the assumption of a common temperature is justified.
To lowest order Eq. (\ref{Tem3}) becomes
\begin{equation}
\frac{ d \ln{T_{\gamma}}}{d\ln{a}} = - \frac{\lambda + 4}{\lambda + 2},\qquad T_{\gamma} \simeq a^{- 1 - \frac{2}{\lambda}},
\label{Tem4}
\end{equation}
where the second relation is obtained from the first one after expanding the obtained result in powers of 
$1/\lambda$.  Equation (\ref{Tem4}) shows that, indeed, $T_{\gamma}$ evolves in a way which is 
intermediate between $a^{-1}$ (as implied in the absence of electrons and protons) and $a^{-2}$.
The differences between the $T_{\gamma}$, $T_{\mathrm{e}}$ and $T_{\mathrm{p}}$ depend upon the 
Thompson and Coulomb rates. The rate of gain of energy per electron as well as the rates of gain of electron and proton thermal energies can be written as 
\begin{eqnarray}
&& \frac{1}{a^2}  \frac{d (a^2 T_{\mathrm{e}})}{d t} = - \Gamma_{\mathrm{Th}}  (T_{\mathrm{e}} - T_{\gamma}) - \Gamma_{\mathrm{Coul}}( T_{\mathrm{e}} - T_{\mathrm{p}}),
\nonumber\\
&&\frac{1}{a^2}  \frac{d (a^2 T_{\mathrm{p}})}{d t} = - \Gamma_{\mathrm{Coul}}( T_{\mathrm{p}} - T_{\mathrm{e}}),
\qquad \frac{\lambda}{a} \frac{d (a T_{\gamma})}{d t} = - \Gamma_{\mathrm{Th}} (T_{\gamma} - T_{\mathrm{e}}),
\label{Tem5}.
\end{eqnarray}
Equation (\ref{Tem5}) implies, as expected, that 
\begin{equation}
\frac{T_{\gamma} - T_{\mathrm{e}}}{T} \simeq 2 \frac{H}{\Gamma_{\mathrm{Th}}},\qquad 
\frac{T_{\mathrm{e}} - T_{\mathrm{p}}}{T} \simeq \frac{H}{\Gamma_{\mathrm{Coul}}}.
\label{Tem6}
\end{equation}
The same hierarchy between Coulomb and Thompson scattering rates also 
determines the small temperature differences between electrons, protons and photons. 
Thus electron-proton collisions are sufficiently 
fast to assess that electrons and protons have indeed the same putative temperature.

\subsection{The baryon-lepton-photon fluid}
The strength of the Coulomb coupling implies that, effectively, there is a unique 
velocity field which is the centre of mass velocity of the electron-proton
fluid, i.e.
\begin{equation}
\vec{v}_{\mathrm{b}} = \frac{m_{\mathrm{e}} \vec{v}_{\mathrm{e}} + m_{\mathrm{p}} \vec{v}_{\mathrm{p}}}{m_{\mathrm{e}} + m_{\mathrm{p}}}.
\label{BL9}
\end{equation}
The velocity $\vec{v}_{\mathrm{b}}$ is the bulk velocity of the plasma \cite{spitzer,krall}.
The evolution equation of $\vec{v}_{\mathrm{b}}$ can be obtained by summing up the 
evolution equations of electrons and ions \cite{krall}, as it happens in the usual MHD treatment of the problem. 
In the synchronous gauge, the baryon velocity and the baryon 
density contrast obey, respectively, the following pair of equations
\begin{eqnarray}
&& \delta_{\mathrm{b}}' = - \theta_{\mathrm{b}} + \frac{h'}{2} + \frac{\vec{E}\cdot\vec{J}}{a^4 \rho_{\mathrm{b}}},
\label{BL10}\\
&& \theta_{\mathrm{b}}' + {\mathcal H} \theta_{\mathrm{b}} = \frac{4}{3} \frac{\rho_{\gamma}}{\rho_{\mathrm{b}}} \epsilon' 
(\theta_{\gamma} - \theta_{\mathrm{b}}) + \frac{ \vec{\nabla} \cdot [ \vec{J} \times \vec{B}] }{a^4 \rho_{\mathrm{b}}}.
\label{BL11}
\end{eqnarray}
where, the divergence of the baryon velocity and the differential optical depth have been introduced as
\begin{equation}
\theta_{\mathrm{b}} = \vec{\nabla} \cdot\vec{v}_{\mathrm{b}},\qquad \epsilon' = x_{\mathrm{e}} \frac{a}{a_{0}} \sigma_{\mathrm{T}} n_{\mathrm{e}}.
\label{BL11a}
\end{equation}
Equation (\ref{BL11}) has been written, unlike the analog equations for CDM and neutrinos, not in Fourier space but in real space to
emphasize the presence of a new term which is nothing but the MHD form of the Lorentz force 
given by $\vec{J}\times \vec{B}$ where $\vec{J}$ is the Ohmic current and $\vec{B}$ is the magnetic field.
The electric field appearing in Eq. (\ref{BL10}) is negligible since, in the plasma frame, the conductivity 
effectively suppresses the Ohmic electric fields.

According to Eq. (\ref{BL11}) the baryon-lepton fluid exchanges momentum with 
the photons. The lowest two multipoles of the Boltzmann hierarchy of the photons, namely 
the density contrast (i.e. the monopole) and the three-divergence of the velocity field (related 
to the dipole of the intensity of the brightness perturbations) are:
\begin{eqnarray}
\delta_{\gamma}' &=& - \frac{4}{3} \theta_{\gamma} + \frac{2}{3} h' ,
\label{PHOT1}\\
\theta_{\gamma}' &=& -\frac{1}{4} \nabla^2 \delta_{\gamma} + \epsilon' (\theta_{\mathrm{b}} - \theta_{\gamma}).
\label{PHOT2}
\end{eqnarray}
 While the sum of the 
electron and proton equations leads to Eq. (\ref{BL11}), their difference 
leads to the Ohm law \cite{spitzer,krall} which relates the total current $\vec{J}$ to the 
electric field through the conductivity, i.e.
\begin{equation}
\vec{J} = \sigma ( \vec{E} + \vec{v}_{\mathrm{b}} \times \vec{B}), \qquad \sigma = \sigma_{\mathrm{c}} a,\qquad 
\vec{J} = a^3 \vec{j},
\label{Ohm}
\end{equation}
where $\sigma_{\mathrm{c}}$ denotes the flat-space conductivity; furthermore, as already mentioned, 
$\vec{B} = a^2 \vec{{\mathcal B}}$ and $\vec{E} = a^2 \vec{{\mathcal E}}$. The usefulness of the latter 
rescalings can be understood by looking also at the other MHD equations, namely
\begin{eqnarray}
&& \vec{\nabla} \cdot \vec{E} =0,\qquad \vec{\nabla}\cdot\vec{B} =0,
\label{BL12}\\
&& \vec{\nabla}\times \vec{B} = 4\pi \vec{J},\qquad \frac{\partial \vec{B}}{\partial \tau} + \vec{\nabla}\times 
\vec{E}=0.
\label{BL13}
\end{eqnarray}
Equations (\ref{BL12}) and (\ref{BL13}) have the same form they would have in flat space. 
The space-time is however curved and with line element given by Eq. (\ref{line}).  Maxwell 
equations in conformally flat backgrounds are known to be invariant 
under a Weyl rescaling of the metric. Consequently 
the corresponding evolution equations have exactly the same form they would have in Minkowskian 
space-time provided the field are appropriately rescaled and provided the conformal time 
coordinate $\tau$ is consistently employed. 

The displacement current does not appear in Eq. (\ref{BL13}). Indeed MHD is a description that 
holds for typical length-scales that are larger than the Debye length and for 
typical time-scales that are much larger than the inverse of the plasma frequency. In other 
words, if we are interested to study the high frequency branch of the spectrum of plasma 
excitations we should resort to a full kinetic (Vlasov-Landau) description \cite{krall}. 
The Ohmic current can then be related to the magnetic field, i.e. 
\begin{equation}
\vec{J} = \frac{1}{4\pi} \vec{\nabla}\times \vec{B}, \qquad \vec{\nabla}\cdot\vec{J} =0.
\label{BL14}
\end{equation}
Thus, the total current, the electric field and the magnetic field are all 
solenoidal albeit for rather different physical reasons. 
Equation (\ref{BL14}) can be used to compute explicitly the Ohmic electric field, i.e. 
\begin{equation}
\vec{E} = - \vec{v}_{\mathrm{b}} \times \vec{B} + \frac{\vec{\nabla}\times\vec{B}}{4\pi \sigma},
\label{BL15}
\end{equation}
which shows that electric fields vanish, at finite conductivity, in the baryon rest frame. They are therefore 
smaller than the magnetic fields since, as 
previously shown explicitly, the pre-decoupling plasma is an excellent 
conductor. The latter statement defines the plasma frame, i.e. the frame 
where, thanks to the large value of the conductivity, the electric fields vanish 
while the magnetic field are not dissipated, by conductivity, at large 
scales. 

Also magnetic fields are affected by conductivity but to a lesser extent and only 
at sufficiently short scales (which are already erased by the finite value 
of the thermal diffusivity scale, i.e. Silk damping).
The typical magnetic diffusivity scale (i.e. the length-scale below which the magnetic
field is dissipated by the finite value of the conductivity) can be understood from the corresponding
magnetic diffusivity equation. Inserting Eq. (\ref{BL15}) into the second relation 
of Eq. (\ref{BL13}) the magnetic diffusivity equation can be written as:
\begin{equation}
\frac{\partial \vec{B}}{\partial \tau} = \vec{\nabla} \times (\vec{v}_{\mathrm{b}}\times\vec{B}) + \frac{\nabla^2 \vec{B}}{4\pi\sigma}.
\label{BL16}
\end{equation}
According to Eq. (\ref{BL16}) the magnetic field power spectrum will be diffused for typical 
wave-numbers $k > k_{\sigma} \simeq \sqrt{4\pi \sigma_{\mathrm{c}} H}$. Thus,
only sufficiently short length-scales $L < L_{\sigma} \simeq k_{\sigma}^{-1}$ are 
dissipated. The ratio of $L_{\sigma}$ to the Hubble radius, i.e. $L_{\sigma} H$,
 being suppressed by $(T/M_{\mathrm{P}})$, is always minute. Around 
equality we can estimate that $L_{\sigma} H \simeq 3.9\times 10^{-17} (T/\mathrm{eV})^{1/4}$ where 
$\sigma_{\mathrm{c}}$ is given by  Eq. (\ref{BL6}).

For $T< \mathrm{MeV}$ the kinetic Reynolds 
number is smaller than one. This property is not verified, for instance, during the 
life of spiral galaxies where, effectively, the kinetic energy of the plasma can be converted into magnetic energy 
by means of the first term of Eq. (\ref{BL16}) which is often dubbed dynamo term \cite{review1}. When 
the kinetic Reynolds number is small (i.e. in the absence of kinetic turbulence) the plasma description following from MHD can be also phrased 
in terms of the conservation of two interesting quantities, i.e. 
the magnetic flux and the magnetic helicity \cite{biskamp}:
\begin{eqnarray}
&& \frac{d}{d t} \biggl(\int_{\Sigma} \vec{B} \cdot d\vec{\Sigma} \biggr) = - \frac{1}{4\pi \sigma }\int \vec{\nabla} \times\vec{\nabla}\times \vec{B} \cdot d\vec{\Sigma},
\label{fluxcons}\\
&& \frac{d}{dt} \biggl(\int_{V} d^{3} x \vec{A} \cdot \vec{B} \biggr) = - \frac{1}{4\pi \sigma} \int_{V} d^{3} x \vec{B} \cdot \vec{\nabla} \times \vec{B}.
\label{helcons}
\end{eqnarray}
In Eq. (\ref{fluxcons}), $\Sigma$ is an arbitrary finite surface that 
moves with the plasma. In the ideal MHD limit (i.e. $\sigma = a\sigma_{\mathrm{c}} \to \infty$) the magnetic flux is 
conserved. In the same limit also the magnetic helicity is 
conserved.  In the resistive limit the magnetic flux and helicity are dissipated 
with a rate proportional to $1/\sigma$ which is small provided the conductivity 
is sufficiently high. The term appearing at the right hand side of Eq. (\ref{helcons}) 
is called magnetic gyrotropy. Since, at high temperatures, the conductivity grows with $T$
the ideal limit is always verified better and better as we go back in time.

The conservation of the magnetic helicity is a statement on the conservation 
of the topological  properties of the 
magnetic flux lines. If the magnetic field is completely 
stochastic, the magnetic flux lines will be closed loops 
evolving independently in the plasma and the helicity 
will vanish. There could be, however, 
more complicated topological situations
where a single magnetic loop is twisted (like some 
kind of M\"obius stripe) or the case where 
the magnetic loops are connected like the rings of a chain.
In both cases the magnetic helicity will not be zero 
since it measures, essentially, the number of links and twists 
in the magnetic flux lines.  The magnetic helicity 
will have no impact on our considerations 
since the scalar fluctuations of the geometry are not affected 
by the helical features of the magnetic fields. On the 
contrary in the vector and tensor cases the situation 
can be different \cite{hel}.

In the resistive MHD approximation the electric components of the energy-momentum 
tensor can be neglected, while the magnetic components are present only at sufficiently 
large scales $L> L_{\sigma}$:
\begin{equation}
{\mathcal T}_{0}^{0}(\vec{x},\tau) = \frac{B^2}{8\pi a^4}, 
\qquad {\mathcal T}_{i}^{j}(\vec{x},\tau)= \frac{1}{4\pi a^4}\biggl[ B_{i} B^{j} - \frac{B^2}{2}\delta_{i}^{j}\biggr]
\label{BB1}
\end{equation}
where $B^2 = B_{i}B^{i}$. In Eq. (\ref{BB1}) the contribution 
of the electric terms can be neglected since they are all suppressed by two powers of the conductivity.
The Poynting vector can be also neglected since, at finite conductivity 
is suppressed as $\sigma^{-1}$ 
\begin{equation}
\qquad {\mathcal T}_{0}^{i}(\vec{x},\tau) = \frac{1}{4\pi a^4}\vec{E}\times \vec{B} \simeq 
 \frac{1}{4\pi a^4 \sigma} \vec{J} \times \vec{B},
\label{BB3}
\end{equation}
where Eq. (\ref{BL13}) has been used in the second equality.
The spatial components 
of the energy momentum tensor can be phrased in terms of the magnetic pressure and of the 
anisotropic stress, i.e. 
\begin{equation}
{\mathcal T}_{i}^{j}(\vec{x},\tau) = - \delta p_{\mathrm{B}}(\vec{x},\tau) \delta_{i}^{j} + \tilde{\Pi}_{i}^{j}(\vec{x},\tau),\qquad 
{\mathcal T}_{0}^{0}(\vec{x},\tau) = \delta\rho_{\mathrm{B}}(\vec{x},\tau),
\label{BB5}
\end{equation}
where, with standard notations:
\begin{eqnarray}
&& \delta\rho_{\mathrm{B}}(\vec{x},\tau) = \frac{B^2(\vec{x})}{8\pi a^4},\qquad \delta p_{\mathrm{B}} = 
\frac{\delta\rho_{\mathrm{B}}}{3},
\label{BB6}\\
&& \tilde{\Pi}_{i}^{j}(\vec{x},\tau) = \frac{1}{4\pi a^4}\biggl[ B_{i} B^{j} - \frac{B^2}{3} \delta_{i}^{j}\biggr].
\label{BB7}
\end{eqnarray}
It is practical to refer the 
magnetic fields to the photon background by means of the following 
rescaling 
\begin{equation}
\Omega_{\mathrm{B}}(\vec{x},\tau) = \frac{\delta\rho_{\mathrm{B}}(\vec{x},\tau)}{\rho_{\gamma}(\tau)}
\equiv\frac{B^2(\vec{x},\tau)}{8\pi \overline{\rho}_{\gamma}}, 
\qquad \partial_{j}\partial^{i} \tilde{\Pi}_{i}^{j} = (p_{\gamma}+ \rho_{\gamma}) \nabla^2 \sigma_{\mathrm{B}}.
\label{BB8}
\end{equation}
where $\overline{\rho}_{\gamma} = a^4 \rho_{\gamma}$. 
With the notations of Eq. (\ref{BB8}) the  identity
\begin{equation}
\nabla^2 \sigma_{\mathrm{B}} = \frac{3}{16\pi \overline{\rho}_{\gamma}} \partial_{i} B_{j} \partial^{j} B^{i} - 
\frac{1}{2} \nabla^2 \Omega_{\mathrm{B}}
\label{BB9}
\end{equation}
 allows to express the three-divergence of the Lorentz force 
appearing in Eq. (\ref{BL11}) in terms of $\Omega_{\mathrm{B}}$ and $\sigma_{\mathrm{B}}$:
\begin{equation}
\frac{3}{4} \frac{\vec{\nabla}\cdot [ \vec{J} \times \vec{B}]}{\overline{\rho}_{\gamma}} = 
\nabla^2 \sigma_{\mathrm{B}} - \frac{1}{4} \nabla^2 \Omega_{\mathrm{B}}.
\label{BB10}
\end{equation}

Stochastically distributed large-scale magnetic fields 
 do not break the spatial isotropy of the background geometry 
introduced in Eq. (\ref{line}). Nearly all magnetogenesis mechanisms suggest 
indeed that the large-scale magnetic fields should be stochastically distributed 
and characterized by their two-point function. Defining the Fourier amplitude 
of the magnetic fields as 
\begin{equation}
B_{i}(\vec{x}) = \frac{1}{(2\pi)^{3/2}} \int d^{3} k B_{i}(k) e^{-i \vec{k}\cdot\vec{x}},
\label{BB11}
\end{equation}
their two-point function can be expressed as 
\begin{equation}
\langle B_{i}(\vec{k}) B_{j}(\vec{p}) \rangle = \frac{2\pi^2}{k^3} P_{ij}(k) P_{\mathrm{B}}(k) \delta^{(3)}(\vec{k} + \vec{p}),
\label{BB12}
\end{equation}
where 
\begin{equation}
P_{ij}(k) = \biggl(\delta_{ij} - \frac{k_{i} k_{j}}{k^2}\biggr),\qquad P_{\mathrm{B}}(k) = 
{\mathcal A}_{\mathrm{B}} \biggl(\frac{k}{k_{\mathrm{L}}}\biggr)^{n_{\mathrm{B}} -1}.
\label{BB13}
\end{equation}
In Eq. (\ref{BB13}) $n_{\mathrm{B}}$ is the magnetic spectral index and ${\mathcal A}_{\mathrm{B}}$ is the amplitude 
\footnote{Note that, according to Eq. (\ref{BB11}), the dimensions of $B_{i}(\vec{x})$ are of $L^{-2}$. Thus,
dimensionally, $[B_{i}(k)] = L$.  But then it is easy to see (taking into account the dimensions of the 
three-dimensional Dirac delta function) that, dimensionally, $[{\mathcal A} _{\mathrm{B}}] =L^{-4}$, i.e. 
${\mathcal A}_{\mathrm{B}}$ (and hence $P_{\mathrm{B}}(k)$) has the same dimensions of the 
magnetic energy density in real space. This is another good reason, unlike previous studies (see \cite{vt2} last 
three references), to 
follow the conventions expressed by Eq. (\ref{BB12}).}
of the magnetic power spectrum referred to the magnetic pivot scale $k_{\mathrm{L}}$. In terms of this definition 
the two-point function of the magnetic fields in real space can be written as 
\begin{equation}
\langle B_{i}(\vec{x}) B_{j}(\vec{y})\rangle = \int d\ln{k} P_{ij}(k) P_{\mathrm{B}}(k) \frac{\sin{kr}}{kr},\qquad r= |\vec{x} - \vec{y}|
\label{BB14}
\end{equation}
Different conventions exist in the literature for assigning the magnetic power spectrum.
For instance in \cite{vt1,vt2} the $k^{-3}$ (appearing at the right hand side of Eq. (\ref{BB12})) was 
included in the definition of $P_{\mathrm{B}}(k)$. Those authors, indeed, only dealt with tensor 
and vector modes and were not confronted with the necessity of assigning power spectra 
according to the standards of CMB physics. We are 
forced on Eqs. (\ref{BB11})--(\ref{BB13}) 
since these are the conventions used to 
define the power spectrum of curvature perturbations (see for instance \cite{WMAP1,WMAP2,WMAP3,WMAP4}) and it would be strange to 
use  normalizations and  definitions of the spectral indices 
that may differ from the ones which are commonly established when 
presenting theoretical and observational studies of the parameter space of CMB anisotropies.

\subsection{Gravitating magnetic fields}
All the species introduced so far 
gravitate and, therefore, affect the evolution of the metric perturbations $\xi$ and $h$. 
In Fourier space, the Hamiltonian and the momentum constraints stemming from the $(00)$ and $(0i)$ 
components of the perturbed Einstein equations are, respectively
\begin{eqnarray}
&& 2k^2 \xi - {\mathcal H} h' = 8\pi Ga^2 [\delta\rho_{\mathrm{t}} + \delta\rho_{\mathrm{B}}],
\label{HAM1}\\
&& k^2 \xi' = - 4\pi G a^2 ( p_{\mathrm{t}} + \rho_{\mathrm{t}}) \theta_{\mathrm{t}}.
 \label{MOM1}
 \end{eqnarray}
 In Eqs. (\ref{HAM1}) and (\ref{MOM1}) $\delta\rho_{\mathrm{t}}$ and $\theta_{\mathrm{t}}$ are 
 the global density fluctuation of the plasma and the total velocity field defined as 
 \begin{eqnarray}
 \delta\rho_{\mathrm{t}} &=& \delta \rho_{\mathrm{c}} + \delta\rho_{\nu} + \delta\rho_{\gamma} + \delta\rho_{\mathrm{b}},
 \label{TOTD}\\
 (p_{\mathrm{t}} + p_{\mathrm{t}}) \theta_{\mathrm{t}} &=& \sum_{a} (p_{\mathrm{a}} + \rho_{\mathrm{a}})\theta_{\mathrm{a}} \equiv \frac{4}{3}\rho_{\nu} \theta_{\nu} + \frac{4}{3} \rho_{\gamma} \theta_{\gamma} + 
 \rho_{\mathrm{c}} \theta_{\mathrm{c}} + \rho_{\mathrm{b}} \theta_{\mathrm{b}}.
 \label{TOTV}
 \end{eqnarray}
 The spatial components of the perturbed Einstein equations (i.e., respectively, $(i=j)$ and $(i\neq j)$)
  lead instead to:
 \begin{eqnarray}
 && h'' + 2 {\mathcal H} h' - 2 k^2 \xi = 24 \pi Ga^2 [\delta p_{\mathrm{t}} + \delta p_{\mathrm{B}}],
\label{SP1}\\
&& (h + 6 \xi)'' + 2 {\mathcal H} ( h + 6 \xi)' - 2 k^2 \xi = 24 \pi G a^2 [ (p_{\nu} + \rho_{\nu}) \sigma_{\nu} + 
(p_{\gamma} + \rho_{\gamma}) \sigma_{\mathrm{B}}].
\label{SP2}
\end{eqnarray}
In Eq. (\ref{SP2}) the neutrino anisotropic stress (also appearing in Eqs. (\ref{NUS2}) and (\ref{NUS3})) 
has been consistently included.
For analytical estimates  it is also useful to write the 
evolution equation for the total density contrast which reads, in the synchronous gauge, 
\begin{equation}
\delta \rho_{\mathrm{t}}' + 3 {\mathcal H} ( c_{\mathrm{st}}^2 + 1) \delta \rho_{\mathrm{t}} + 3 {\mathcal H} \delta p_{\mathrm{nad}} + (p_{\mathrm{t}} + \rho_{\mathrm{t}}) \theta_{\mathrm{t}} - (p_{\mathrm{t}} + \rho_{\mathrm{t}}) \frac{h'}{2} =0,
\label{TDC1}
\end{equation}
where the total sound speed $c_{\mathrm{st}}^2$ and the non-adiabatic pressure 
fluctuation $\delta p_{\mathrm{nad}}$ have been introduced.
Their respective definitions can be extracted from the following pair of relations:
\begin{equation}
\delta p_{\mathrm{t}} = c_{\mathrm{st}}^2 \delta\rho_{\mathrm{t}} + \delta p_{\mathrm{nad}}, \qquad 
c_{\mathrm{st}}^2 = \frac{p_{\mathrm{t}}'}{\rho_{\mathrm{t}}'}.
\label{TDC2}
\end{equation}
Equation (\ref{TDC2}) implies that the pressure fluctuations can be generated either 
by inhomogeneities  in the energy density or by fluctuations of the sound speed itself. The latter 
fluctuations are non-adiabatic in nature since they arise, physically, as a fluctuation of the 
specific entropy, i.e. the entropy density of the photon gas measured in units of the concentration 
of another given species. This property is customarily used to classify the initial conditions 
of CMB anisotropies which are therefore divided into adiabatic and non-adiabatic. 

In the case of the adiabatic mode, by definition, the fluctuations in the entropy density vanish over typical scales 
larger than the Hubble radius at recombination. The opposite holds for the non-adiabatic 
modes. In the case of the CDM-radiation mode the specific entropy is just 
given by ${\varsigma} = T^3/n_{\mathrm{c}}$ where $n_{\mathrm{c}}$ is the concentration 
of the CDM particles. The entropy fluctuations (i.e. 
the relative fluctuations in the specific entropy) are given by 
\begin{equation}
{\mathcal S} = \frac{\delta {\varsigma}}{\varsigma} = \frac{3}{4} \delta_{\gamma} - \delta_{\mathrm{c}}.
\label{TDC3}
\end{equation}
 More generally, given two species of the plasma the entropy fluctuations 
 are defined as \cite{kodama,malik,gioentro}
 \begin{equation}
 {\mathcal S}_{\mathrm{ij}} = -\biggl(\frac{\delta_{\mathrm{i}}}{w_{\mathrm{i}} + 1} - 
 \frac{\delta_{\mathrm{j}}}{w_{\mathrm{j}} + 1}\biggr).
 \label{TDC4}
 \end{equation}
 The non-adiabatic pressure fluctuations can then be written as\footnote{Owing to their definitions, both ${\mathcal S}_{\mathrm{ij}}$ and $\delta p_{\mathrm{nad}}$ are 
 gauge-invariant.}:
 \begin{equation}
 \delta p_{\mathrm{nad}}= \frac{1}{6 {\mathcal H} \rho_{\mathrm t}'} \sum_{{\mathrm i}\,{\mathrm j}} 
\rho_{\mathrm i}'\,\rho_{\mathrm j}' (c_{\mathrm{ s\,i}}^2 -c_{\mathrm{ s\,j}}^2) {\cal S}_{\mathrm{ i\,j}},\qquad c_{\mathrm{ s\,i}}^2 = 
\frac{p_{\mathrm i}'}{\rho_{\rm i}'},
\label{TDC5}
\end{equation}
where $c_{\mathrm{ s\,i}}^2$ and  $c_{\mathrm{ s\,j}}^2$ are sound speeds of two (generic) species.
Thus, according to Eq. (\ref{TDC5}),  $\delta p_{\mathrm{nad}}$ measures, indeed,  the degree 
of compositeness of the plasma: if more species are present, more non-adiabatic modes are possible and 
$\delta p_{\mathrm{nad}}$ receives more contributions. 
In the pre-equality plasma there are four (regular) non-adiabatic modes (i.e. the CDM-radiation mode, the baryon-radiation mode, the 
 neutrino density and the neutrino isocurvature modes). There is one adiabatic mode whose 
 presence is  strongly suggested by the analysis of cosmological data in the framework of a $\Lambda$CDM 
 scenario. This does not exclude the presence of a dominant adiabatic mode with a subdominant 
 non-adiabatic component (see, for instance, \cite{h1}).
 
 The analysis will be here limited to the case of a single adiabatic mode in the presence 
 of large-scale magnetic fields. This is the minimal situation compatible 
 with the $\Lambda$CDM framework. Our code can accommodate also 
 non-adiabatic initial conditions in the presence of magnetic fields. The 
 initial conditions discussed in \cite{mg2}, if appropriately translated to the synchronous frame, allow for this 
 possibility. Mixed initial conditions involve a dominant magnetized adiabatic mode 
 and a number of magnetized non-adiabatic modes with subdominant amplitude. 

 Two variables are customarily used  to parametrize the power spectrum of the metric fluctuations.
They will be denoted by ${\mathcal R}$ and by $\zeta$. 
In terms of the synchronous degrees of freedom, they 
can be defined  as 
\begin{equation}
{\mathcal R} = \xi + \frac{{\mathcal H} \xi'}{{\mathcal H}^2 - {\mathcal H}'},\qquad
\zeta = \xi - \frac{{\mathcal H} (\delta \rho_{\mathrm{t}} + \delta \rho_{\mathrm{B}})}{\rho_{\mathrm{t}}'}.
\label{TDC6}
\end{equation}
Even if both  ${\mathcal R}$ and $\zeta$ are gauge-invariant, their physical interpretation is obtained by expressing the two variables 
in specific gauges: ${\mathcal R}$ is often dubbed curvature perturbation since it corresponds, in the 
comoving orthogonal gauge, to the perturbations of the spatial curvature. In analog terms, $\zeta$ 
is interpreted as the curvature perturbation in the gauge where the density contrast vanishes 
(also called uniform density gauge). 
Taking the difference of ${\mathcal R}$ and $\zeta$ and using the Hamiltonian constraint 
(\ref{HAM1}), the following equation can be obtained:
\begin{equation}
 \zeta - {\mathcal R} = - \frac{2 k^2\xi - ( h + 6 \xi)'}{24 \pi G a^2 (p_{\mathrm{t}} + \rho_{\mathrm{t}})}
\label{TDC7}
\end{equation}
The quantity at the right hand side of Eq. (\ref{TDC7}) is ${\mathcal O}(k^2 \tau^2)$ and, therefore,
 it is negligible when the relevant wavelengths are larger than the Hubble radius, in particular 
 around equality. This property can be immediately understood by 
 expressing the combination at the right hand side of Eq. (\ref{TDC7}) in terms 
 of longitudinal gauge variables and, most notably, $\psi$ which denotes 
 the spatial fluctuation of the metric in the longitudinal gauge:
 \begin{equation}
\zeta - {\mathcal R} = - \frac{ k^2 \psi}{12\pi G a^2 (p_{\mathrm{t}} + \rho_{\mathrm{t}})},\qquad 
\psi = - \xi + \frac{h' + 6 \xi'}{2 k^2}.
\label{TDC8}
\end{equation}
In the longitudinal gauge $\psi$ is constant, to lowest order, when the relevant wavelengths 
are larger than the Hubble radius (see Appendix A).
The evolution of $\zeta$ on scales 
larger than the Hubble radius translates immediately in the evolution of ${\mathcal R}$. 
The evolution of $\zeta$ can be simply obtained by inserting Eq. (\ref{TDC6}) 
into Eq. (\ref{TDC1}). The logic is to trade $\delta \rho_{\mathrm{t}}$ in favor of 
$\zeta$. The result of this manipulation, after the use of the covariant conservation 
equation of the total fluid (i.e. Eq. (\ref{FL3})), is:
\begin{equation}
\zeta' = - \frac{{\mathcal H}}{p_{\mathrm{t}} + \rho_{\mathrm{t}}} \delta p_{\mathrm{nad}} + 
\frac{{\mathcal H}}{p_{\mathrm{t}} + \rho_{\mathrm{t}}} \biggl( c_{\mathrm{st}}^2 - \frac{1}{3}\biggr)
\delta\rho_{\mathrm{B}} - \frac{\overline{\theta}_{\mathrm{t}}}{3},\qquad \overline{\theta}_{\mathrm{t}} =
\theta_{\mathrm{t}} - \frac{h' + 6 \xi'}{2},
\label{TDC9}
\end{equation}
where $\overline{\theta}_{\mathrm{t}}$ is the three-divergence of the velocity field 
in the longitudinal gauge (see the Appendix, Eq. (\ref{L2})). If $\delta p_{\mathrm{nad}} =0$ (as contemplated in the 
present paper), then Eq. (\ref{TDC9}) can be explicitly integrated 
with the result that 
\begin{equation}
{\zeta}(k,\tau) = \zeta_{*}(k) - \frac{3 R_{\gamma} \Omega_{\mathrm{B}}(k)\alpha}{4 ( 3\alpha + 4)},
\label{TDC10}
\end{equation}
where $\alpha = a/a_{\mathrm{eq}}$ and $R_{\gamma}= 1 - R_{\nu}$ (see Eq. (\ref{rnu})). So the spectrum of the primordial adiabatic mode 
will be given in terms of $\zeta_{*}(k)= {\mathcal R}_{*}(k)$. The two-point function 
in Fourier space will be, for ${\mathcal R}$
\begin{equation}
\langle {\mathcal R}_{*}(\vec{k},\tau) {\mathcal R}_{*}(\vec{p},\tau) \rangle = \frac{2\pi^2 }{k^3} 
{\mathcal P}_{\mathcal R}(k) \, \delta^{(3)}(\vec{k} + \vec{p}), \qquad {\mathcal P}_{{\mathcal R}}(k) = 
{\mathcal A}_{\mathcal R} \biggl(\frac{k}{k_{\mathrm{p}}}\biggr)^{n_{\mathrm{s}} -1},
\label{TDC11}
\end{equation}
where $n_{\mathrm{s}}$ is the scalar (adiabatic) spectral index, $k_{\mathrm{p}} = 0.002 \, \mathrm{Mpc}^{-1}$ 
is the so-called pivot scale and ${\mathcal A}_{\mathcal R}$ is, by definition, the amplitude of the power 
spectrum at the pivot scale.
With these conventions the two-point function in real space becomes 
\begin{equation}
\langle {\mathcal R}_{*}(\vec{x},\tau) {\mathcal R}(\vec{y},\tau)\rangle = 
\int d\ln{k} {\mathcal P}_{{\mathcal R}}(k) \frac{\sin{kr}}{kr},\qquad r = |\vec{x}-\vec{y}|
\label{TDC12}
\end{equation}

So far the evolution equations of the lowest multipoles 
of the Boltzmann hierarchy have been introduced. It is relevant to recall also 
the brightness perturbations of the radiation field which are related to
the inhomogeneities of the Stokes parameters.
In the synchronous coordinate system the evolution equations of the brightness perturbations can be 
written as 
\begin{eqnarray}
&&  \Delta_{\mathrm{I}}' + i k \mu \Delta_{\mathrm{I}} = - \biggl[ \xi' - \frac{\mu^2}{2}( h' + 6 \xi')\biggr] +
\epsilon' \biggl[  - \Delta_{\mathrm{I}} + \Delta_{\mathrm{I}0} + \mu v_{\mathrm{b}} - \frac{1}{2} P_{2}(\mu) S_{\mathrm{Q}}\biggr],
\label{BR1}\\
&& \Delta_{\mathrm{Q}}' + i k \mu \Delta_{\mathrm{Q}} = \epsilon' \biggl[- \Delta_{\mathrm{Q}} + \frac{1}{2} ( 1 - P_{2}(\mu)) S_{\mathrm{Q}}\biggr],
\label{BR2}\\
&& \Delta_{\mathrm{U}}' + i k \mu \Delta_{\mathrm{U}} = - \epsilon'  \Delta_{\mathrm{U}},
\label{BR3}\\
&& v_{\mathrm{b}}' + {\mathcal H} v_{\mathrm{b}} + \frac{\epsilon'}{R_{\mathrm{b}}} ( 3 i \Delta_{\mathrm{I}1} + v_{\mathrm{b}}) + 
i k \frac{\Omega_{\mathrm{B}} - 4 \sigma_{\mathrm{B}}}{4 R_{\mathrm{b}}}=0,
\label{BR4}
\end{eqnarray}
where $R_{\mathrm{b}}$ has been defined in Eq. (\ref{Rbdef}) and where we defined $v_{\mathrm{b}} = \theta_{\mathrm{b}}/(i k)$. Moreover, in Eqs. (\ref{BR1}) and (\ref{BR2}):
\begin{equation}
S_{\mathrm{Q}} = \Delta_{\mathrm{I}2} + \Delta_{\mathrm{Q}0} + \Delta_{\mathrm{Q}2}.
\end{equation}
The notations $\Delta_{\mathrm{I}\ell}$ and $\Delta_{\mathrm{Q}\ell}$ denote the $\ell$-th multipole
of $\Delta_{\mathrm{I}}$ and $\Delta_{\mathrm{Q}}$. In Eqs. (\ref{BR1}) and (\ref{BR2}) 
$P_{2}(\mu) = (3\mu^2 -1)/2$ is the second Legendre polynomial. Equations (\ref{BR1})--(\ref{BR4}) constitute the basis of the semi-analytical approach used to estimate the 
magnetized temperature autocorrelations \cite{mg1,mg2,mg3}. 
In particular, the aforementioned equations have been solved 
in the tight-coupling approximation to first and second order \cite{mg1,mg2}. The physical 
information contained in Eqs. (\ref{BR1})--(\ref{BR4}) can be summarized by noticing 
that to zeroth-order in the tight-coupling expansion the CMB is not polarized in the baryon rest frame 
so that $\Delta_{\mathrm{Q}}$ and $\Delta_{\mathrm{U}}$ will be zero. To first-order in the tight-coupling expansion 
the quadrupole of the polarization (i.e. $\Delta_{\mathrm{Q}2}$) is proportional to the zeroth-order dipole. 
Since the zeroth-order dipole feels the Lorentz force, the polarization is also affected by the presence 
of large-scale magnetic fields.

\renewcommand{\theequation}{3.\arabic{equation}}
\setcounter{equation}{0}
\section{Magnetized initial conditions for the Boltzmann hierarchy}
\label{sec3}
At early times, close to the moment when initial conditions are set, 
the evolution equations for baryons and photons are integrated in the tight-coupling approximation. 
Otherwise this would represent a stiff problem 
owing to the largeness of the Thompson rate. 
Consider, first of all, the difference between the baryon velocity equation (i.e. Eq. (\ref{BL11})) and the photon velocity equation (i.e. Eq. (\ref{PHOT2})); the result of this manipulation is:
\begin{equation}
(\theta_{\gamma} - \theta_{\mathrm{b}})' + \frac{\epsilon'}{R_{\mathrm{b}}} ( 1 + R_{\mathrm{b}}) (\theta_{\gamma} - \theta_{\mathrm{b}}) =
k^2 \frac{\delta_{\gamma}}{4}  + {\mathcal H} \theta_{\mathrm{b}} - 
\frac{k^2}{4 R_{\mathrm{b}}} (\Omega_{\mathrm{B}} - 4 \sigma_{\mathrm{B}}),
\label{INC1}
\end{equation}
where Eq. (\ref{BB10}) has been also used to express the Lorentz force in terms of $\Omega_{\mathrm{B}}$ and 
$\sigma_{\mathrm{B}}$. Owing to the presence of $\epsilon'$, Eq. 
(\ref{INC1}) stipulates that any initial difference in the baryon-photon velocity is quickly washed out.
Consequently, at early times $\theta_{\gamma} \simeq \theta_{\mathrm{b}}$. 

Denoting by $\theta_{\gamma\mathrm{b}}$ the common value of the photon-baryon velocity field, the 
corresponding evolution equation can be obtained by combining Eqs. (\ref{BL11}) and (\ref{PHOT2}) in such a way that the scattering terms exactly cancel at the price of introducing explicitly $R_{\mathrm{b}}$ i.e. 
the baryon-to-photon ratio of Eq. (\ref{Rbdef}). The net result of this procedure is:
\begin{equation}
\theta_{\gamma\mathrm{b}}' + \frac{{\mathcal H} R_{\mathrm{b}}}{1 + R_{\mathrm{b}}} \theta_{\gamma\mathrm{b}} + \frac{\eta}{\rho_{\gamma} (R_{\mathrm{b}} + 1)} k^2 \theta_{\gamma\mathrm{b}}= 
\frac{k^2}{ 4 ( 1 + R_{\mathrm{b}})} \delta_{\gamma} + \frac{k^2 (\Omega_{\mathrm{B}} - 4 \sigma_{\mathrm{B}})}{4 ( 1 + R_{\mathrm{b}})},
\label{INC2}
\end{equation}
where we have also taken into account the shear viscosity contribution (proportional to $\eta$) which 
is responsible of the diffusion damping:
\begin{equation}
\eta = \frac{4}{15} \rho_{\gamma} \lambda_{\mathrm{Th}},\hspace{2cm} \lambda_{\mathrm{Th}} = \frac{1}{\epsilon'}.
\label{INC2a}
\end{equation}
The shear viscosity term (to a given order in the tight-coupling expansion)
allows for the estimate of diffusive effects. Standard considerations related 
to the zeroth-order in the tight coupling expansion imply that 
\begin{equation}
\frac{1}{k^2_{\mathrm{D}}} = \frac{2}{5} \int_{0}^{\tau} c_{\mathrm{sb}}(\tau') \frac{ a_{0} d\tau'}{a(\tau')\,\,x_{\mathrm{e}} n_{\mathrm{e}} \sigma_{\mathrm{Th}}}.
\label{INC2b}
\end{equation}
To second order in the tight-coupling expansion the inclusion of the polarization allows to estimate 
\cite{zal}:
\begin{equation}
\frac{1}{k_{\mathrm{D}}^2 }= \int_{0}^{\tau} \frac{d\tau'}{6 (R_{\mathrm{b}} + 1) \epsilon'} \biggl[\frac{16}{15} + \frac{R_{\mathrm{b}}^2}{R_{\mathrm{b}} + 1}\biggr].
\label{INC2c}
\end{equation}
The  factor $16/15$ arises since the polarization fluctuations are taken consistently 
into account in the derivation. This difference is physically relevant. Grossly speaking we can indeed say that
more polarization implies more anisotropy (and vice versa); more polarization implies a faster damping by diffusion. 
Note that $k_{\mathrm{D}}$ provides 
an effective ultra-violet cut-off for the magnetic energy spectra and will be used later on.

Correspondingly Eqs. (\ref{BL10}) and (\ref{PHOT2}) can be written as 
\begin{equation}
\delta_{\gamma}' = \frac{2}{3} h'- \frac{4}{3} \theta_{\gamma\mathrm{b}},\qquad \delta_{\mathrm{b}}' = \frac{h'}{2} - \theta_{\gamma\mathrm{b}}.
\label{INC3}
\end{equation}
Equations (\ref{INC2}) and (\ref{INC3}) can be further combined to get a single equation
for the density contrast of the radiation field $\delta_{\gamma}$ with the result that
\begin{equation}
 \delta_{\gamma}'' + \frac{{\mathcal   H} R_{\rm b}}{R_{\rm b} + 1} \delta_{\gamma}' + 
k^2 c_{\mathrm{sb}}^2 \delta_{\gamma} =\frac{2}{3} \frac{[ (R_{\mathrm{b}} + 1) h' ]'}{R_{\mathrm{b}} + 1}
+ \frac{k^2 }{3 ( R_{\rm b} + 1)} [ 4 \sigma_{\rm B} - \Omega_{\rm B}],
\label{INC4}
\end{equation}
where the Silk damping has been neglected.

The same equation can be obtained by systematically expanding the brightness 
perturbations of the radiation field (see, in particular, Eqs. (\ref{BR1}) and (\ref{BR4})) to zeroth 
order in the tight coupling expansion and by recalling that the precise relation between 
the monopole of the radiation intensity and the photon density contrast is given by 
 $4 \Delta_{\mathrm{I}0} = \delta_{\gamma}$. In Eq. (\ref{INC4})  the baryon-photon 
 sound speed $c_{\mathrm{sb}}$ has been introduced (see Eq. (\ref{SEE3})). 
 In the absence of magnetic fields, the second  source term
 in Eq. (\ref{INC4}) vanishes. The resulting equation (in different gauges) can be again employed for the 
 semi-analytical estimates of the temperature autocorrelations \cite{peebles1,pavel1,pavel2,seljak,hu1}.

The whole Einstein-Boltzmann hierarchy will now be solved 
to zeroth-order in the tight-coupling expansion and for typical wavelengths larger 
than the Hubble radius before equality (i.e. $k\tau < 1$ for $\tau< \tau_{\mathrm{eq}}$).
The Hamiltonian 
constraint of Eq. (\ref{HAM1}) and Eq. (\ref{SP1}) will first be solved. The obtained solution, parametrized 
in terms of a suitable number of arbitrary constants, will be inserted into the other equations. 
The final solution will only depend upon the spectrum of the adiabatic mode and upon the power 
spectra of $\Omega_{\mathrm{B}}$ and $\sigma_{\mathrm{B}}$.
Equations (\ref{HAM1}) and (\ref{SP1}) are solved provided:
\begin{eqnarray}
&& \xi(k,\tau) = - 2 C(k) + A_{\xi}(k) k^2 \tau^2,\qquad h(k,\tau) = - C(k) k^2 \tau^2 - A_{h}(k) k^4 \tau^4,
\label{INC5}\\
&& \delta_{\gamma}(k,\tau) = - R_{\gamma} \Omega_{\mathrm{B}}(k) - A_{\gamma}(k) k^2 \tau^2, \qquad 
\delta_{\nu}(k,\tau) = - R_{\gamma} \Omega_{\mathrm{B}}(k) - A_{\nu}(k) k^2 \tau^2.
\label{INC6}
\end{eqnarray}
To lowest order in $k\tau$, $h(k,\tau)$ does not have a constant term whose presence would entail 
a spurious gauge mode which must be projected out by exploiting the remaining gauge freedom \cite{PV1}. The compatibility of Eqs. (\ref{INC5}) 
and (\ref{INC6}) with Eqs. (\ref{HAM1}) and (\ref{SP1}) leads to the following 
condition 
\begin{equation}
3 (R_{\gamma} A_{\gamma}(k) + R_{\nu} A_{\nu}(k)) = 2 C(k),
\label{INC7}
\end{equation}
which guarantees that Eqs. (\ref{HAM1}) and (\ref{SP1}) are satisfied with corrections which are 
${\mathcal O}(k^4\tau^2)$.
In Eq. (\ref{INC7})  $R_{\gamma} = 1 - R_{\nu}$ (see Eq. (\ref{rnu})). The neutrino and photon fractions arise 
since, prior to equality, the Hamiltonian constraint of Eq. (\ref{HAM1}) 
can be simply written, after using Eq. (\ref{FL1}), as $4 k^2 \xi - 2{\mathcal H} h' =3{\mathcal H}^2 [ R_{\nu} \delta_{\nu} + R_{\gamma}\delta_{\gamma}]$. Equation (\ref{SP1}) can be also recast in a similar form.

The evolution equations of the velocity fields can be solved with a similar technique. In particular 
Eqs. (\ref{NUS1})--(\ref{NUS2}) and (\ref{INC2})--(\ref{INC3}) imply that 
\begin{equation}
\theta_{\gamma\mathrm{b}}(k,\tau) = D_{\gamma\mathrm{b}}^{(1)}(k) k^2 \tau + D_{\gamma\mathrm{b}}^{(2)}(k) k^4\tau^3,\qquad \theta_{\nu}(k,\tau) = D_{\nu}^{(1)}(k) k^2 \tau + D_{\nu}^{(2)}(k) k^4\tau^3,
\label{INC8}
\end{equation}
where the various constants must satisfy:
\begin{eqnarray}
&& 3A_{\gamma}(k) = 2 D_{\gamma\mathrm{b}}^{(1)}(k) + 2 C(k),\qquad  3A_{\nu}(k) = 2 D_{\nu}^{(1)}(k) + 2 C(k),
\label{INC9}\\
&& D_{\gamma\mathrm{b}}^{(1)}(k) = \frac{R_{\nu}}{4}\Omega_{\mathrm{B}}(k) - \sigma_{\mathrm{B}}(k),\qquad 
 D_{\nu}^{(1)}(k)=  \frac{R_{\gamma}}{R_{\nu}} \sigma_{\mathrm{B}}(k) - \frac{R_{\gamma}}{4} \Omega_{\mathrm{B}}(k).
 \label{INC10}
\end{eqnarray}
The compatibility of the obtained solution with the momentum constraint of Eq. (\ref{MOM1}) imposes, moreover,
the following pair of conditions
\begin{equation}
R_{\nu} D_{\nu}^{(1)}(k) + R_{\gamma} D_{\gamma\mathrm{b}}^{(1)}(k) =0,\qquad 
R_{\nu} D_{\nu}^{(2)}(k) + R_{\gamma} D_{\gamma\mathrm{b}}^{(2)}(k) + A_{\xi}(k)=0.
\label{INC11}
\end{equation}
The evolution equations involving the neutrino anisotropic stress (i.e. Eqs. (\ref{NUS3}) and (\ref{SP2}))
are solved by 
\begin{equation}
\sigma_{\nu}(k,\tau) = - \frac{R_{\gamma}}{R_{\nu}} \sigma_{\mathrm{B}}(k) + A_{\sigma}(k) k^2 \tau^2.
\label{INC12}
\end{equation}
As in the previous cases there are non-trivial conditions to be satisfied and they are, in the case of 
Eqs. (\ref{NUS3}) and (\ref{SP2}), 
\begin{equation}
3 [ 6 A{\xi}(k) - C(k)] + 2 C(k) = 6 R_{\nu} A_{\sigma}(k),\qquad A_{\sigma}(k) = \frac{2}{15} D_{\nu}^{(1)}(k) + 
\frac{2}{15} C(k) - \frac{4}{5} A_{\xi}(k).
\label{INC13}
\end{equation}
We are left with the evolution equations of the baryon and CDM density contrasts and with the velocity 
field of the CDM perturbations. The solutions for these quantities, as they emerge, respectively, 
from Eqs. (\ref{INC3}) and (\ref{CDMS1}) are:
\begin{eqnarray}
&& \theta_{\mathrm{c}}(k,\tau) =0,\qquad \delta_{\mathrm{c}}(k,\tau) = - \frac{3}{4} R_{\gamma} \Omega_{\mathrm{B}}(k) - \frac{C(k)}{2} k^2 \tau^2,
\nonumber\\
&& \delta_{\mathrm{b}}(k,\tau) = - \frac{3}{4} R_{\gamma} \Omega_{\mathrm{B}}(k) - \frac{1}{2}\biggl[ C(k) - \sigma_{\mathrm{B}}(k) + \frac{R_{\nu}}{4} \Omega_{\mathrm{B}}(k)\biggr] k^2 \tau^2.
\label{INC14}
\end{eqnarray}
All the compatibility conditions constraining the form of the solution can be solved, and, after 
some algebra, the full solution for the initial conditions of the lowest multipoles 
of the Einstein- Boltzmann hierarchy becomes:
\begin{eqnarray}
 \xi(k,\tau) &=& - 2 C(k) + \biggl[\frac{4 R_{\nu} + 5}{6 ( 4 R_{\nu} + 15)} C(k) + \frac{R_{\gamma} ( 4 \sigma_{\mathrm{B}}(k) - R_{\nu} \Omega_{\mathrm{B}}(k))}{ 6 ( 4 R_{\nu} + 15)}\biggr] k^2 \tau^2,
\label{S1}\\
h(k,\tau) &=& - C(k) k^2 \tau^2 - \frac{1}{36} \biggl[ \frac{8 R_{\nu}^2 - 14 R_{\nu} - 75}{(2 R_{\nu} + 25)(4 R_{\nu} + 15)} C(k) 
\nonumber\\
&+& \frac{R_{\gamma} ( 15 - 20 R_{\nu})}{10( 4 R_{\nu} + 15) ( 2 R_{\nu} + 25)} (R_{\nu}\Omega_{\mathrm{B}}(k) - 4 \sigma_{\mathrm{B}}(k))\biggr] k^4 \tau^4,
\label{S2}\\
\delta_{\gamma}(k,\tau) &=& - R_{\gamma} \Omega_{\mathrm{B}}(k) - \frac{2}{3} \biggl[ C(k) - \sigma_{\mathrm{B}}(k) + \frac{R_{\nu}}{4} \Omega_{\mathrm{B}}(k)\biggr] k^2 \tau^2,
\label{S3}\\
\delta_{\nu}(k,\tau) &=& - R_{\gamma} \Omega_{\mathrm{B}}(k) - \frac{2}{3} \biggl[ C(k) + \frac{R_{\gamma}}{4 R_{\nu}}\biggl( 4\sigma_{\mathrm{B}}(k) - R_{\nu} \Omega_{\mathrm{B}}(k)\biggr)\biggr] k^2 \tau^2,
\label{S4}\\
\delta_{\mathrm{c}}(k,\tau) &=& - \frac{3}{4}R_{\gamma} \Omega_{\mathrm{B}}(k) - \frac{C(k)}{2} k^2 \tau^2,
\label{S5}\\
\delta_{\mathrm{b}}(k,\tau) &=& - \frac{3}{4}R_{\gamma} \Omega_{\mathrm{B}}(k) - \frac{1}{2} \biggl[ C(k) - \sigma_{\mathrm{B}}(k)+ \frac{R_{\nu}}{4} \Omega_{\mathrm{B}}(k) \biggr] k^2\tau^2,
\label{S6}\\
\theta_{\gamma\mathrm{b}}(k,\tau) &=& \biggl[ \frac{R_{\nu}}{4} \Omega_{\mathrm{B}}(k) - \sigma_{\mathrm{B}}\biggr] 
k^2 \tau -\frac{1}{36} \biggl[ 2 C(k) + \frac{R_{\nu} \Omega_{\mathrm{B}}(k) - 4 \sigma_{\mathrm{B}}(k)}{2}\biggr] k^4\tau^3,
\label{S7}\\
\theta_{\nu}(k,\tau) &=& \biggl[ \frac{R_{\gamma}}{R_{\nu}} \sigma_{\mathrm{B}}(k) - \frac{R_{\gamma}}{4} \Omega_{\mathrm{B}}(k)\biggr] k^2 \tau
- \frac{1}{36}\biggl[\frac{2 ( 4 R_{\nu} + 23)}{4 R_{\nu} + 15} C(k) 
\nonumber\\
&+& \frac{R_{\gamma}( 4 R_{\nu} + 27)}{2 R_{\nu} ( 4 R_{\nu} + 15)}( 4 \sigma_{\mathrm{B}}(k) - R_{\nu} \Omega_{\mathrm{B}}(k))\biggr] k^4 \tau^3,
\label{S8}\\
 \theta_{\mathrm{c}}(k,\tau) &=& 0,
\label{S9}\\
\sigma_{\nu}(k,\tau) &=& - \frac{R_{\gamma}}{R_{\nu}} \sigma_{\mathrm{B}}(k) + \biggl[ \frac{4 C(k)}{3( 4 R_{\nu} + 15)} + \frac{R_{\gamma}( 4 \sigma_{\mathrm{B}}(k) - R_{\nu} \Omega_{\mathrm{B}})}{ 2 R_{\nu}(4 R_{\nu} + 15)}\biggr] k^2 \tau^2.
\label{S10}
\end{eqnarray}
In the limit $\Omega_{\mathrm{B}}(k) \to 0$ and $\sigma_{\mathrm{B}}(k)\to 0$ 
the solution corresponds to the usual adiabatic mode. To lowest order, in fact, we can appreciate 
that the non-adiabatic pressure fluctuations introduced in Eqs. (\ref{TDC4}) and (\ref{TDC5}) vanish 
since 
\begin{equation}
\delta_{\gamma}(k,\tau) \simeq \delta_{\nu}(k,\tau)\simeq \frac{4}{3} \delta_{\mathrm{c}}(k,\tau) \simeq 
\frac{4}{3} \delta_{\mathrm{b}}(k,\tau).
\label{S11}
\end{equation}
The adiabatic mode is parametrized in terms of $C(k)$. 
The spectrum of $C(k)$ is related to the spectrum of ${\mathcal R}_{*}(k)$ introduced 
in Eq. (\ref{TDC11}).   Equation (\ref{S1}) can be inserted into Eq. (\ref{TDC6}) with the result
that, to lowest order in $k\tau$, ${\mathcal R}(k) = - 2 C(k)$. 
According to Eq. (\ref{S9}), the CDM velocity field 
is set exactly to $0$. This requirement avoids the presence of the second gauge mode (the first 
one was projected out by excluding the constant solution for $h(k,\tau)$).
For infinitesimal diffeomeorphisms (see also the Appendix) the metric fluctuations 
change by the Lie derivative in the direction parametrized by the two gauge parameters 
$\tilde{\epsilon}_{0}$ and $\tilde{\epsilon}$. By imposing the synchronous gauge the $\tilde{\epsilon}_{0}$ and $\tilde{\epsilon}$
are determined only up to two functions which are constant in time but not in space.
This is the ultimate rationale for the possible persistence, in the synchronous gauge, of two spurious 
gauge modes. Such a  possibility is avoided by gauging away the two unphysical 
solutions and this achieved, at a practical level, by setting $\theta_{\mathrm{c}}(k,\tau)=0$ 
and by requiring that $h(k,\tau)$ does not have a constant mode.
A complementary way of addressing this issue is to work out the same solution in a frame
where the gauge freedom is completely removed. The solution obtained in this 
gauge must match with the results obtained in the synchronous gauge and transformed 
to the new gauge. In the case of the present problem, also following previous works, 
it is productive to cross-check the results in the longitudinal coordinate system. The main 
ingredients for this analysis are reported in the Appendix where the synchronous and longitudinal 
solutions are explicitly connected.

The initial spectrum of the magnetic fields is encoded in $\Omega_{\mathrm{B}}$ and 
$\sigma_{\mathrm{B}}$. The spectrum of $\Omega_{\mathrm{B}}(k)$ 
and $\sigma_{\mathrm{B}}(k)$ must then be computed in terms of the spectrum of the magnetic field 
introduced in Eqs. (\ref{BB12}) and (\ref{BB13}). 
This is a rather lengthy calculation and here only the main steps will be outlined. 

Since both $\Omega_{\mathrm{B}}(\vec{x})$ and $\sigma_{\mathrm{B}}(\vec{x})$ are 
quadratic in the magnetic field intensities (see, for instance,  Eqs. (\ref{BB8}) and (\ref{BB9})) 
their expressions in Fourier space will lead to the following 
two convolutions 
\begin{eqnarray}
\Omega_{\mathrm{B}}(\vec{q}) &=& \frac{1}{(2\pi)^{3/2}} 
\frac{1}{8\pi \overline{\rho}_{\gamma}} \int d^{3} k B_{i}(k) B^{i}(\vec{q} - \vec{k}),
\label{CORR1}\\
\sigma_{\mathrm{B}}(\vec{q}) &=& \frac{1}{(2\pi)^{3/2}} \frac{1}{16 \pi \overline{\rho}_{\gamma}} \int 
d^{3} k \biggl[ \frac{3 ( q^{j} - k^{j}) k^{i}}{q^2} B_{j}(k)B_{i}(\vec{q} - \vec{k}) - B_{i}(\vec{q} - \vec{k}) B^{i}(\vec{k})\biggr].
\label{CORR2}
\end{eqnarray}
The correlation functions for $\Omega_{\mathrm{B}}(\vec{k})$ and $\sigma_{\mathrm{B}}(\vec{k})$ 
are then defined as 
\begin{equation}
\langle \Omega_{\mathrm{B}}(\vec{q}) \Omega_{\mathrm{B}}(\vec{p})\rangle = \frac{2\pi^2}{q^3} {\mathcal P}_{\Omega}(q) \delta^{(3)}(\vec{q} + \vec{p}),\qquad \langle \sigma_{\mathrm{B}}(\vec{q}) \sigma_{\mathrm{B}}(\vec{p})\rangle = \frac{2\pi^2}{q^3} {\mathcal P}_{\sigma}(q) \delta^{(3)}(\vec{q} + \vec{p}),
\label{CORR3}
\end{equation}
To compute ${\mathcal P}_{\Omega}(q)$ and ${\mathcal P}_{\sigma}(q)$ in terms of the magnetic power 
spectra we must go through the  straightforward  (but rather lengthy) procedure of expressing the stochastic 
averages of four fields in terms of the two-point function of Eqs. (\ref{BB12}) and (\ref{BB13}).
Then the obtained results must be integrated over the momenta.
After performing the first of the previously mentioned steps we obtain 
\begin{eqnarray}
{\mathcal P}_{\Omega}(q) &=& \frac{q^{3}}{(2\pi)} \frac{1}{8\pi \overline{\rho}_{\gamma}} \int 
d^{3} k \frac{P_{\mathrm{B}}(k)}{k^3} \frac{P_{\mathrm{B}}(|\vec{q} - \vec{k}|)}{|\vec{q} - \vec{k}|^3} \biggl\{ 1 + \frac{[\vec{k} \cdot (\vec{q} - \vec{k})]^2}{k^2 |\vec{q} - \vec{k}|^2}\biggr\}
\label{CORR4}\\
 {\mathcal P}_{\sigma}(p) &=& \frac{p^3}{(2\pi) }\frac{1}{(16\pi \overline{\rho}_{\gamma})^2} 
\int d^{3} k \frac{P_{\mathrm{B}}(k)}{k^3} \frac{P_{\mathrm{B}}(|\vec{p} - \vec{k}|)}{|\vec{p} - \vec{k}|^3}
\biggl\{ 1 + \frac{[\vec{k}\cdot (\vec{p} - \vec{k})]^2}{k^2 |\vec{p} - \vec{k}|^2}  
\nonumber\\
&+&\frac{6}{p^2} \biggl[ \vec{k}\cdot( \vec{p} - \vec{k}) - \frac{[\vec{k}\cdot (\vec{p} - \vec{k})]^3}{k^2 | \vec{p} - \vec{k}|^2} \biggr]
\nonumber\\
&+& \frac{9}{p^4} \biggl[ k^2 |\vec{p} - \vec{k}|^2 - 2 [ \vec{k}\cdot(\vec{p} - \vec{k})|]^2 + 
\frac{[\vec{k}\cdot (\vec{p} - \vec{k})]^4}{k^2 |\vec{p} - \vec{k}|^2} \biggr]\biggr\}.
\label{CORR5}
\end{eqnarray}
In Eqs. (\ref{CORR4}) and (\ref{CORR5}) the notation $k = |\vec{k}|$ has been 
employed. To complete the calculation the angular integration and then 
the radial integration must be performed. The integration measure can be written, in spherical coordinates,
as $d^3 k = k^2 d k d\cos{\vartheta} d\varphi$. The integration over $\varphi$ is trivial and leads just 
to a factor $2\pi$. However the integration over $d\cos{\vartheta}$ (between $-1$ and $1$) 
is rather cumbersome (but doable in exact terms). Indeed all scalar products arising in Eqs. (\ref{CORR4}) and (\ref{CORR5}) induce 
a factor $\cos{\vartheta}$. So calling $x = \cos{\vartheta}$, the expressions like $|\vec{q} - \vec{k}|$ 
(which appear ubiquitously in Eqs. (\ref{CORR4}) and (\ref{CORR5})) become 
$|\vec{q} - \vec{k}| = \sqrt{ q^2 + k^2 - 2 q k x}$. The combinations of all similar factors 
must then be integrated over $x$. The physically interesting region of spectra is realized when $ 1< n_{\mathrm{B}} < 5/2$. 
We will conventionally refer to this case as to the one of
blue spectral indices. It is also interesting to discuss in some detail the case of red spectra 
(i.e. $n_{\mathrm{B}} < 1$). Violet spectra (i.e. $n_{\mathrm{B}} \gg 1$) are 
mainly constrained by the diffusion scale $k_{\mathrm{D}}$ (see Eq. (\ref{INC2c})). 
Simplistic estimates of the Silk damping scale lead to $k_{\mathrm{D}}^{-2} \simeq 0.3 (a/a_{\mathrm{dec}})^{5/2}/[\sqrt{\omega_{M}} \omega_{\mathrm{b}}] \, \mathrm{Mpc}^2$.
The exactly scale-invariant case leads to a logarithmically divergent 
power spectrum. 

The magnetic power spectra 
are usually defined within an appropriate regularization of the magnetic energy density \cite{maxrev1}:
\begin{equation}
\langle B_{i}(\vec{x}) B^{i}(\vec{y}) \rangle = 2 \int d\ln{k} P_{\mathrm{B}}(k) \frac{\sin{kr}}{kr} W(k), \qquad r = |\vec{x} - \vec{y}|,
\label{CORR8}
\end{equation}
where $W(k)$ is an appropriate window function.
Consider first the case of a blue spectrum. The energy density can be 
regularized over a typical comoving scale $L$ (which is related to the magnetic pivot 
scale in Fourier space) by means of a Gaussian window function $W(k) = e^{-k^2L^2}$. 
Equation (\ref{BB14}) then implies:
\begin{equation}
B_{\mathrm{L}}^2(r) = (2\pi)^{1 - n_{\mathrm{B}}} {\mathcal A}_{\mathrm{B}} \Gamma\biggl(\frac{n_{\mathrm{B}} -1}{2}\biggr)  F_{11}\biggl(\frac{n_{\mathrm{B}}-1}{2},\frac{3}{2}, - \frac{r^2 k_{\mathrm{L}}^2}{16\pi^2}\biggr),
\label{CORR8a}
\end{equation}
where $F_{11}(a,b, z)$ is the Kummer confluent hypergeometric function \cite{abr,grad}.
Since $\lim_{z\to 0} F_{11}(a,b, z) = 1$, 
\begin{equation}
B_{\mathrm{L}}^2 = \lim_{r\to0} \langle B_{i}(\vec{x}) B_{j}(\vec{y})\rangle = {\mathcal A}_{\mathrm{B}}^2 (2\pi)^{1 - n_{\mathrm{B}}} \Gamma\biggl(\frac{n_{\mathrm{B}} -1}{2}\biggr).
\label{CORR9}
\end{equation}
In the radial integrals of Eqs. (\ref{CORR4}) and (\ref{CORR5}),  ${\mathcal A}_{\mathrm{B}}$ can be traded for 
$B_{\mathrm{L}}^2$ so that ${\mathcal P}_{\Omega}(k)$ and ${\mathcal P}_{\sigma}(k)$ can be 
written, respectively, as:
\begin{equation}
{\mathcal P}_{\Omega}(k) = \overline{\Omega}_{\mathrm{BL}}^2  \biggl(\frac{k}{k_{\mathrm{L}}}\biggr)^{2(n_{\mathrm{B}} -1)} {\mathcal F}(n_{\mathrm{B}}),\qquad {\mathcal P}_{\sigma}(k) = \overline{\Omega}_{\mathrm{BL}}^2  \biggl(\frac{k}{k_{\mathrm{L}}}\biggr)^{2(n_{\mathrm{B}} -1)} {\mathcal G}(n_{\mathrm{B}}),
\label{CORR10}
\end{equation}
where 
\begin{eqnarray}
\overline{\Omega}_{\mathrm{BL}} &=& \frac{B_{\mathrm{L}}^2}{8\pi \overline{\rho}_{\gamma}} = 7.5 \times 10^{-9} \biggl(\frac{B_{\mathrm{L}}}{\mathrm{nG}}\biggr)^{2},
\label{CORR11}\\
{\mathcal F}(n_{\mathrm{B}}) &=&  \frac{(2\pi)^{2(n_{\mathrm{B}} -1)}}{\Gamma^2\biggl(\frac{n_{\mathrm{B}}-1}{2}\biggr)}\biggl[\frac{4( 7 - n_{\mathrm{B}})}{3 (n_{\mathrm{B}} -1) ( 5 - 2 n_{\mathrm{B}})} 
+  \frac{4}{(2 n_{\mathrm{B}} - 5)} \biggl( \frac{k}{k_{\mathrm{D}}}\biggr)^{5 - 2 n_{\mathrm{B}}} \biggr]
\label{CORR12}\\
{\mathcal G}(n_{\mathrm{B}}) &=& \frac{(2\pi)^{2(n_{\mathrm{B}} -1)}}{\Gamma^2\biggl(\frac{n_{\mathrm{B}}-1}{2}\biggr)}\biggl[ \frac{ n_{\mathrm{B}} + 29}{15 ( 5 - 2 
n_{\mathrm{B}})( n_{\mathrm{B}} -1)} 
+ \frac{7}{5} \frac{1}{(2 n_{\mathrm{B}} - 5)} \biggl( \frac{k}{k_{\mathrm{D}}}\biggr)^{5 - 2 n_{\mathrm{B}}} \biggr],
\label{CORR13}
\end{eqnarray}
 It should be remarked that when $1<n_{\mathrm{B}} < 5/2$, we can formally send the diffusion scale to infinity 
 (i.e. $k_{\mathrm{D}}\to \infty$) and the final result will still be convergent. Consequently,
as already remarked in related contexts \cite{vt1} the diffusion damping only enters the case when 
the spectral slopes are violet (i.e. $n_{\mathrm{B}} \gg 5/2$).

For $n_{\mathrm{B}}<1$ the window function appearing in Eq. (\ref{CORR8}) can be chosen as a simple
step function $W(k)=\theta(k-k_0)$. If this is the case ${\mathcal P}_{\Omega}(k)$ and ${\mathcal P}_{\sigma}(k)$ can be formally written exactly as in Eq. (\ref{CORR10}) but with two slightly different 
pre-factors which shall be denoted by $\overline{{\mathcal F}}(n_{\mathrm{B}})$ and $\overline{\mathcal G}(n_{\mathrm{B}})$:
\begin{equation}
{\mathcal P}_{\Omega}(k) = \overline{\Omega}_{\mathrm{BL}}^2  \biggl(\frac{k}{k_{0}}\biggr)^{2(n_{\mathrm{B}} -1)} \overline{{\mathcal F}}(n_{\mathrm{B}}),\qquad {\mathcal P}_{\Omega}(k) = \overline{\Omega}_{\mathrm{BL}}^2  \biggl(\frac{k}{k_{0}}\biggr)^{2(n_{\mathrm{B}} -1)} \overline{{\mathcal G}}(n_{\mathrm{B}}),
\label{CORR14}
\end{equation}
where 
\begin{eqnarray}
\overline{{\mathcal F}}(n_{\mathrm{B}}) &=&\frac{16}{3}(1-n_{\mathrm{B}})^2
\left[\frac{n_{\mathrm{B}}-7}{(n_{\mathrm{B}}-1)(2n_{\mathrm{B}}-5)}+\frac{2}{1-n_{\mathrm{B}}}\left(\frac{k_0}{k}\right)^{n_{\mathrm{B}}-1}\right],
\label{CORR15}\\
\overline{{\mathcal G}}(n_{\mathrm{B}}) &=&(1-n_{\mathrm{B}})^2
\left[\frac{4n_{\mathrm{B}}+116}{15(5-2n_{\mathrm{B}})(n_{\mathrm{B}}-1)}+\frac{8}{3}\frac{1}{1-n_{\mathrm{B}}}
\left(\frac{k_0}{k}\right)^{n_{\mathrm{B}}-1}\right].
\label{CORR16}
\end{eqnarray}
where $k_0$ is of the order of (but smaller than) the Hubble rate.

For normalization purposes it is useful to have an explicit expression of the Sachs-Wolfe plateau 
which includes the magnetic energy density.
This estimate can be performed by solving Eq. (\ref{BR1}) with the line integration 
method. The result of this procedure is 
\begin{eqnarray}
&& \Delta_{\mathrm{I}}(\vec{k}, \hat{n}, \tau) = \int_{0}^{\tau_{0}} e^{ - i k \mu (\tau_{0} - \tau)} e^{- \epsilon(\tau,\tau_{0})} \biggl[ - \xi' + 
\frac{\mu^2}{2} ( h' + 6 \xi')\biggr]+
\nonumber\\
&&  \int_{0}^{\tau_{0}} e^{ - i k \mu (\tau_{0} - \tau)} {\mathcal K}(\tau) \biggl[ \Delta_{\mathrm{I}0} + \mu v_{\mathrm{b}} - 
\frac{1}{2} P_{2}(\mu) S_{\mathrm{Q}}\biggr],
\label{LSI1}
\end{eqnarray}
where ${\mathcal K}(\tau)$ is the visibility function and 
\begin{equation}
\epsilon(\tau,\tau_{0}) = \int_{\tau}^{\tau_{0}} \frac{a}{a_{0}} \sigma_{\mathrm{Th}} n_{\mathrm{e}},\qquad {\mathcal K}(\tau) = \epsilon' 
e^{-\epsilon(\tau,\tau_{0})}.
\label{LSI2}
\end{equation}
The term $\mu^2$ appearing in Eq. (\ref{LSI1}) can be integrated by parts and, subsequently, the visibility function 
can be approximated by a Dirac delta function centered at the decoupling time. 
Neglecting the integrated Sachs-Wolfe contribution and the Doppler term:
\begin{equation}
\Delta_{\mathrm{I}}^{(\mathrm{SW})}(\vec{k}, \hat{n}, \tau_{\mathrm{dec}}) = \biggl[\frac{\delta_{\gamma}}{4} - 
\frac{(h +6 \xi)''}{2 k^2}\biggr]_{\tau_{\mathrm{dec}}} e^{- i k \mu \tau_{0}},
\label{LSI3}
\end{equation}
where $\tau_{\mathrm{dec}}$ has been neglected in comparison with $\tau_{0}$ in the argument of the exponential factor.
To evaluate Eq. (\ref{LSI3}) we need to know the value of the combination $(h + 6\xi)'$ after equality when the relevant modes 
have wavelengths larger than the Hubble radius. Let us notice that, for the mentioned wavelengths, Eq. (\ref{INC3})
implies that $\delta_{\gamma}' \simeq 2 h' /3$. Thus, denoting by $\delta_{\gamma}^{(\mathrm{f})}$ and $\delta_{\gamma}^{(i)}$ 
the final (i.e. at the decoupling) and initial (i.e. before equality) values of the density contrast we will have 
\begin{equation}
\delta_{\gamma}^{(\mathrm{f})} = \delta_{\gamma}^{(\mathrm{i})} + \frac{2}{3}(h^{(\mathrm{f})} - h^{(\mathrm{i})}).
\label{LSI4}
\end{equation}
The evolution of $\xi$ across equality can be obtained from Eqs. (\ref{TDC6}), (\ref{TDC9}) and (\ref{TDC10}) and it is given
by solving the following equation:
\begin{equation}
\frac{d \xi}{d\alpha} + \frac{3 \alpha + 4}{2 \alpha (\alpha +1)} \xi = \frac{4 + 3 \alpha}{2 \alpha (\alpha + 1)} \biggl[ {\mathcal R}_{*}(k) - 
\frac{3 R_{\gamma} \Omega_{\mathrm{B}}(k) \alpha}{4( 3\alpha + 4)} \biggr],
\label{LSI5}
\end{equation}
where $\alpha = a/a_{\mathrm{eq}}$. Once $\xi$ is known we can easily deduce $(h + 6 \xi)'$ from Eq. (\ref{SP2}) which implies 
(neglecting the anisotropic stress when the corresponding wavelengths are larger than the Hubble radius):
\begin{equation}
[ (h + 6 \xi)' a^2]' \simeq 2 k^2 a^2 \xi.
\label{LSI6}
\end{equation}
The final result for the ordinary Sachs-Wolfe term can then be written as 
\begin{equation}
\Delta_{\mathrm{I}}^{(\mathrm{SW})}(\vec{k}, \hat{n}, \tau_{\mathrm{dec}}) = \biggl[- \frac{{\mathcal R}_{*}(k)}{5} + \frac{R_{\gamma}}{20} \Omega_{\mathrm{B}}(k)\biggr]e^{- i k \mu \tau_{0}}
\label{LSI7}
\end{equation}
Expanding the plane wave in series of Legendre polynomials the $\Delta_{\mathrm{I}\ell}(k,\tau_{0})$ can be easily extracted and the 
$C_{\ell}$ estimated with standard integration over the comoving wave-number $k$. The result is:
\begin{equation}
C^{(\mathrm{SW})}_{\ell} = \biggl[ \frac{{\cal A}_{{\mathcal R}}}{25} \,{\mathcal Z}_{1}(n_{\mathrm{s}},\ell)  +
\frac{1}{400} \, R_{\gamma}^2  \overline{\Omega}^{2}_{{\rm B}\,L} {\mathcal Z}_{2}(n_{\mathrm{B}},\ell) - 
\frac{1}{50} \sqrt{ {\cal A}_{{\mathcal R}}} \, R_{\gamma} \,\overline{\Omega}_{{\rm B}\, L}\,{\mathcal Z}_{3} 
(n_{\mathrm{s}},n_{\mathrm{B}}, \ell) \cos{\gamma_{br}}\biggr],
\label{SWP}
\end{equation}
where 
\begin{eqnarray}
{\mathcal Z}_{1}(n_{\mathrm{s}},\ell) &=& \frac{\pi^2}{4} \biggl(\frac{k_0}{k_{\rm p}}\biggl)^{n-1} 2^{n_{\mathrm{s}}} \frac{\Gamma( 3 - n_{\mathrm{s}}) \Gamma\biggl(\ell + 
\frac{ n_{\mathrm{s}} -1}{2}\biggr)}{\Gamma^2\biggl( 2 - \frac{n_{\mathrm{s}}}{2}\biggr) \Gamma\biggl( \ell + \frac{5}{2} - \frac{n_{\mathrm{s}}}{2}\biggr)},
\label{Z1}\\
{\mathcal Z}_{2}(n_{\mathrm{B}},\ell) &=& \frac{\pi^2}{2} 2^{2(n_{\mathrm{B}} -1)} {\cal F}(n_{\mathrm{B}}) \biggl( \frac{k_{0}}{k_{L}}\biggr)^{ 2(n_{\mathrm{B}} -1) } \frac{ \Gamma( 4 - 2 n_{\mathrm{B}}) 
\Gamma(\ell + n_{\mathrm{B}} -1)}{\Gamma^2\biggl(\frac{5}{2} - n_{\mathrm{B}}\biggr) \Gamma(\ell + 3 - n_{\mathrm{B}})},
\label{Z2}\\
{\mathcal Z}_{3}(n_{\mathrm{s}},n_{\mathrm{B}},\ell) &=&\frac{\pi^2}{4} 2^{n_{\mathrm{B}} -1} 2^{\frac{n_{\mathrm{s}} +1}{2}} \,\sqrt{{\cal F}(n_{\mathrm{B}})}\, \biggl(\frac{k_{0}}{k_{L}}\biggr)^{n_{\mathrm{B}} -1} \biggl(\frac{k_{0}}{k_{\rm p}}\biggr)^{\frac{n_{\mathrm{s}} + 1}{2}} 
\nonumber\\
&\times&\frac{ \Gamma\biggl(\frac{7}{2} - n_{\mathrm{B}} - \frac{n_{\mathrm{s}}}{2}\biggr) \Gamma\biggl( \ell + 
\frac{n_{\mathrm{B}}}{2} + \frac{n_{\mathrm{s}}}{4} - \frac{3}{4}\biggr)}{\Gamma^2\biggl(\frac{9}{4} - \frac{n_{\mathrm{B}}}{2} - \frac{n_{\mathrm{s}}}{4}\biggr)
\Gamma\biggl( \frac{11}{4} + \ell - \frac{n_{\mathrm{B}}}{2} - \frac{n_{\mathrm{s}}}{4} \biggr)}.
\label{Z3}
\end{eqnarray}
In Eq. (\ref{Z3}) $\gamma_{\mathrm{br}}$ is the correlation angle that has been included to keep 
the expressions as general as possible. In what follows the main focus will however be on the case where 
the adiabatic mode of curvature perturbations is not correlated with the magnetized 
contribution (i.e. $\gamma_{\mathrm{br}} = \pi/2$). Note, however, that if $\cos{\gamma_{\mathrm{br}}}>0$ then the 
cross-correlation between the adiabatic component and the magnetic component will lower the Sachs-Wolfe 
plateau allowing for a magnetized contribution which is comparatively larger than in the case where $\gamma_{\mathrm{br}}= \pi/2$.
We leave this possibility for future studies \cite{far6}.

\renewcommand{\theequation}{4.\arabic{equation}}
\setcounter{equation}{0}
\section{Magnetized temperature autocorrelations}
\label{sec4}
\begin{figure}
\begin{center}
\begin{tabular}{|c|c|}
      \hline
      \hbox{\epsfxsize = 7.6 cm  \epsffile{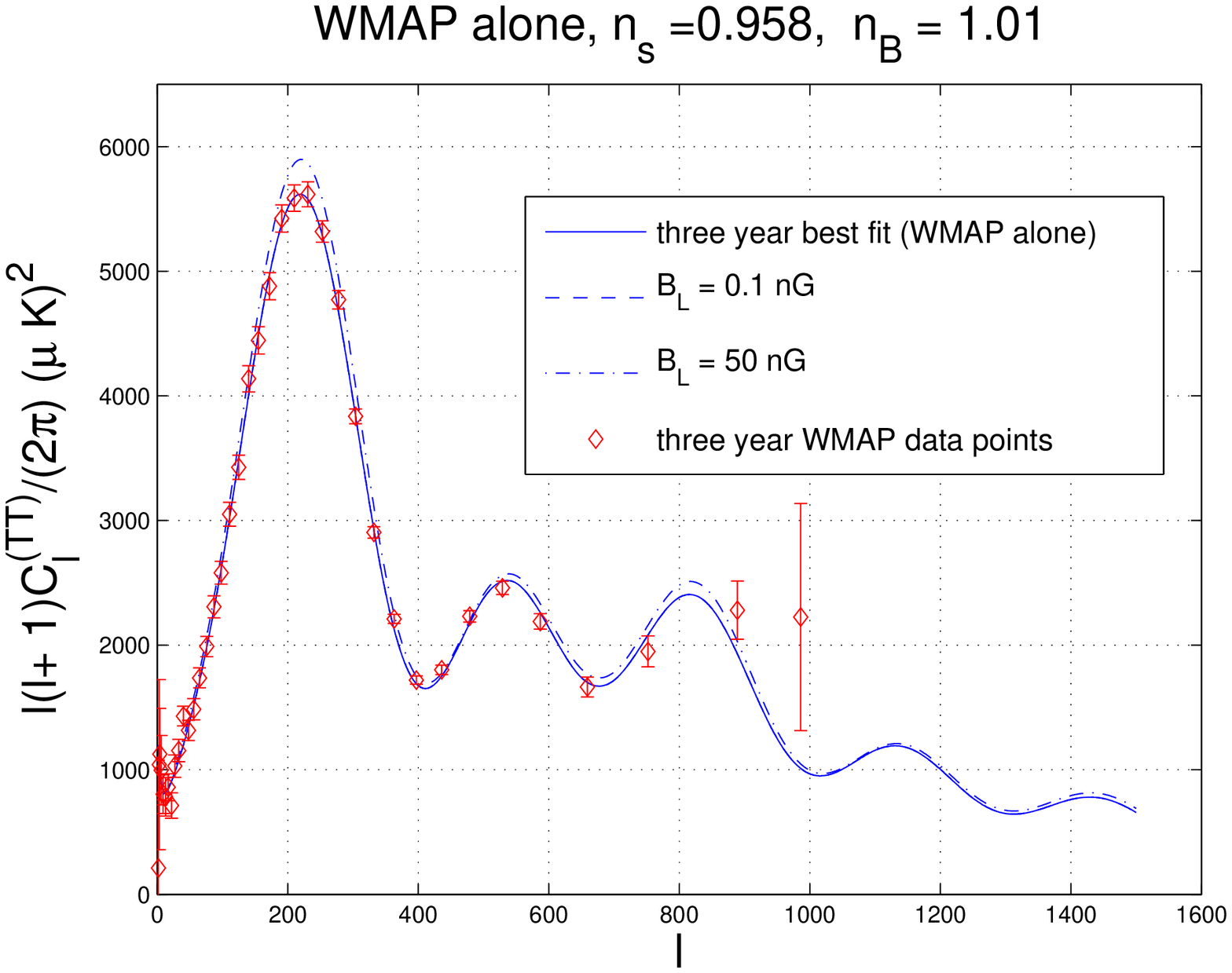}} &
     \hbox{\epsfxsize = 7.6 cm  \epsffile{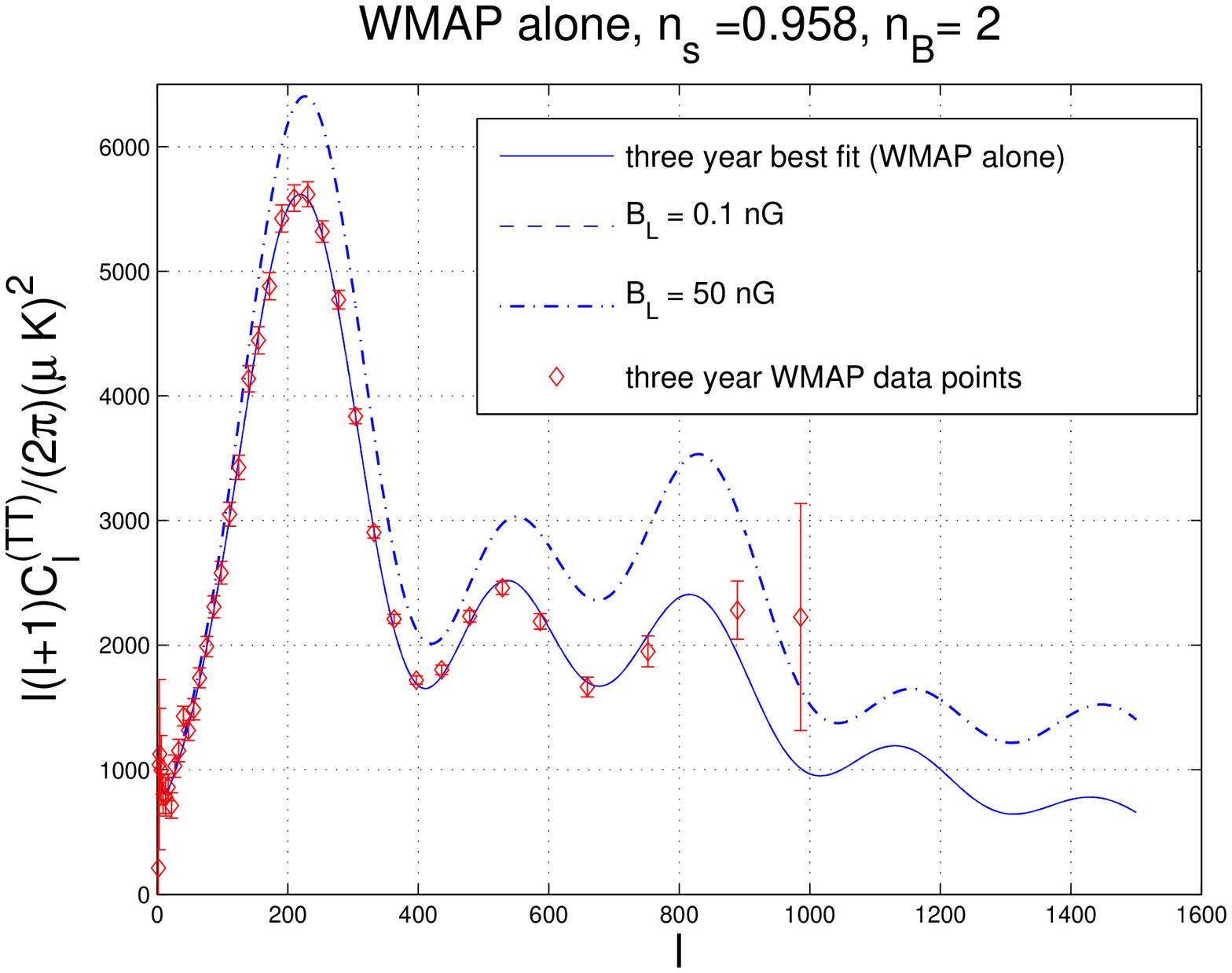}}\\
      \hline
\end{tabular}
\end{center}
\caption[a]{The temperature autocorrelations for blue magnetic spectral indices 
are compared with the best fit model arising from the WMAP alone 
analyzed in terms of a pure $\Lambda$CDM model with no tensors. The value of the magnetic pivot 
scale is $k_{\mathrm{L}} = 1\, \mathrm{Mpc}^{-1}$.}
\label{Figure1}
\end{figure}

 In Fig. \ref{Figure1}  the 
results of the numerical integration are illustrated in terms of the temperature autocorrelations.
The parameters are fixed to the best fit of the WMAP data alone implying
that the value of the scalar spectral index is $n_{\mathrm{s}} =0.958$. The 
full set of cosmological parameters used to compute the models in Fig. \ref{Figure1}
is given as follows\footnote{Consistently with our notations we should denote 
with $\epsilon_{\mathrm{re}}$ the optical depth to reionization which is 
conventionally denoted by $\tau$ and which we use to indicate the conformal time. However, since in the present and in the following sections the optical 
depth and the conformal time coordinate are never mentioned together, we will stick to the conventional terminology 
and denote with $\tau$ the optical depth.}:
\begin{equation}
(\Omega_{\mathrm{b}0},\, \Omega_{\mathrm{c}0}, \,\Omega_{\mathrm{\Lambda}}, \,h_{0}, \,n_{\mathrm{s}},\,\tau)= (0.042,\, 0.198,\, 0.76,\, 0.732,\, 0.958,\,0.089).
\label{best1}
\end{equation}
Moreover the tensors are absent from the fit 
and  $R_{\nu} =0.408$ (i.e., according to Eq. (\ref{rnu}), $N_{\nu} =3.04$).
In Fig. \ref{Figure1} (full curve in both plots) the $C_{\ell}^{(\mathrm{TT})}$ are illustrated for the best 
fit parameters reported in Eq. (\ref{best1}). The magnetic spectral indices 
are, in both cases blue, i.e., according to the terminology of the previous section, 
$1<n_{\mathrm{B}} < 5/2$. With the dashed line
the magnetic fields corresponding to $B_{\mathrm{L}} = 0.1 $ nG is reported. The 
dashed curve cannot be distinguished from the best fit curve. If the regularized 
magnetic field intensity is $ B_{\mathrm{L}} \leq {\mathcal O}(0.1\,\mathrm{nG})$, then 
the difference of the TT correlations (as well as the EE and TE correlations) 
with respect to the three year best fit is below the accuracy of the code. The latter 
statement depends, of course, on the spectral index and on the range 
of multipoles. Indeed, as argued in Section \ref{sec6}, the large-multipole 
region (i.e. $\ell \gg 1500$) is more sensitive to regularized fields 
of nG strengths.

The dot-dashed curve denotes, in both plots, the temperature autocorrelations 
computed in the case $B_{\mathrm{L}} = 50$ nG. 
The results illustrated in Fig. \ref{Figure1} 
are qualitatively similar for different choices of the parameters close 
to the best fit values.
The inclusion of a magnetized background has a threefold 
effect on the temperature autocorrelations. The height of the 
first peak gets increased. The second peak is distorted and it eventually 
turns into a hump for sufficiently large values of $B_{\mathrm{L}}$ (or of $n_{\mathrm{B}}$). The third 
peak is, at the same time, distorted and raised. 
In Fig. \ref{Figure1} as we move from the plot at the left to the plot at the right 
the spectral index increases. The increase of the spectral slope entails 
also an increase in the distortions. The latter trend, however, is not monotonic at least 
in the case of the first acoustic peak.
A more thorough illustration of this feature will be provided in Section \ref{sec6}. 
It should be borne in mind that, within the conventions established in Section \ref{sec3}
the scale-invariant limit of the magnetic power spectra is realized for $n_{\mathrm{B}} \to 1$. 
In analog terms, the Harrison-Zeldovich limit for the power spectrum of curvature 
perturbations occurs when $n_{\mathrm{s}} \to 1$.

The features illustrated in Fig. \ref{Figure1} and scrutinized 
in the previous paragraph do not depend upon the data sets. 
The same qualitative patterns can be observed if the pivotal model 
is taken to be the best fit inferred from the combination of the WMAP data 
with all the other data. In this case the central values of the cosmological parameters 
are slightly changed \cite{WMAP1,WMAP2,WMAP3} according to:
\begin{equation}
(\Omega_{\mathrm{b}0},\, \Omega_{\mathrm{c}0}, \,\Omega_{\mathrm{\Lambda}}, \,h_{0}, \,n_{\mathrm{s}},\,\tau)= (0.044,\, 0.223,\, 0.733,\, 0.704,\,0.947,\,0.073).
\label{best2}
\end{equation}
By fitting the WMAP data with the ones 
of the gold sample of type Ia supernovae \cite{SN2}
the central values of the cosmological parameters are yet a bit different from the ones reported in Eqs. (\ref{best1}) and (\ref{best2}):
\begin{equation}
(\Omega_{\mathrm{b}0},\, \Omega_{\mathrm{c}0}, \,\Omega_{\mathrm{\Lambda}}, \,h_{0}, \,n_{\mathrm{s}},\,\tau)= (0.045,\, 0.231,\, 0.724,\, 0.701,\, 0.946,\,0.079).
\label{best3}
\end{equation}
\begin{figure}
\begin{center}
\begin{tabular}{|c|c|}
      \hline
      \hbox{\epsfxsize = 7.6 cm  \epsffile{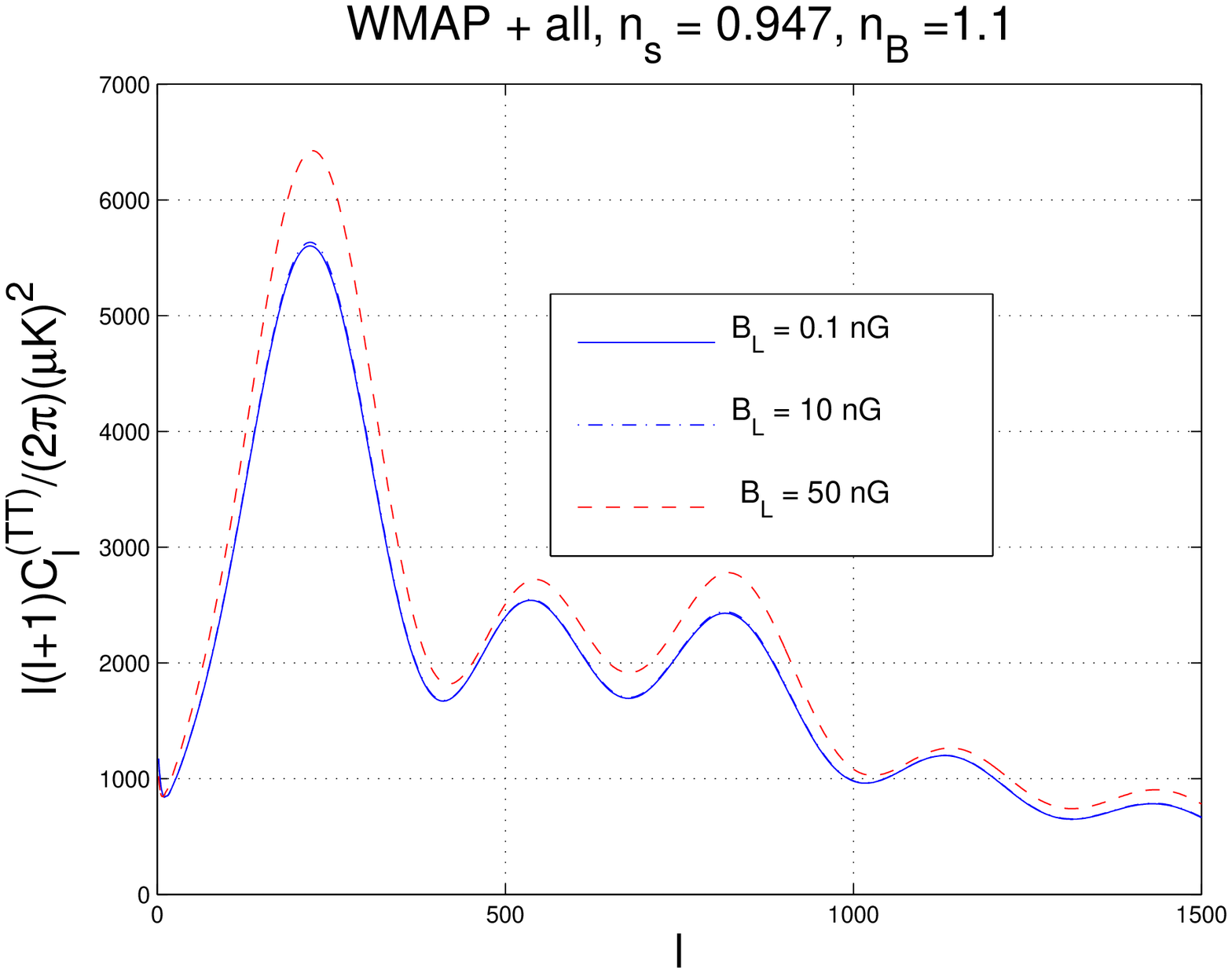}} &
     \hbox{\epsfxsize = 7.6 cm  \epsffile{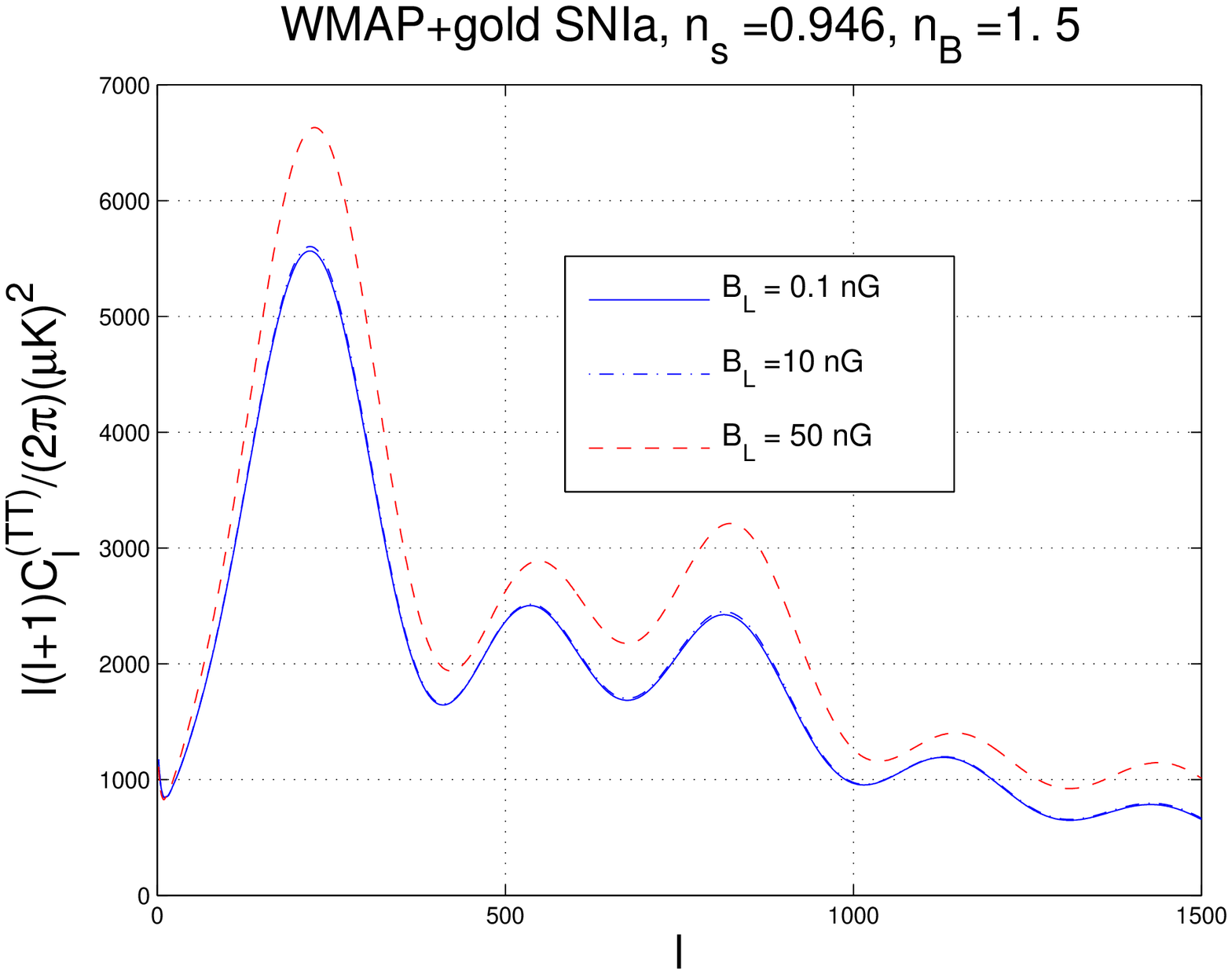}}\\
      \hline
\end{tabular}
\end{center}
\caption[a]{The temperature autocorrelations for blue magnetic spectral indices.
In the plot at the left the cosmological parameters are fixed to the central values of the best fit
when the WMAP data are combined with 
all the cosmological data sets (see Eq. (\ref{best2})). In the plot at the right 
the cosmological parameters are fixed to the central values of the best fit when the 
WMAP data are combined with the gold sample of type Ia supernovae (see Eq. (\ref{best3})). 
As in Fig. \ref{Figure1} the pivotal model is 
 $\Lambda$CDM scenario with no tensors. The value of the magnetic pivot 
scale is $k_{\mathrm{L}} = 1\, \mathrm{Mpc}^{-1}$.}
\label{Figure2}
\end{figure}
In Fig. \ref{Figure2} the temperature autocorrelations are computed when the cosmological 
parameters are fixed as in Eq. (\ref{best2}). As the magnetic field strength 
increases from $10$ to $50$ nG the distortion patterns already illustrated in Fig. \ref{Figure1} become 
more pronounced. 
By comparing Figs. \ref{Figure1} and \ref{Figure2} the same distortion patterns can be 
observed.  In Fig. \ref{Figure2} the spectral tilt increases from the left to the right 
plot. As in Fig. \ref{Figure1}, also in Fig. \ref{Figure2}  the plot at the left is close to the scale-invariant 
limit of the magnetic power spectrum while the plot at the right illustrates a spectral slope 
which is bluer.

The distortion patterns arising in Figs. \ref{Figure1} and \ref{Figure2} have a semi-analytical interpretation.
 In \cite{mg3} 
the effects of magnetic fields on the temperature autocorrelations 
have been discussed in a semi-analytical perspective and for blue spectral indices
 (i.e. $1<n_{\mathrm{B}} < 5/2$). 
A consistent use of the  tight-coupling 
approximation allowed for the estimate of the 
$C^{\mathrm{(TT)}}_{\ell}$ at low multipoles (i.e. $\ell <30$) and also at large multipoles (i.e. 
$\ell >100$).  Using the large-order expansion of the spherical Bessel functions (and of their derivatives)
the shape of the TT correlations has been reduced to the numerical calculation 
of four integrals \cite{mg3}. The semi-analytical approach described in \cite{mg3} seems, a posteriori, 
rather brutal. Nonetheless it is  amusing that the essential patterns of the distortions 
in the acoustic region have been correctly captured.

The Doppler region is sensitive to 
the relative phases and amplitudes of the monopole and dipole terms of the brightness perturbations.
The contribution of the Lorentz force provides an extra source to the monopole equation (see Eq. (\ref{INC4})). A computable difference in the relative amplitudes of the monopole and dipole terms then arises also analytically and it is reflected in the overall distortion. The reasonable results obtainable within the 
tight-coupling expansion (improved to second order) represent a powerful cross-check for the consistency 
both of the numerical approach and of the semi-analytical calculation.
\begin{figure}
\begin{center}
\begin{tabular}{|c|c|}
      \hline
      \hbox{\epsfxsize = 7.6 cm  \epsffile{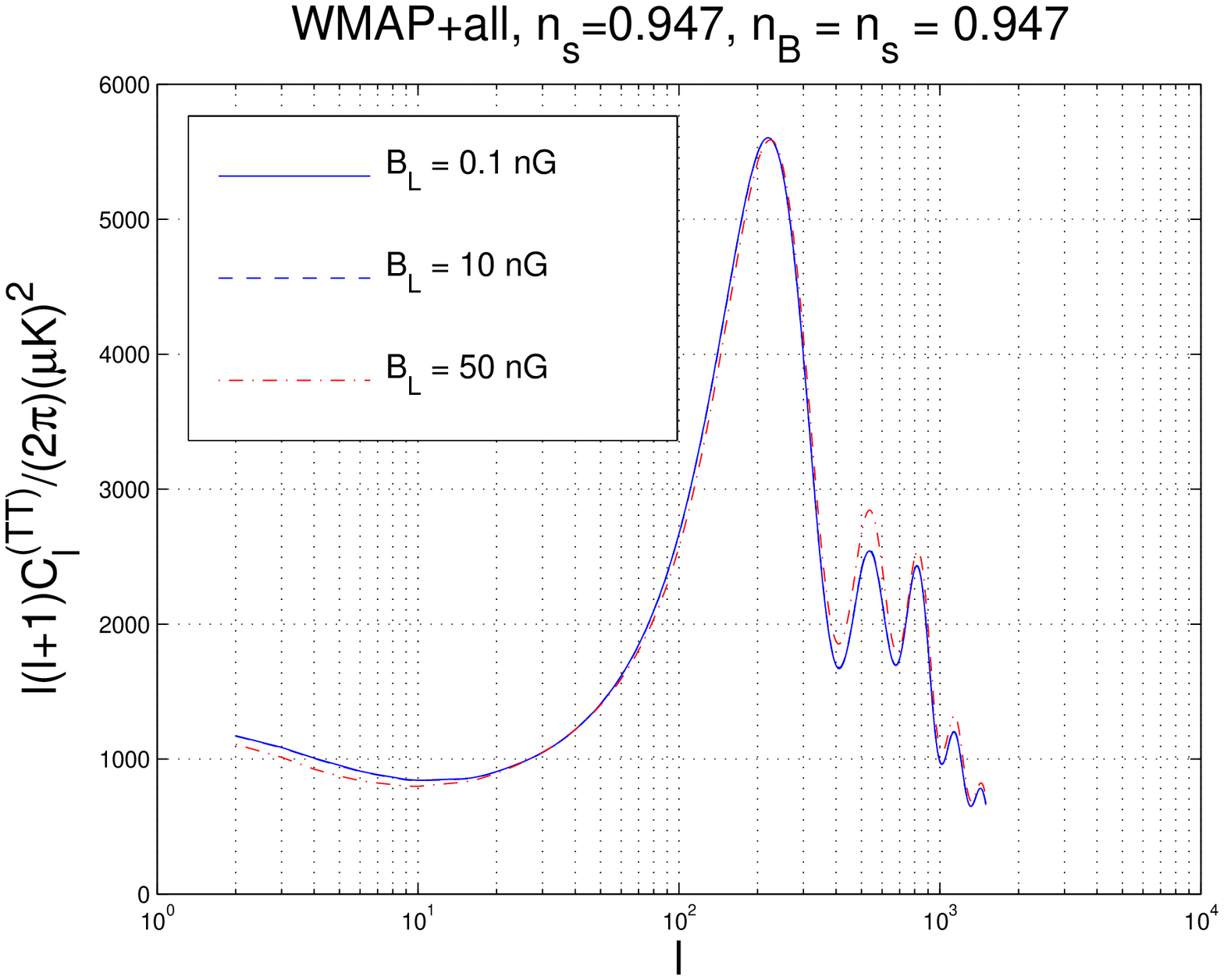}} &
     \hbox{\epsfxsize = 7.6 cm  \epsffile{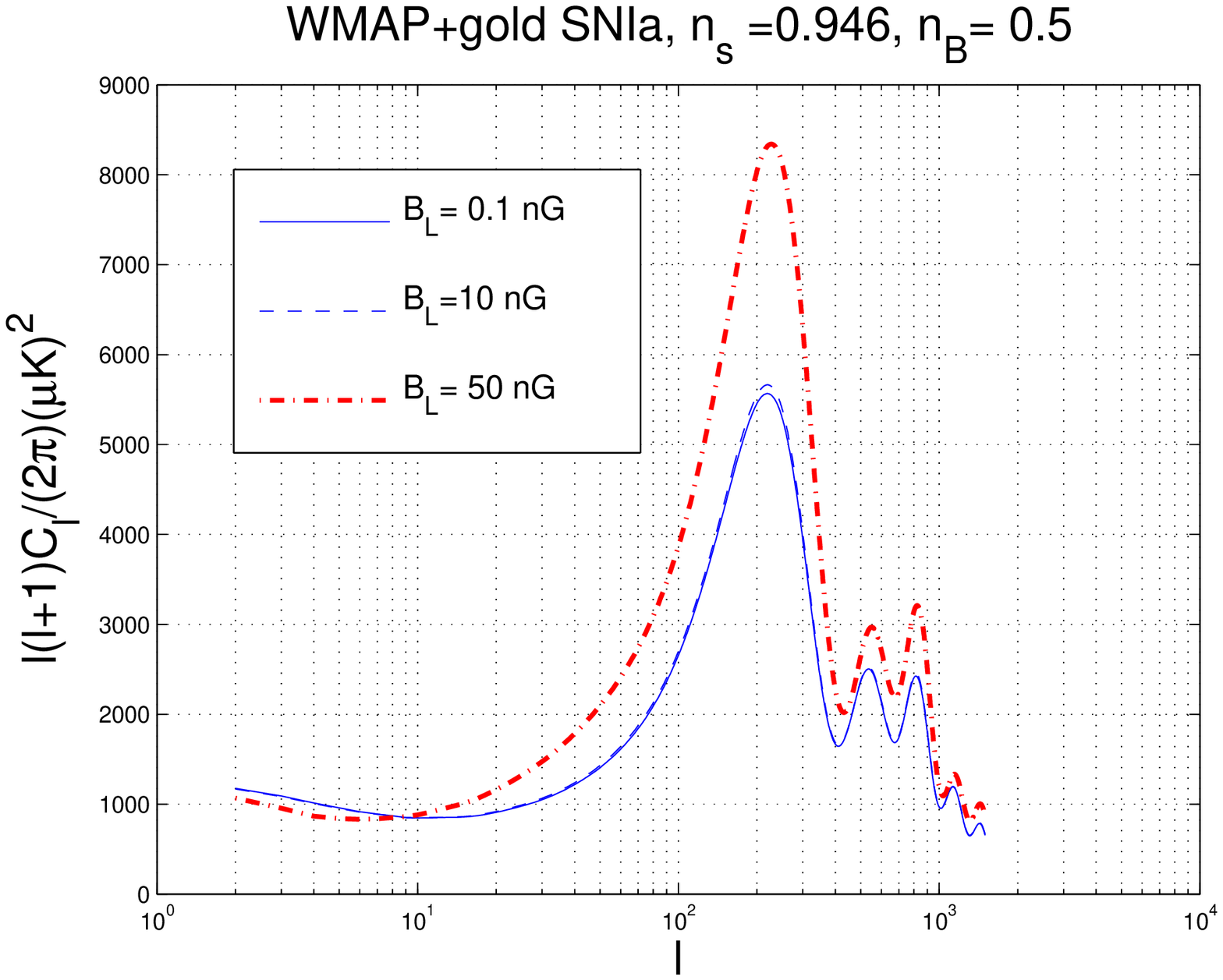}}\\
      \hline
\end{tabular}
\end{center}
\caption[a]{The temperature autocorrelations are illustrated in the case of red tilt. The values 
of the cosmological parameters in the left and right plots have been 
fixed, respectively, as in Eqs. (\ref{best2}) and (\ref{best3}).}
\label{Figure3}
\end{figure}
In spite of the encouraging agreement of the numerical results with the semi-analytical evaluations, a weaker dependence upon the amplitude of the magnetic fields has to be admitted. Indeed, the reionization effects have been neglected in \cite{mg3}  and the 
recombination has been treated within a Gaussian parametrization of the visibility function. 
Furthermore, always in \cite{mg3}, the overall amplitude of the angular power spectrum was determined by matching the low-$\ell$ regime with the Doppler regime where the Bessel functions have been basically replaced with their asymptotic expressions for $\ell  \gg 1$.  Numerically these approximations have been dropped.

The range of spectral indices $1 < n_{\mathrm{B}} <5/2$, on a theoretical ground, is well motivated. 
The two-point function of the magnetic fields decreases, in this 
case at large distances. The diffusive effects are negligible since, as explained in the 
previous section, the two-point functions of the energy density (and of the Lorentz force) is insensitive, in this case, to the ultra-violet cut-off. Finally various magnetogenesis models predict this kind of spectra\footnote{See, for instance, \cite{review1} and \cite{maxrev1} for some reviews on this subject. It would be impossible to refer 
to all the attempts along this direction. For recent results see \cite{kerst} and references 
therein. As specifically discussed in the introduction, the purpose her is not to endorse a particular 
model but to develop the tools which will allow to assess the primordial nature of the magnetic field. In this sense the goal of the present analysis is more modest.}. 

The last motivation, however, is just accidental.
The true question behind these considerations is slightly different and can be phrased by asking:
which is the spectrum of magnetic fields at the onset of gravitational collapse of the protogalaxy?
As we can in principle measure the matter power spectrum it would 
not be insane to think that, in a future, also the magnetic power spectra of different objects 
could be measured. Indeed there are attempts to characterize, for instance, the present 
features of our galactic magnetic field in terms of an appropriate power spectrum 
\cite{han}. In a related perspective one could observe that it is equally plausible 
to study the mean-squared fluctuation of the Faraday Rotation Measure (RM), as it was proposed
in Refs. \cite{goldshmidt,crusius}. One of the key projects 
of the radio-astronomy community is the celebrated Square Kilometer Array (SKA)\footnote{The collecting 
area of the instrument, as the name suggest, will be of $10^{6}\, {\rm m}^2$.
The specifications for the SKA require an angular resolution of $0.1$
arcsec at $1.4$ GHz, a frequency capability of $0.1$--$25$ GHz, and a field of view of at 
least $1\,{\rm deg}^2$ at $1.4$ GHz \cite{SKA,feretti}.  The number of independent beams is expected to be larger than $4$ and the number of instantaneous 
pencil beams will be roughly 100 with a maximum 
primary beam separation of about $100$ $\mathrm{deg}$ at low frequencies 
(becoming $1$ $\mathrm{deg}$ at high frequencies, i.e. of the order 
of $1$ GHz).
These specifications will probably allow full sky surveys of Faraday Rotation.}. This 
instrument will allow to obtain full sky survey of the RM and, in that 
context, it will be even more plausible to collect valuable informations 
on the magnetic power spectra at large scales. These measurements, even if 
feasible in the future, will not provide direct indications on the protogalactic 
field but rather on the present field. Still it is not excluded that the morphological 
features of the observed field could be connected with the protogalactic features.
If the magnetic field does not flip its sign 
from one spiral arm to the other, then a strong dynamo action 
can be suspected \cite{beck1}. In the opposite case the magnetic field of spiral galaxies should 
be primordial i.e. present already at the onset of gravitational collapse.
 An excellent review on the evidence of magnetism in nearby galaxies can be found in \cite{beck2}.
\begin{figure}
\begin{center}
\begin{tabular}{|c|c|}
      \hline
      \hbox{\epsfxsize = 7.6 cm  \epsffile{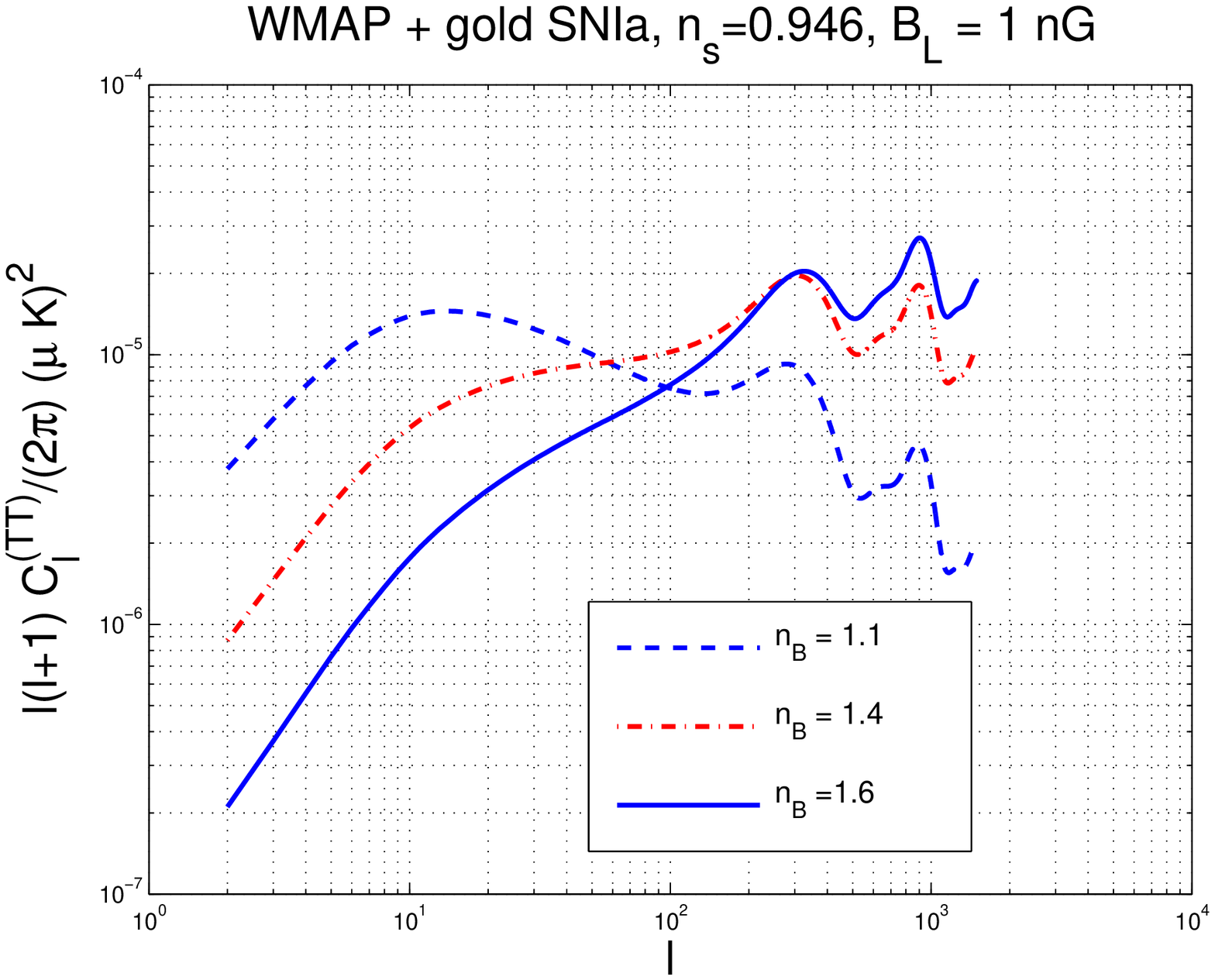}} &
     \hbox{\epsfxsize = 7.6 cm  \epsffile{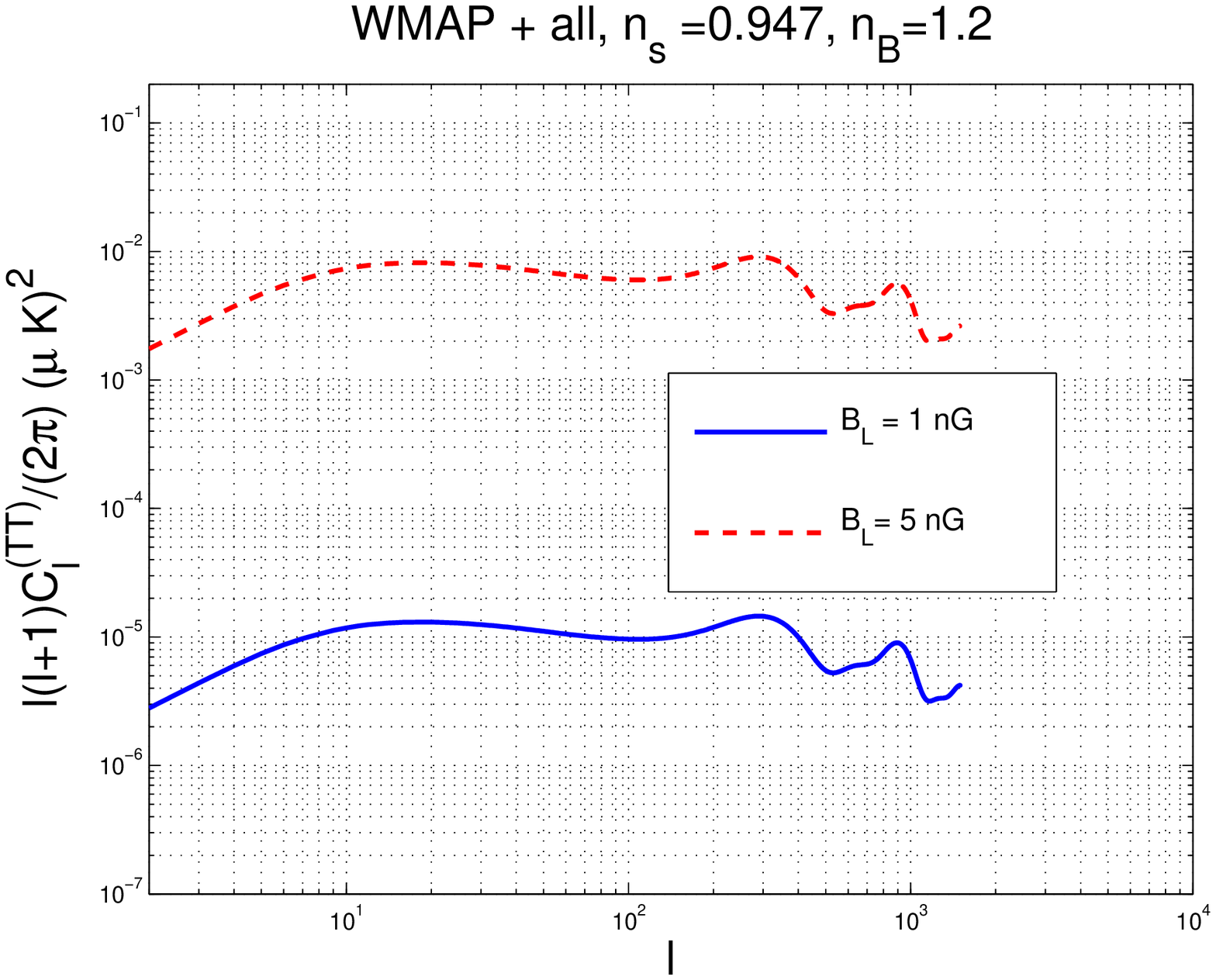}}\\
      \hline
\end{tabular}
\end{center}
\caption[a]{The temperature autocorrelations are illustrated in the case of vanishing adiabatic mode 
for fixed magnetic field strength (plot at the left) and for fixed magnetic spectral index (plot at the right). The magnetic pivot scale is taken to be $k_{\mathrm{L}} = 1 \mathrm{Mpc}^{-1}$ and the 
other values of the cosmological parameters are fixed to their best fit values 
as in Eq. (\ref{best2}) (plot at the left) and as in Eq. (\ref{best3}) (plot at the right).}
\label{Figure4}
\end{figure}
In the model-independent approach followed in the present paper it is natural to ask what happens if 
the magnetic power spectra have a red tilt. 
In Fig. \ref{Figure3} (left plot) the spectral tilts of the magnetic power spectrum 
and of the spectrum of curvature perturbations coincide. In other words 
$n_{\mathrm{s}} = n_{\mathrm{B}}= 0.947<1$. The specific figure (i.e. $0.947$) is dictated 
by the adoption, as fiducial set of data, of the best fit to the WMAP data combined 
with all the other data (see Eq. (\ref{best2}) and also the titles of the plots in Fig. \ref{Figure3}). 
In the plot at the right of Fig. \ref{Figure3} the spectral indices of the magnetic energy density 
and of the anisotropic stress decreases (i.e. $n_{\mathrm{B}} = 0.5 \ll n_{\mathrm{s}}$). 
So we can say that, in Fig. \ref{Figure3} the magnetic spectral index is redder at the right 
than at the left. The two plots in Fig. \ref{Figure3} have been presented in semi-logarithmic 
coordinates and this choice allows to scrutinize in more depth the main feature associated 
with red spectral indices: as the spectral index becomes redder, a systematic 
decrease of the low multipoles is observed. This trend has been investigated 
with other examples (which will not be reported for reasons of space). The conclusion 
of this analysis is that, indeed, red spectra can lead to a lower quadrupole 
and, only apparently, improve the agreement with the data. We say 
{\em only apparently} because every time the TT correlation diminishes 
at low multipoles, the first pair of acoustic peaks is raised and distorted 
to an unacceptable degree. This aspect can be appreciated, in a rather 
extreme case, in the plot at the right of Fig. \ref{Figure3}. For $B_{\mathrm{L}} = 50$ nG 
the low multipoles would indeed represent better the experimental points. However, the 
Doppler peak explodes to, roughly, $9000$$(\mu\mathrm{K})^2$.

 When the spectra have a red tilt 
 the magnetic pivot scale coincides 
effectively with the infra-red cut-off of the spectrum. 
If the cut-off is as large as the present 
Hubble patch the magnetic field acquires, for practical purposes a preferred direction. 
One could be tempted to say that this offers an explanation of the lower 
value of the quadrupole. Indeed various proposals have been put forward 
to explain the quadrupole with a specific anisotropic model falling in one of 
the Bianchi classes.  The consistency of our numerical approach shows, 
however, that these kinds of red spectra lowering the quadrupole 
are simply pathological. Indeed, as it appears from Fig. \ref{Figure3}, red spectral slopes 
lowering the quadrupole and spontaneously breaking spatial isotropy are not consistent 
at higher multipoles. We suspect that current attempts of justifying the quadrupole anomaly 
based on anisotropic models would simply fail when confronted with the higher multipoles. 
The considerations reported here are a first quantitative indication in this direction. 
It would be interesting to pursue this analysis in further detail \cite{far6}.
The quadrupole 
anomaly is probably better addressed in the framework of 
pre-inflationary initial conditions such as the ones 
discussed in \cite{ndh1,ndh2} where the low value of the 
quadrupole is attributed to a fast-roll phase.  
\begin{figure}
\begin{center}
\begin{tabular}{|c|c|}
      \hline
      \hbox{\epsfxsize = 7.6 cm  \epsffile{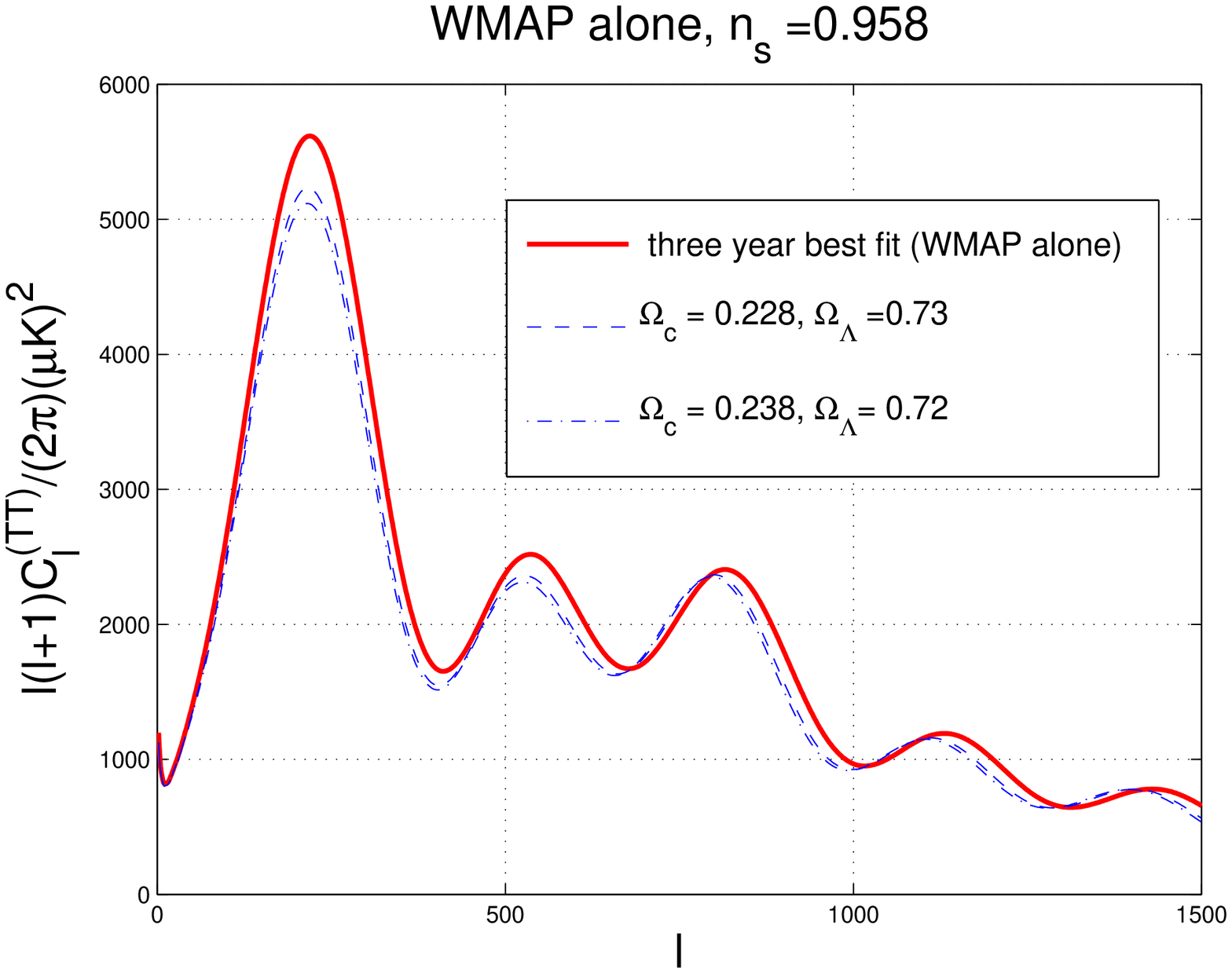}} &
     \hbox{\epsfxsize = 7.6 cm  \epsffile{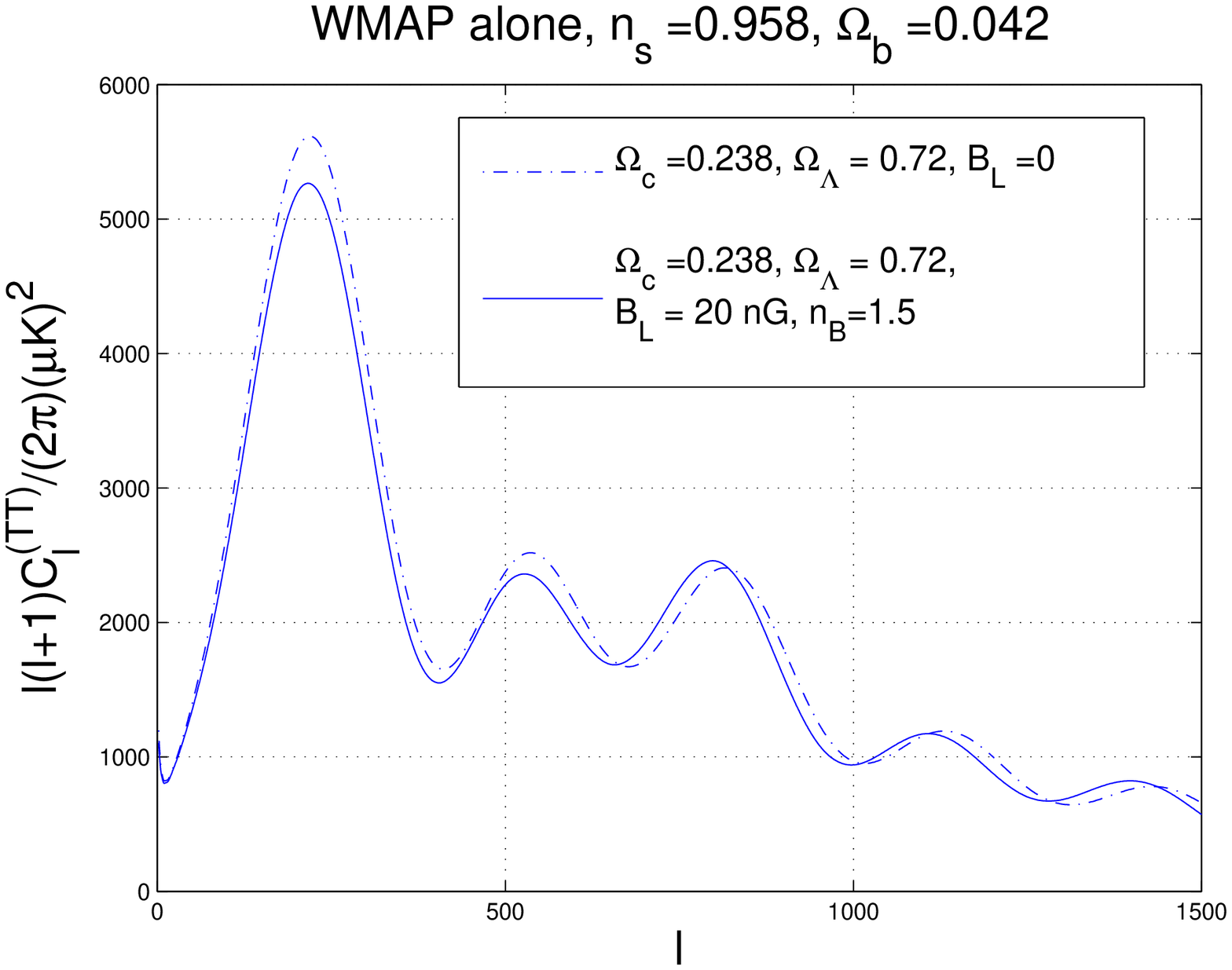}}\\
      \hline
\end{tabular}
\end{center}
\caption[a]{The effect of the magnetic fields is illustrated in the case when 
the CDM fraction is increased by enforcing the spatial flatness and when the baryonic fraction is kept fixed to the best fit value.}
\label{Figure4a}
\end{figure}

The final comment in connection with red spectra deals with regularized magnetic field 
of regularized amplitude much smaller than the nG. In both plots of Fig. \ref{Figure3}
the case of $B_{\mathrm{L}} = 0.1$ nG cannot be distinguished from the 
corresponding best fit model. The latter observation suggests that 
weak magnetic fields (i.e. $B_{\mathrm{L}}  \simeq {\mathcal O}( 0.1 \,\mathrm{nG})$) with red
tilt are not incompatible with current data on the TT correlations.

If the adiabatic mode 
of primeval origin is switched to zero, we can expect, on the basis 
of purely analytical considerations, that the typical amplitude of the temperature autocorrelations will be 
vanishingly small to begin with. Indeed, the overall amplitude will not be controlled
by the power spectrum of the adiabatic mode but by the power spectrum of the
magnetic energy density, i.e. ${\mathcal P}_{\Omega}(k)$. The amplitude of the 
adiabatic power spectrum evaluated at the pivot scale $k_{\mathrm{p}}$ is of the order of $10^{-9}$.
The amplitude of ${\mathcal P}_{\Omega}(k)$, evaluated at the magnetic pivot scale $k_{\mathrm{L}}$ goes as $\overline{\Omega}_{\mathrm{BL}}^2$ 
which is ${\mathcal O}(10^{-18})$ for  $B_{\mathrm{L}} \simeq \mathrm{nG}$. Consequently 
if the adiabatic contribution is switched off we will expect that the temperature autocorrelations 
will be about $9$ orders of magnitude smaller than in the case where the adiabatic mode 
was present. This means that while in the case of, for instance, Fig. \ref{Figure1} 
$\ell(\ell +1)C_{\ell}^{(\mathrm{TT})}/(2\pi) \simeq 10^{3} (\mu\,\mathrm{K})^2$, when the adiabatic mode 
is switched to zero  $\ell(\ell +1)C_{\ell}^{(\mathrm{TT})}/(2\pi) \simeq 10^{-6} (\mu\,\mathrm{K})^2$.

In Fig. \ref{Figure4} the theoretical expectations are confirmed by the numerical results. 
In the plot at the right $B_{\mathrm{L}} = 1\, \mathrm{nG}$. For different values of the spectral indices
the TT correlation exhibit a humpy profile in the acoustic region.  This 
point is also stressed in the right plot of Fig. \ref{Figure4} where for fixed $n_{\mathrm{B}}$ the 
magnetic field strength is enhanced from $1$ nG to $5$ nG. This entails an increase 
of the temperature autocorrelations of a factor $6 \times 10^{2}$ which fits with our expectation 
which would be, in this case, $(5)^{4}$.
According to Fig. \ref{Figure1} the case $B_{\mathrm{L}} = 50 \, \mathrm{nG}$ is 
already excluded by the present data and for the corresponding values 
of the spectral index. This observation helps along two opposite directions.
The nature of the distortion induced by the magnetic fields seems to be hard to reproduce by 
varying the standard CMB parameters. For instance it is known that by lowering 
$\omega_{\mathrm{b}}$ the height of the peaks diminishes. Similar effects (but with a 
different quantitative impact) are observed when $\omega_{\mathrm{c}}$ increases
(always enforcing the flatness of the model). None of these two effects distorts 
the peaks as in the case of magnetic fields. One can also think that by adding spatial 
curvature and by either decreasing $\omega_{\mathrm{b}}$ or increasing $\omega_{\mathrm{c}}$ 
the effects of the magnetic fields can be appropriately mimicked. However, in the case of the 
magnetic fields not only the first peak increases but also the ratio of the second to the first peak
is modified. The shift in the position of the peaks is much more 
severe in non-flat models than in the case of nG magnetic fields.

In Fig. \ref{Figure4a} (plot at the left) the baryonic fraction $\Omega_{\mathrm{b}0}$ has been 
fixed to the best fit value of the WMAP data alone (see Eq. (\ref{best1})). The CDM contribution 
has then been increased (by always keeping the model flat). In the right plot we took the most extreme 
model illustrated in the left plot (i.e. the one labeled by the dot-dashed line) and 
compared it with the same model where, however the magnetic field is included. This 
shows that the kind of correlated distortion induced by the magnetic fields cannot be simply reduced 
to an increase of the peaks (see also Section \ref{sec6} for a more extended scrutiny 
of this statement). We analyzed  the characteristic shapes obtainable by changing 
also other parameters as the Hubble rate and, also in that case,  the magnetic fields
induce distortions which cannot be mimicked by known shape effects. 
Under certain circumstances, a slight increase 
in the CDM fraction (less extreme than those illustrated in Fig. \ref{Figure4a}) can be 
compensated by the presence of a minute magnetized background. 

This type of considerations bring up naturally the need of including the magnetic fields 
as an extra set of parameters in the current strategies of data analysis \cite{giokunz1}.
In its simplest realization the magnetized $\Lambda$CDM paradigm entails the inclusion 
of two new parameters, i.e. the magnetic spectral index and the amplitude 
of the regularized field. It cannot be excluded, in other words, that a combined 
action of different effects will be compensated by a magnetic field leading, ultimately, to a better 
fit. The accuracy of forthcoming data (see also Section \ref{sec6}) seems to suggest 
that we will soon be sensitive to nG magnetic fields and, then, global strategies 
of parameter extraction will allow either to confirm the $\Lambda$CDM paradigm or to improve 
it.

A handy parallel can be drawn with a slightly different physical case which however bears 
some analogy with the one discussed here.
A commonly employed approach 
to the initial conditions is the one we could define as model-independent 
(see, for instance, \cite{h1,h2}).
When analyzing cosmological data a very interesting question is to know 
if the data allow for a sizable non-adiabatic component. It is by now well established how to constrain 
the CDM-isocurvature mode \cite{h3,h4}. This analysis entails, in the simplest case, the addition of two 
extra-parameters, i.e. the amplitude and spectral index of the non-adiabatic mode. However 
there could be even more complicated situations where a cross-corrrelation term is present. This 
term parametrizes a possible correlation between the adiabatic an the non-adiabatic components and 
typically leads to further parameters. Recently interesting results have been reported in this context. For 
instance it has been shown that indeed the addition of an adiabatic component with blue spectrum may improve 
the global fits of cosmological parameters \cite{h5}. 
\begin{figure}
\begin{center}
\begin{tabular}{|c|c|}
      \hline
      \hbox{\epsfxsize = 7.6 cm  \epsffile{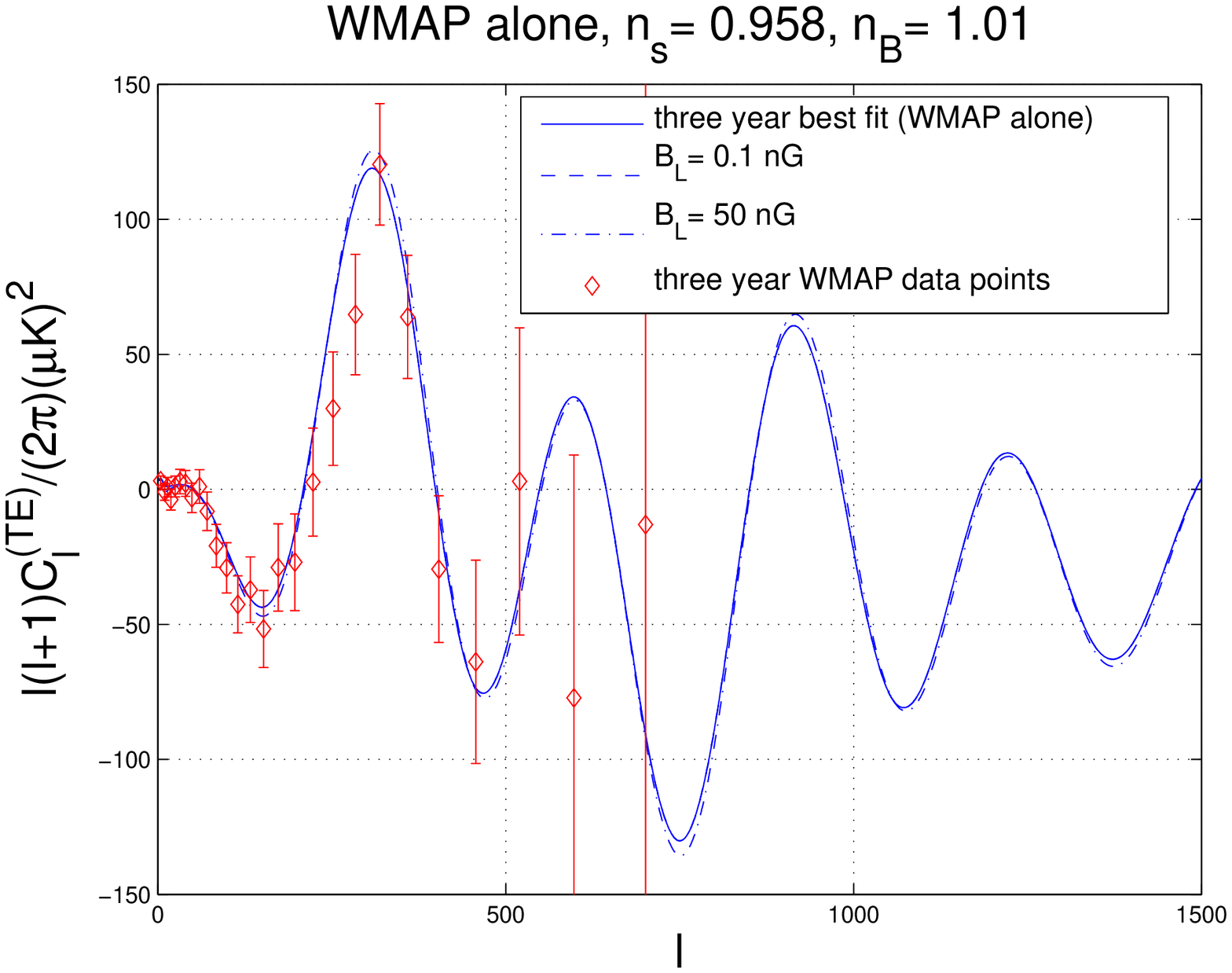}} &
     \hbox{\epsfxsize = 7.6 cm  \epsffile{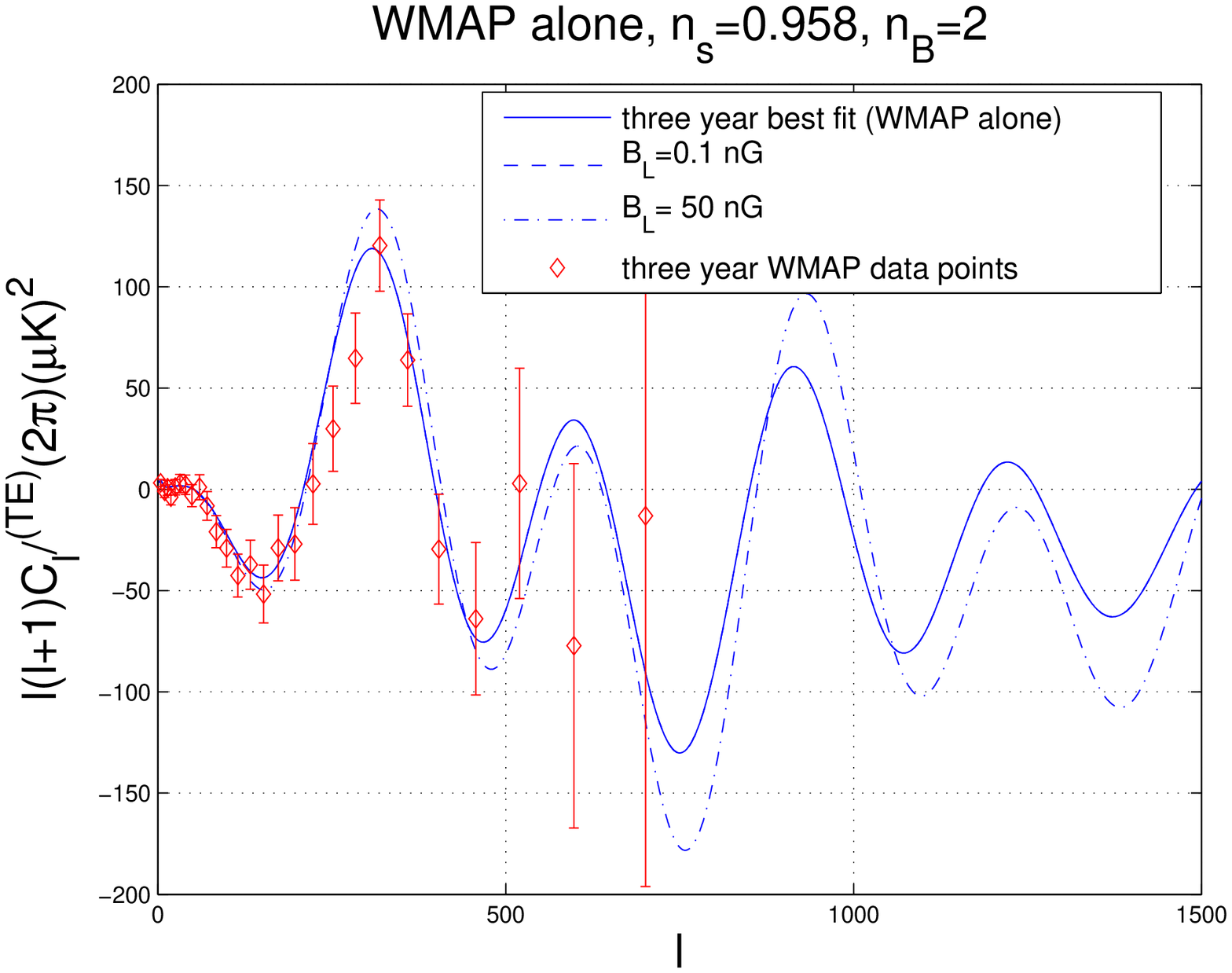}}\\
      \hline
\end{tabular}
\end{center}
\caption[a]{The magnetized TE correlations are illustrated for different values of the spectral indices and different 
values of the magnetic field intensities. The fiducial set of the cosmological parameters is the one 
of Eq. (\ref{best1}).}
\label{Figure5}
\end{figure}
The situation described in the case of the isocurvature modes is similar to what happens in the case 
of large-scale magnetic fields with the crucial difference that, in the present case, not only the initial conditions 
but also the dynamics is affected by the addition of stochastic magnetic fields. As 
customarily done, for other parameters, it will be appropriate to include the magnetic fields 
when confronting all the cosmological data sets. This idea will allow to set bounds and compare fits 
in a way which is less brutal than the one sometimes employed when dealing with large-scale 
magnetic fields. The numerical approach 
developed and applied in the present study is the first step in this direction which 
we plan to investigate throughly in the near future \cite{far6}.

Always in connection with the isocurvature modes we wish to stress 
that large-scale magnetic fields can be included also in the case when the initial conditions 
are not predominantly adiabatic but rather obtained as a mixture of adiabatic and non-adiabatic 
components. 
In this study, for reasons of space, we just focused on the 
magnetized adiabatic mode. 
It is therefore possible to study, with our approach, all the usual situations encountered in 
conventional CMB calculations \cite{far6}. 

\renewcommand{\theequation}{5.\arabic{equation}}
\setcounter{equation}{0}
\section{Polarization correlations and cross-correlations}
\label{sec5}
The TE cross-correlations 
are probably the strongest indicator of the adiabatic nature of the CMB initial 
conditions. Indeed, in the adiabatic case, the $C_{\ell}^{(\mathrm{TE})}$ 
shows a characteristic anticorrelation peak for $\ell \simeq (3/4) \ell_{\mathrm{Doppler}} \simeq 150$ 
where $\ell_{\mathrm{Doppler}}$ denotes the observed position of the Doppler peak \cite{WMAP4,mg2}. The relation between 
Doppler and anticorrelation peaks is a distinctive feature of the adiabaticity of the fluctuations prior to recombination. 
\begin{figure}
\begin{center}
\begin{tabular}{|c|c|}
      \hline
      \hbox{\epsfxsize = 7.6 cm  \epsffile{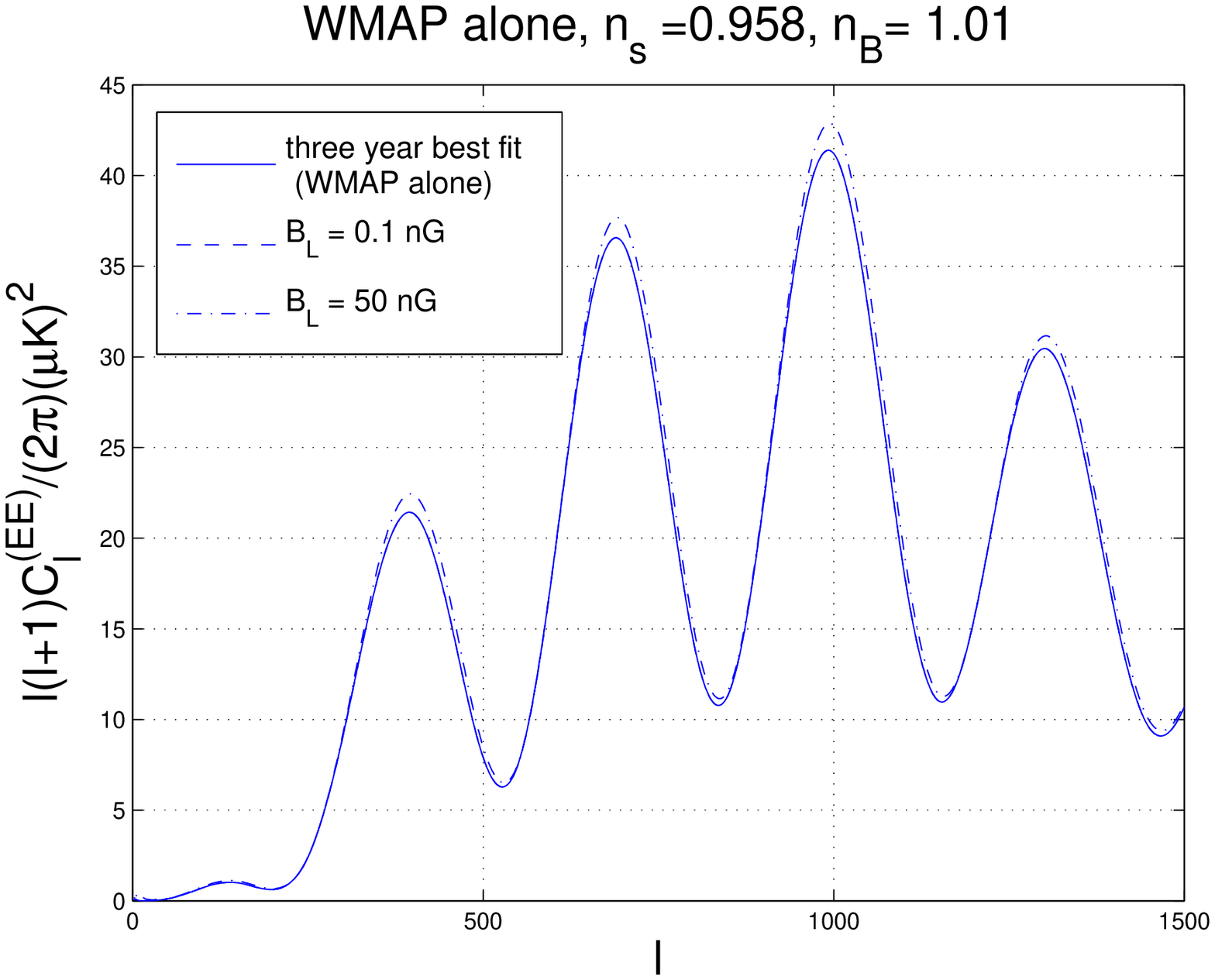}} &
     \hbox{\epsfxsize = 7.6 cm  \epsffile{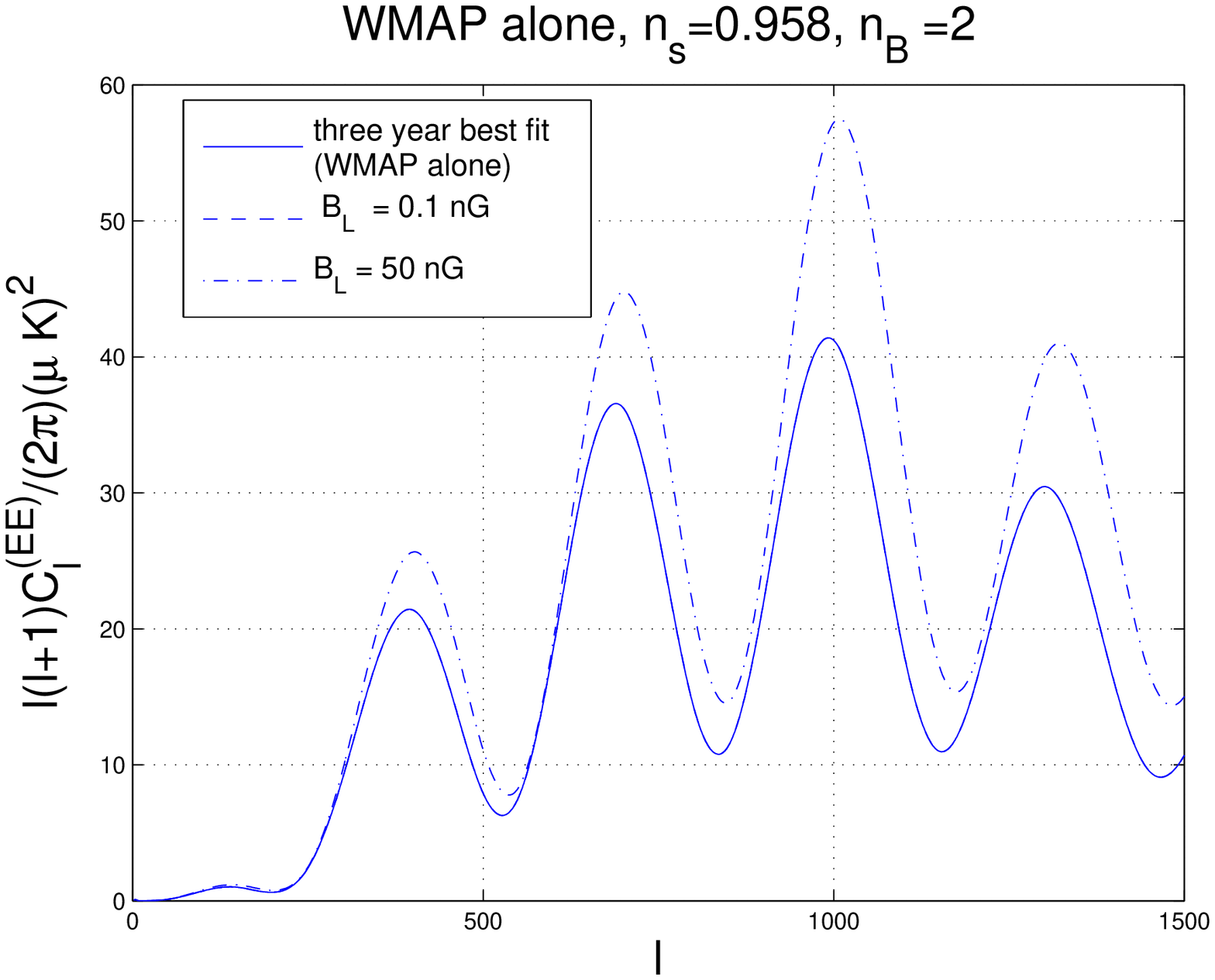}}\\
      \hline
\end{tabular}
\end{center}
\caption[a]{The magnetized EE correlations for blue values of the magnetic spectral tilt. The cosmological 
parameters are the same as in Fig. \ref{Figure5}.}
\label{Figure6}
\end{figure}
The polarization observations are therefore a rather sensitive tool which can be used 
to scrutinize the possible contribution of a magnetized component. 
At the moment various experiments reported a positive detection of the EE and TE correlations.
Besides the three year results of the WMAP collaboration \cite{WMAP2} there are, at the moment,
the three year data of the DASI experiment \cite{dasi3yr}, the CAPMAP results \cite{CAPMAP}, the 
(almost) three year results of CBI \cite{CBI} and the preliminary results of QUAD \cite{Quad}.
In the present version of the code 
 the polarization correlations and cross-correlations (i.e EE and TE power spectra) 
can be explicitly computed since consistent initial conditions have been given for the whole 
Boltzmann hierarchy. The magnetic field can have two distinct effects on the CMB polarization. 
Since gravitating magnetic fields 
modify the structure of the adiabatic mode 
and of the evolution of the baryon-photon fluid, the TE and EE angular power spectra will be different.
The second effect would  be due to the presence of a Faraday rotation term which would couple 
the evolution equations for the two brightness perturbations which are sensitive to polarization. 
In the language of Eqs. (\ref{BR2}) and (\ref{BR3}) 
this term would couple the $U$ and $Q$ Stokes parameters producing, ultimately, a rotation of the 
polarization plane of the CMB. The Faraday coupling can be easily included if the magnetic field 
is uniform \cite{far1}. In the case of stochastic magnetic field this calculation has never been done. There 
are certainly semi-analytical attempts in this direction (see, for instance, \cite{far2,far3,far4}). However, the main problem with these calculations is that they assume that the magnetic fields only rotate 
the polarization without entering in any other place of the evolution equations. 
There are, on the contrary, reasons to believe that, for a stochastic field, the two effects 
can be equally important\footnote{In nearly all Faraday rotation studies it is assumed 
that the description of the plasma is given in terms of a single fluid. This assumption is 
not correct \cite{far5}. Indeed Faraday rotation requires necessarily a kinetic treatment (or at least 
a two fluid treatment). We are investigating the possibility within our numerical approach \cite{far6}.}.
 In this discussion we will not include the Faraday rotation 
term by a uniform field since this would break explicitly the spatial isotropy and concentrate 
on the numerical solution when the magnetic fields are consistently introduced in the initial conditions
and in all the other evolution equations, as it was done for the TT correlations in the previous section.
\begin{figure}
\begin{center}
\begin{tabular}{|c|c|}
      \hline
      \hbox{\epsfxsize = 7.6 cm  \epsffile{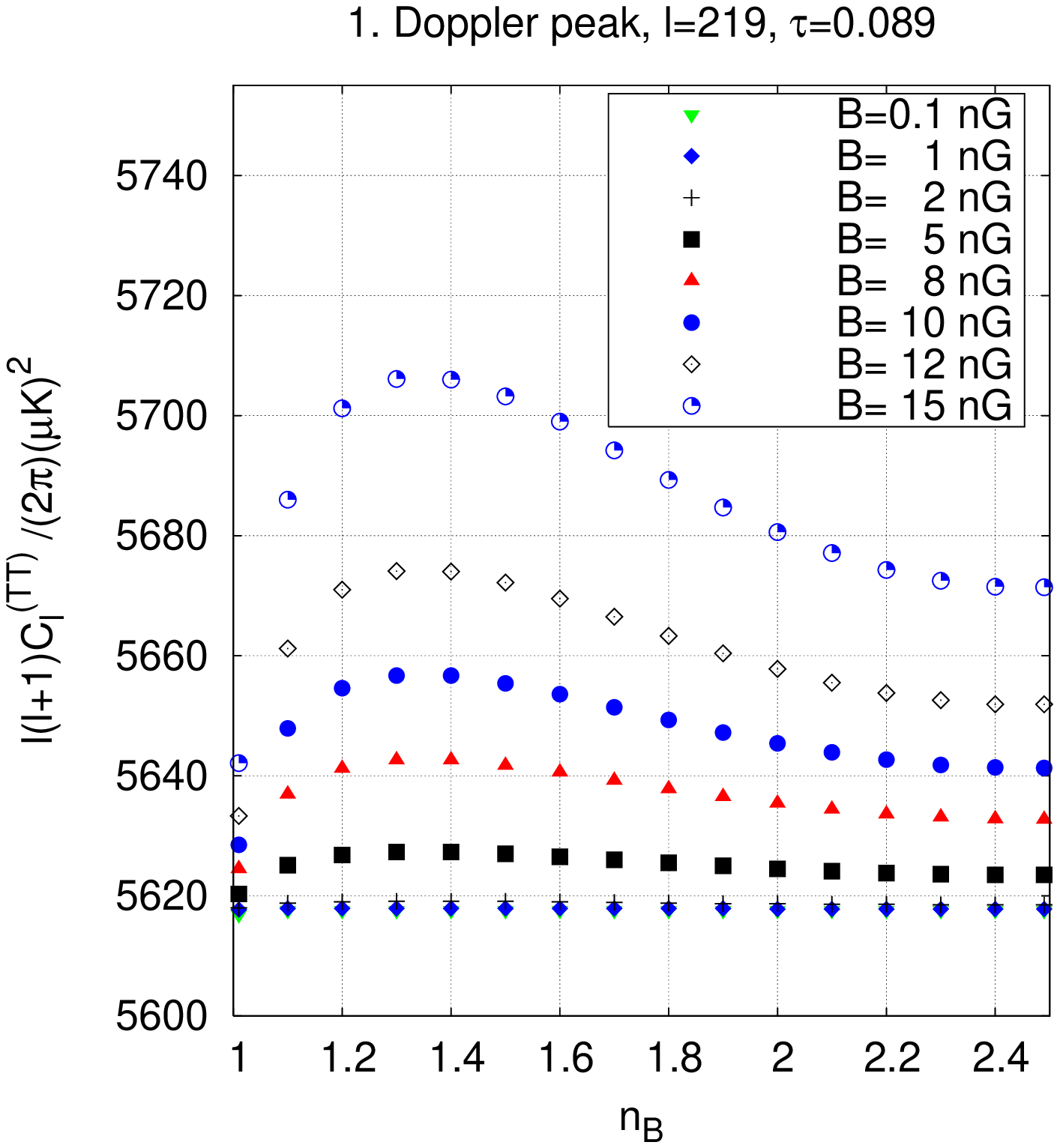}} &
     \hbox{\epsfxsize = 7.6 cm  \epsffile{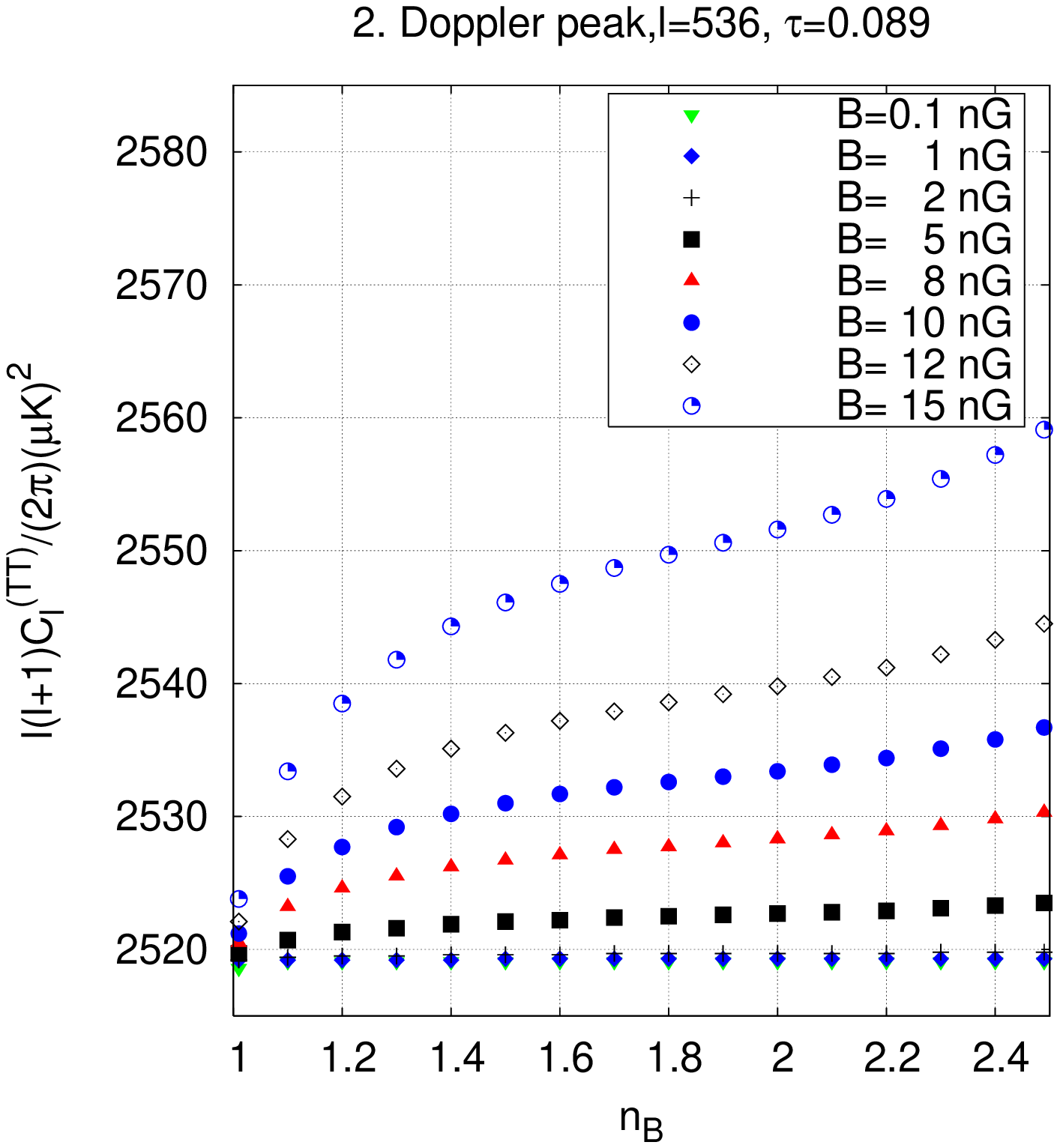}}\\
      \hline
\end{tabular}
\end{center}
\caption[a]{The height of the first and second peaks for different values 
of the magnetic field intensity as a function of the magnetic spectral index. The other 
cosmological parameters have been fixed as in Eq. (\ref{best1}).}
\label{Figure7}
\end{figure}

 In Fig. \ref{Figure5} the TE correlations are illustrated for the case of blue magnetic spectral indices
$1 < n_{\mathrm{B}} < 5/2$.  The parameters 
used in the calculation are exactly the ones employed in Fig. \ref{Figure1}. If the magnetic field is of the order of $0.1$ nG the magnetized
TE correlations  cannot be distinguished from the three year best fit of the WMAP data. 
Unlike the case of the temperature autocorrelations (where the position of the Doppler peak 
cannot be moved by a stochastic magnetic field)  there is an observable shift of the 
second and third (correlation) peaks of the TE spectra. This distortion also entails a shift 
of the position  of the corresponding peaks.
A similar effect can be observed in the magnetized EE correlations which are reported in Fig. \ref{Figure6}. 
Also in this case the peaks are raised and partially shifted. 
Figures \ref{Figure5} and \ref{Figure6} show, a posteriori, that the magnetic fields also affect 
the polarization observables even {\em without} a Faraday rotation term. This observation 
supports our previous statements. The physical reason of the obtained result can be 
understood very simply. To zeroth-order in the tight-coupling expansion, 
the magnetic field affects the dipole of the brightness perturbation 
for the intensity. Always to zeroth order, this contribution is reflected 
in a further source term for the monopole. But both the TE and EE 
power spectra arise to first-order in the tight-coupling expansion 
and are proportional to the first-order dipole through a term which is, up to a numerical 
factor, $k/\epsilon'$ \cite{mg2}. This 
shows why we also get an effect on the polarization observables even if 
the Faraday rotation term is absent. 

The results obtained so far show that it is possible to obtain accurate 
estimates of the temperature autocorrelations and of the polarization correlations 
also in the presence of a magnetized background. Conventional CMB 
calculations have a high level of accuracy and this is due, both, to the 
precise understanding of the initial conditions and to the thorough 
comprehension of the dynamics. At the moment,
it is possible to achieve the same level of accuracy also when a magnetized 
background is included. 

\renewcommand{\theequation}{6.\arabic{equation}}
\setcounter{equation}{0}
\section{Waiting for Planck}
\label{sec6}
The Planck explorer satellite \cite{Planck} will provide high precision measurements of the 
cosmic microwave background. In view of this exciting time the quantitative trends illustrated  in Sections \ref{sec4} and \ref{sec5} will now be scrutinized in more depth.
The Doppler peaks are both distorted and increased. 
For $2<\ell< 2500$ the extrapolated best fit to the WMAP data alone 
predicts, at different locations, $7$ acoustic peaks.
The 
heights and shapes of the $7$ peaks have been monitored for different values of the magnetic field intensity 
and of the magnetic spectral index. For reasons of space we will just focus 
on the first and second and on the sixth and seventh. This will suffice for the quantitative trend we wish 
to illustrate.

\begin{figure}
\begin{center}
\begin{tabular}{|c|c|}
      \hline
      \hbox{\epsfxsize = 7.6 cm  \epsffile{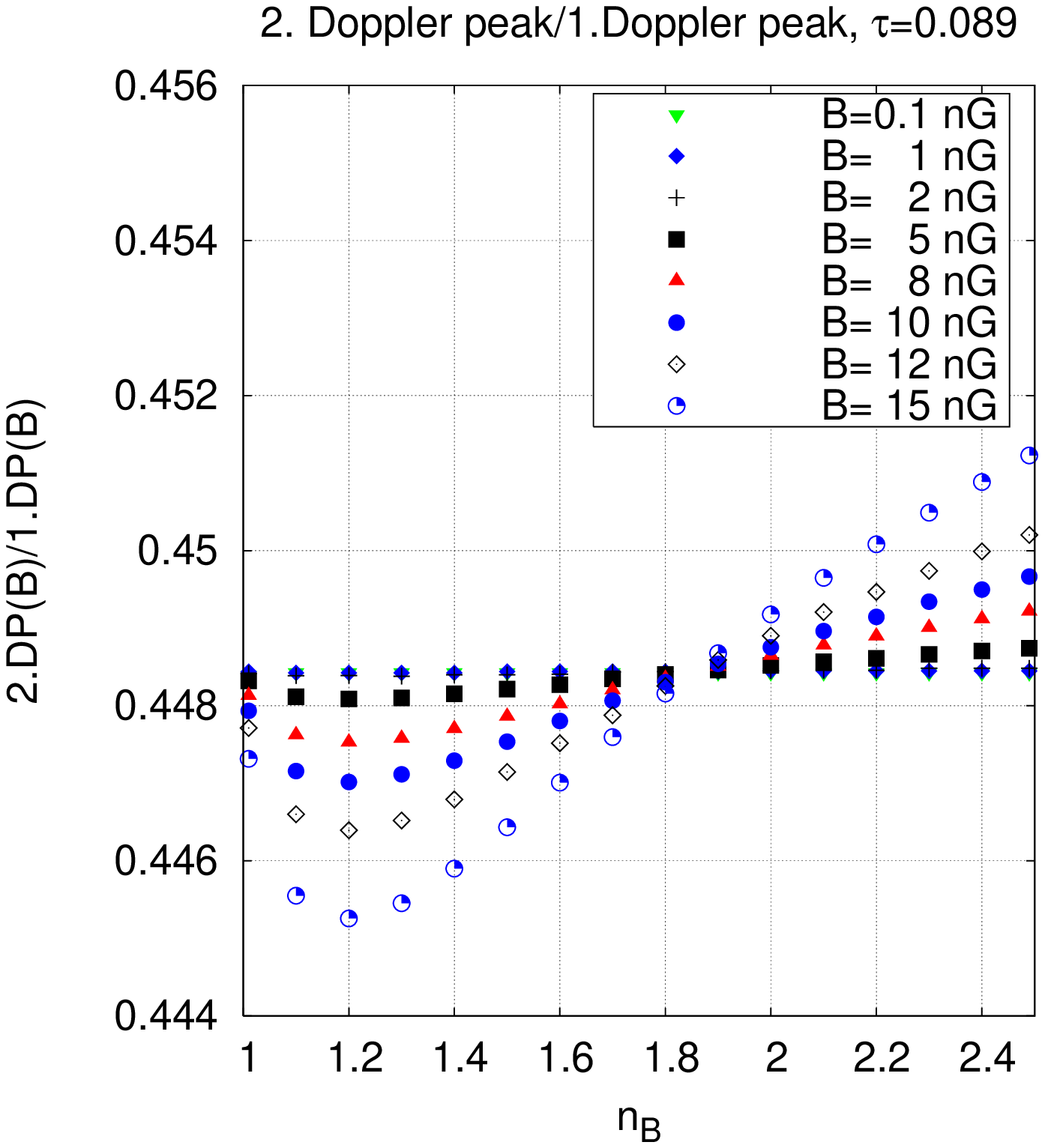}} &
     \hbox{\epsfxsize = 7.6 cm  \epsffile{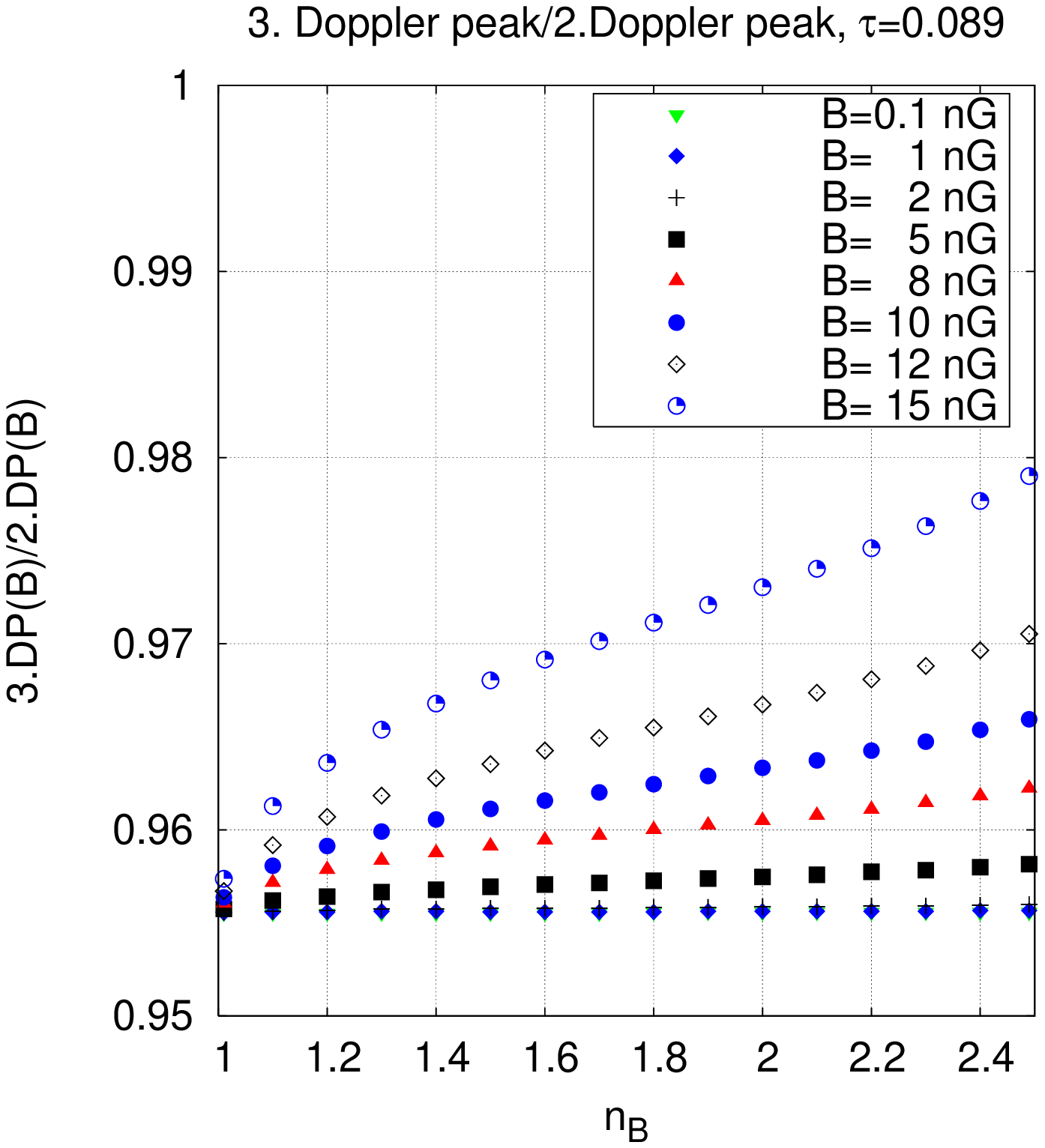}}\\
      \hline
\end{tabular}
\end{center}
\caption[a]{The ratio of the heights of the second to the first peak (plot at the left) 
and of the heights of the third to the first peak (plot at the right). The parameters 
are fixed as in Fig. \ref{Figure7}.}
\label{Figure8}
\end{figure}

\begin{figure}
\begin{center}
\begin{tabular}{|c|c|}
      \hline
      \hbox{\epsfxsize = 7.6 cm  \epsffile{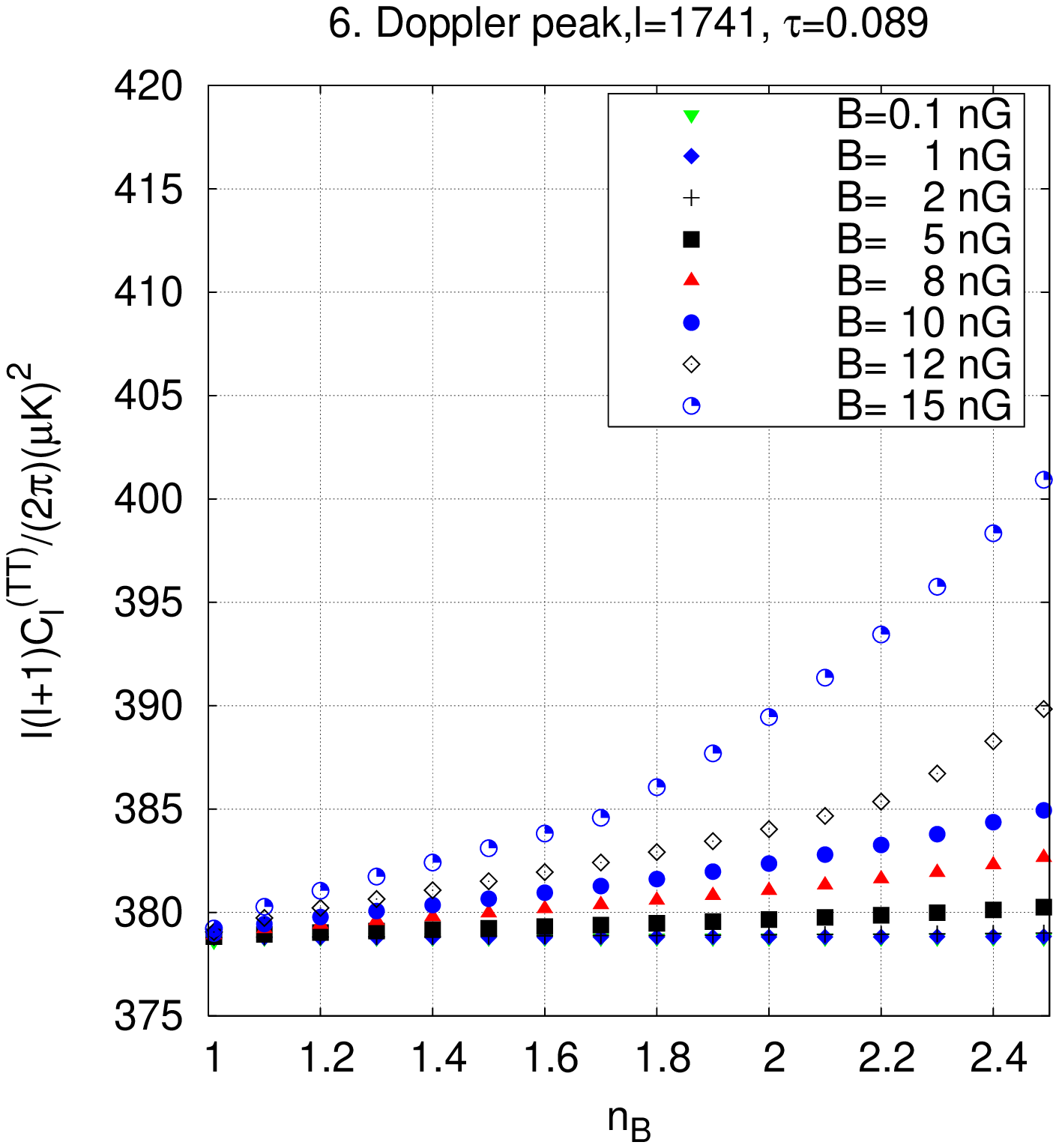}} &
     \hbox{\epsfxsize = 7.6 cm  \epsffile{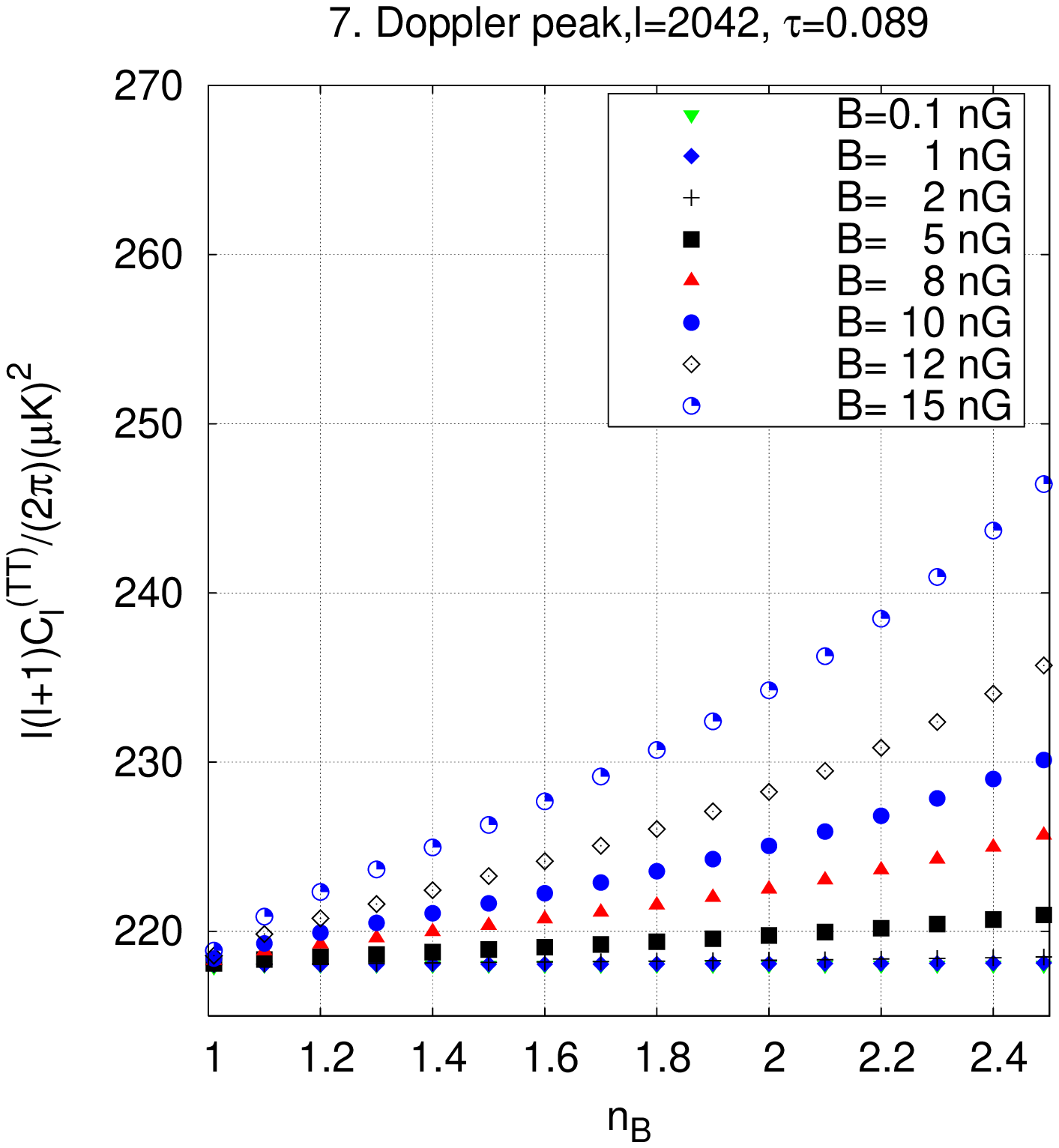}}\\
      \hline
\end{tabular}
\end{center}
\caption[a]{The heights of the sixth (plot at the left) and of the seventh (plot at the right) 
Doppler peaks is illustrated for different values of the regularized magnetic field 
as a function of the spectral index.}
\label{Figure9}
\end{figure}

\begin{figure}
\begin{center}
\begin{tabular}{|c|c|}
      \hline
      \hbox{\epsfxsize = 7.6 cm  \epsffile{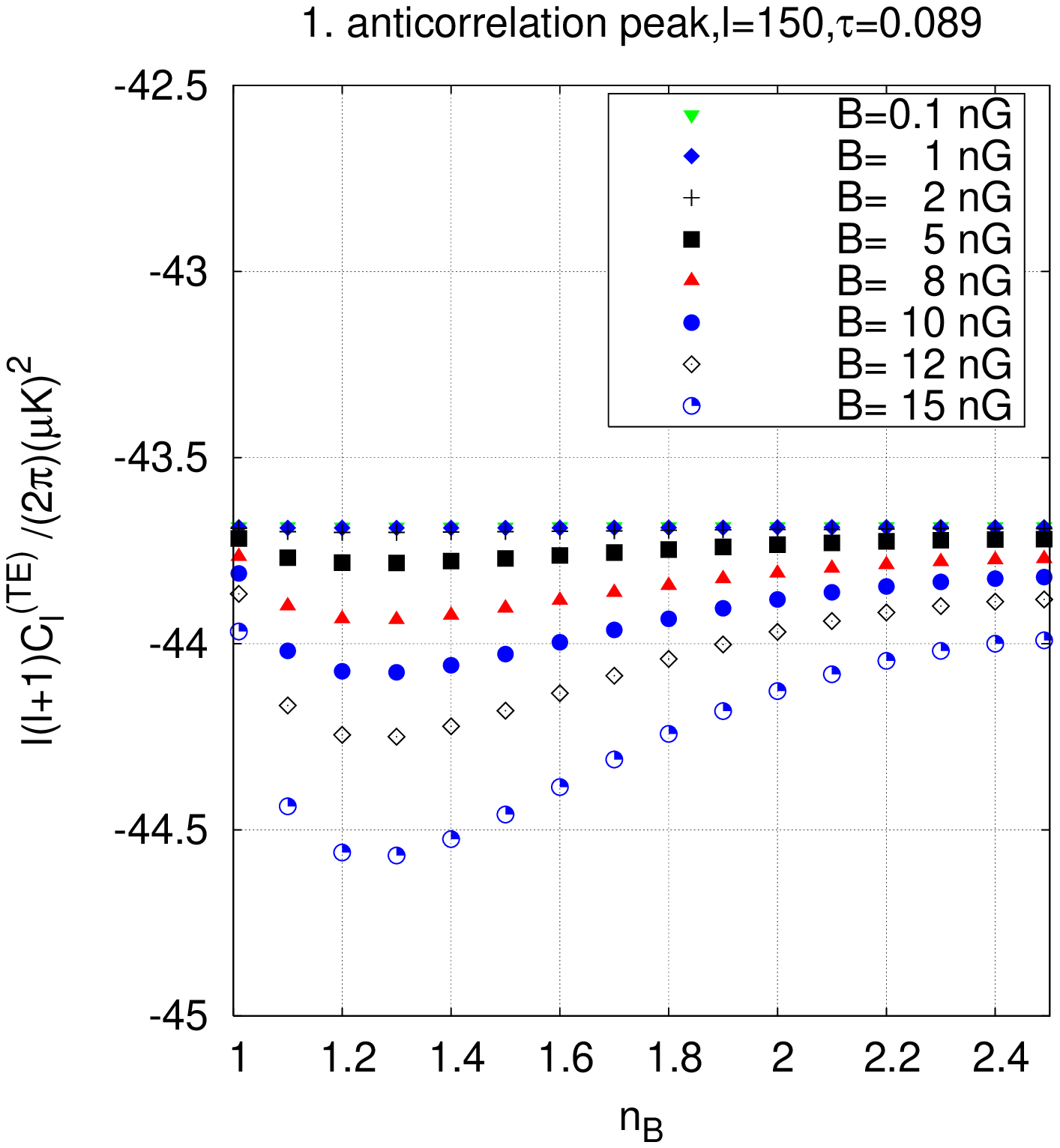}} &
     \hbox{\epsfxsize = 7.6 cm  \epsffile{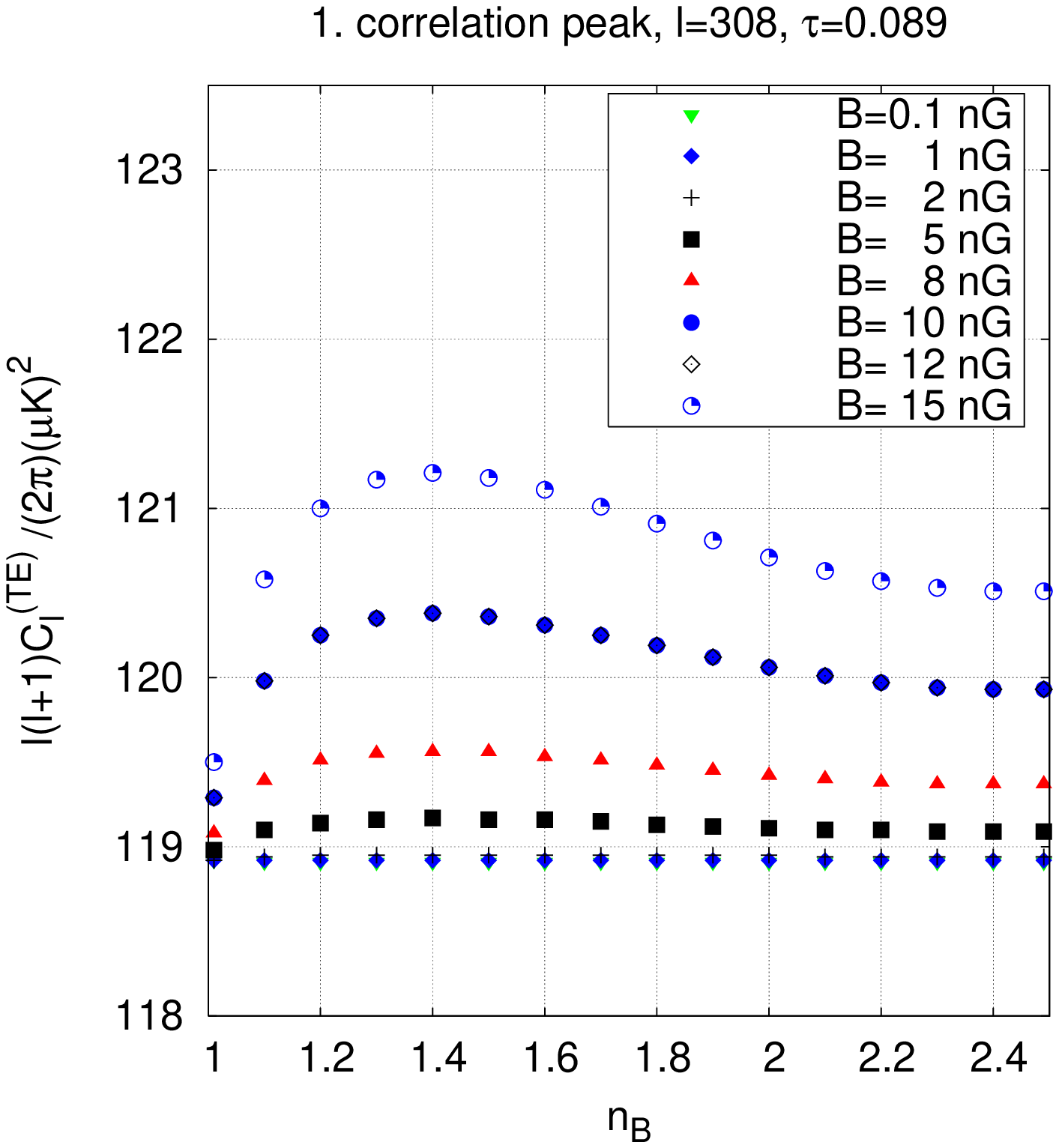}}\\
      \hline
\end{tabular}
\end{center}
\caption[a]{The height of the first anticorrelation and correlation peaks 
is illustrated, respectively, in the left and in the right plots.}
\label{Figure10}
\end{figure}

\begin{figure}
\begin{center}
\begin{tabular}{|c|c|}
      \hline
      \hbox{\epsfxsize = 7.6 cm  \epsffile{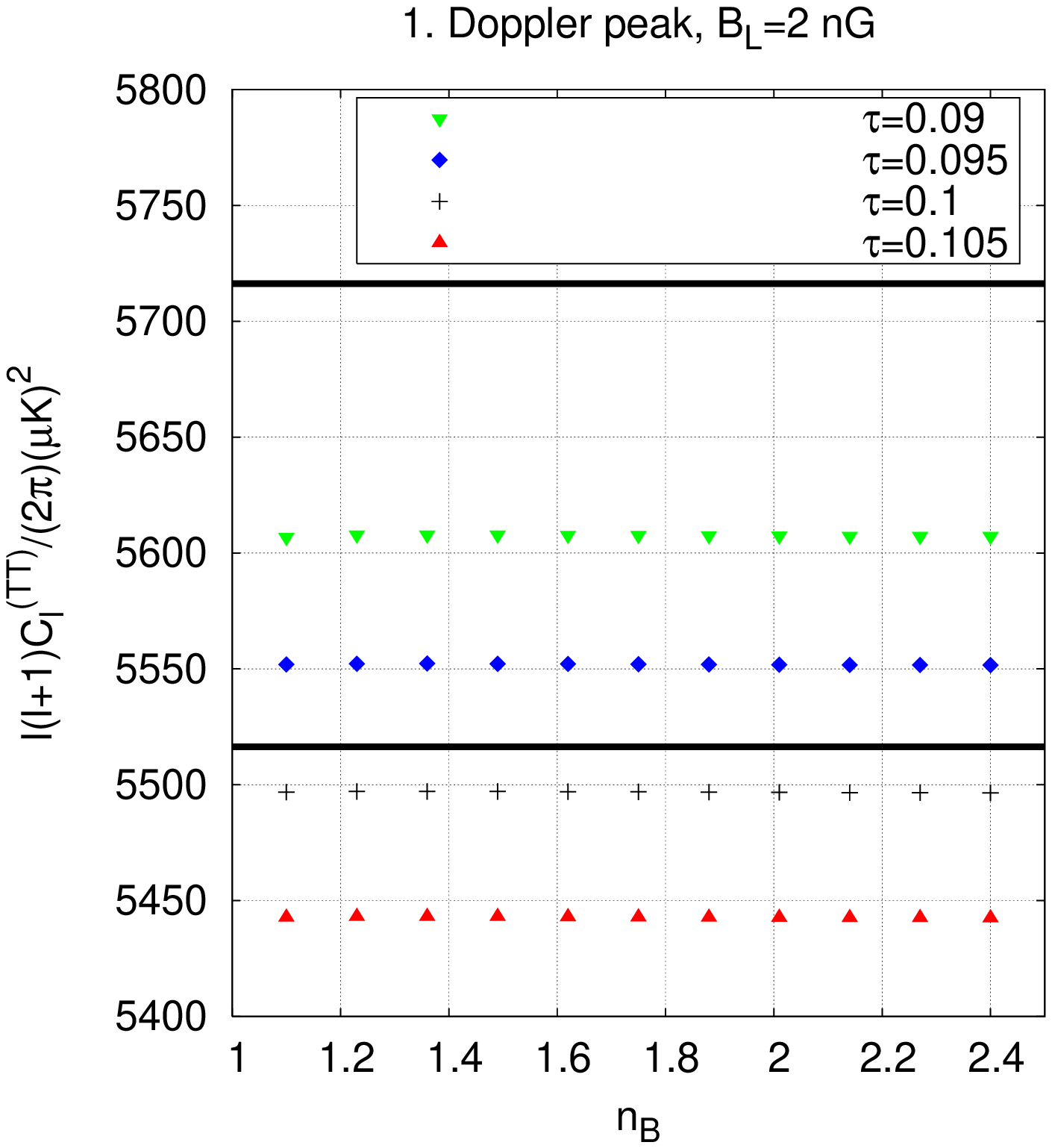}}&
     \hbox{\epsfxsize = 7.6 cm  \epsffile{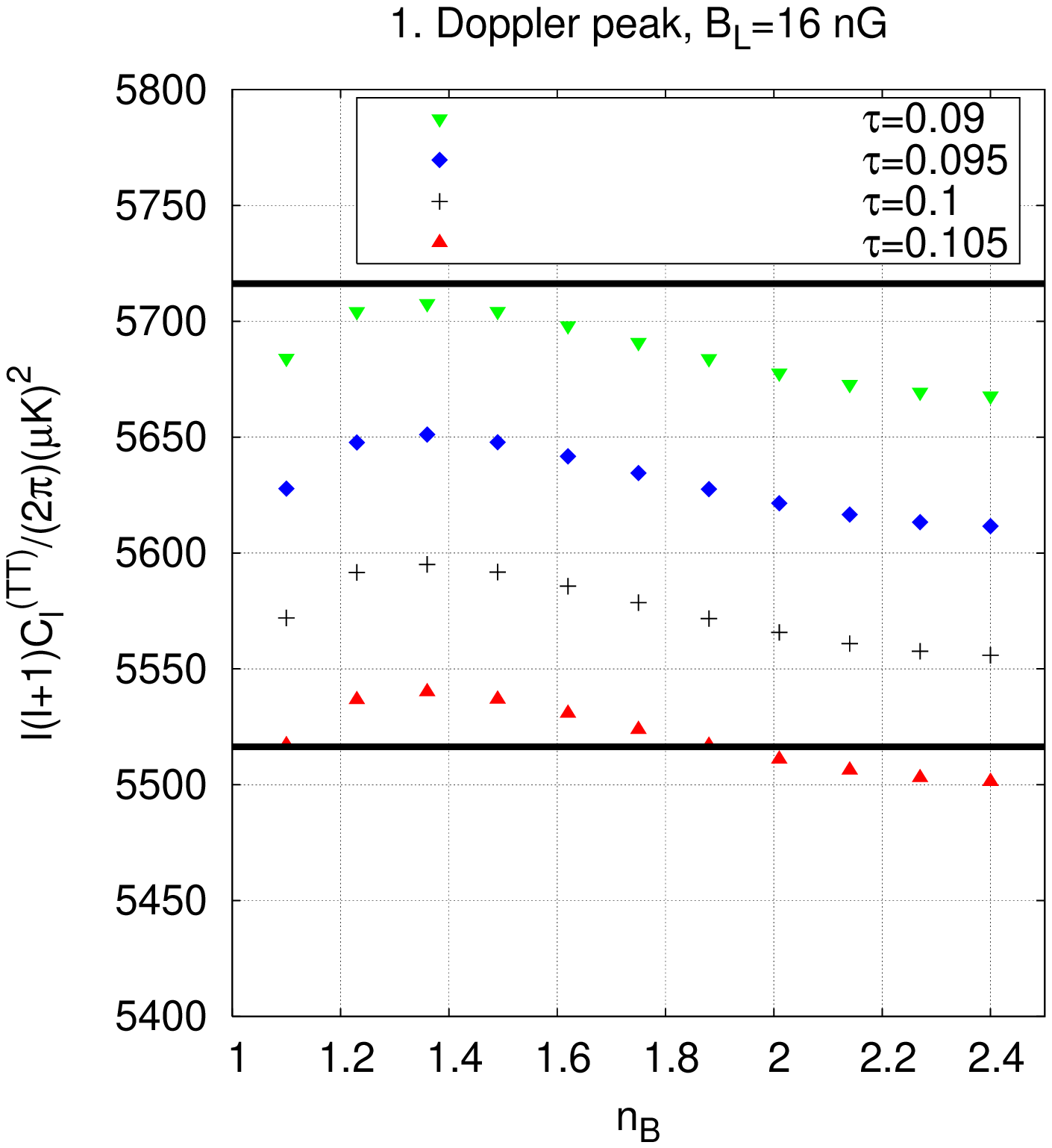}}\\
      \hline
\end{tabular}
\end{center}
\caption{The height of the Doppler peak is presented as a function of the 
magnetic spectral index $n_B$ for different values of the
magnetic field strength and the optical depth. The horizontal lines indicate the 
observational bounds on the Doppler peak from WMAP3.}
\label{Figure11}
\end{figure}

In Fig. \ref{Figure7} the height of the first peak is reported for different values of the regularized magnetic field, 
as a function of the magnetic spectral index. Note that in all the figures from Fig. \ref{Figure7} to \ref{Figure11}
the title of each plot labels the peak. So, for instance, the notation ``$1.$Doppler peak"means that the corresponding 
plot refers to the first Doppler peak. 

The maximal increase of the acoustic peaks always arises for intermediate spectral tilts.
The TT correlations are not only  shifted upwards 
but they are also distorted: this is  evident from Fig. \ref{Figure8} where 
we illustrate, respectively, the ratio of the second peak to the first (plot at the left) 
and the ratio of the third peak to the first. From the left plot of Fig. \ref{Figure8} 
it is clear that an increase in the spectral index entails, for $n_{\mathrm{B}} < 1.8$,
a decrease of the height of the second peak in comparison with the first. When 
$n_{\mathrm{B}} > 1.8$ the opposite behaviour is observed.
For a given value of the magnetic field the correlated distortion of the first peak
is an effect of the order of few percent (as it can be argued from the corresponding plots). 
As we reach into the region 
$\ell > 1500$ the effect becomes more pronounced especially for the seventh peak.
While the first peak is not sensitive to a nG magnetic field, the 
seventh peak can be a reasonable indicator of the presence of large-scale 
magnetic fields in the nG range.
This aspect is illustrated in Fig. \ref{Figure9} in the cases of the sixth and seventh peaks. 
From the plot at the right, for instance, a $15$ nG field has a $10$ percent effect on the shift.  
The precise value depends on the spectral index, as it can be argued from Fig. \ref{Figure9}.

The observed correlated distortion is also  visible 
in the case of the TE angular power spectra. In Fig. \ref{Figure10} 
the first anticorrelation peak (plot at the left) is compared 
with the first correlation peak (plot at the right).
The nominal value of the first correlation peak (appearing in the title of the plot) is the one stemming 
from the three year best fit to the WMAP data alone. 

In Section \ref{sec4} we pointed out that a slight increase in the CDM fraction could be compensated by the presence of a magnetic field. This kind of potential degeneracies 
can only be throughly discussed in the framework of a general parameter estimation which also 
includes, to begin with, the magnetic field parameters. This analysis is beyond the scope 
of this paper, however, it is useful to investigate in an eclectic perspective, also other 
potentially interesting degeneracies which can be only assessed (or even partially resolved)
in more systematic approaches. 

One of these potential degeneracies involves the optical depth to reionization, i.e. $\tau$. 
The increase of $\tau$ in a model without 
magnetic field yields a lower height of the Doppler peak. 
The values of the best fit model for WMAP data alone are used (see Eq. (\ref{best1}))
apart from the optical depth $\tau$ which is assumed to be varying between
0.09 and 0.105.
In Figure \ref{Figure11} the height of the first acoustic peak is shown for different values of the 
magnetic spectral index $n_B$ for different values of the magnetic field strength
and of the optical depth $\tau$.

The dependence on the magnetic spectral index becomes 
more important as $B_{\mathrm{L}}$ increases. While for $B_{\mathrm{L}}=2$ nG the 
height of the acoustic peak is basically independent of the spectral index,
there is,  for $B_{\mathrm{L}}=16$ nG,  a nontrivial functional relation between the height 
of the peak and the magnetic spectral index. The variations in $\tau$ can be partially
compensated by the addition of the magnetic field. Note, indeed, that the full line 
denote the experimental indetermination in the position of the peak.

In the near future the Planck explorer mission with its low frequency and 
high frequency instruments will be able to probe with higher accuracy 
the temperature autocorrelations, the polarization autocorrelations 
and cross-correlations.  At the moment the putative  Planck sensitivity 
can only be inferred from the last version of the Planck blue book \cite{Planck}.
The nominal sensitivity might not be the one effectively achieved by the instruments. 
Given the present specifications of the instruments, 
is not clear in our case what are the best frequency channels to be combined in order 
to be mostly sensitive to the features introduced in the angular power spectra by the 
large-scale magnetic fields. In spite of this we find it interesting to elaborate 
on the possible implications of our endeavors for the Planck measurements 
at high multipoles (i.e. $\ell > 1500$).

To compare the nominal Planck sensitivity with the situation where 
magnetic fields are consistently included in a $\Lambda$CDM paradigm (with no tensors) 
we find it useful to adopt the following measure:
\begin{eqnarray}
{\mathcal D}^{(\mathrm{TT})}_{\mathrm{B},0}\equiv\frac{|C^{(\mathrm{TT})}_{\ell}(B_{\mathrm{L}},n_{\mathrm{B}})-C^{(\mathrm{TT})}_{\ell}(B_{\mathrm{L}}=0)|}
{C^{(\mathrm{TT})}_{\ell}(B_{\mathrm{L}}=0)}.
\label{est1}
\end{eqnarray}
In Eq. (\ref{est1})  $C^{(\mathrm{TT})}_{\ell}(B_{\mathrm{L}}=0)$ is computed from the three year best fit 
to the WMAP data alone; $C^{(\mathrm{TT})}_{\ell}(B_{\mathrm{L}},n_{\mathrm{B}})$ is the TT correlation 
but computed with a magnetic field of regularized intensity $B_{\mathrm{L}}$ and 
characterized by a spectral index $n_{\mathrm{B}}$. In different frameworks 
a similar estimator has been also employed \cite{zalsel,pav4}.

\begin{figure}
\begin{center}
      \epsfxsize = 7.6 cm  \epsffile{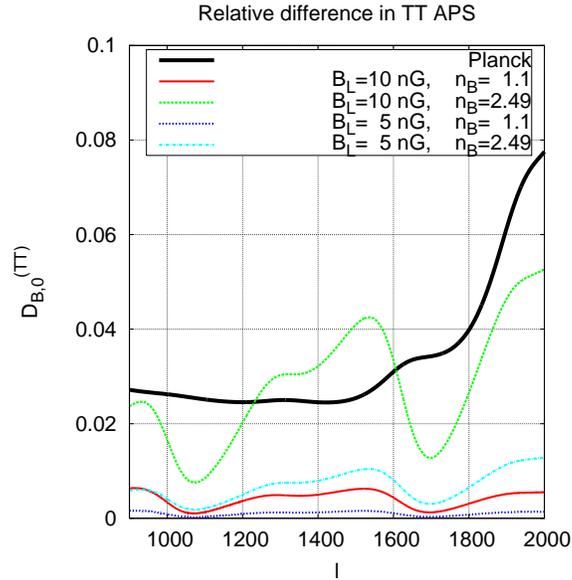}   
\end{center}
\caption[a]{The estimator of Eq. (\ref{est1}) is illustrated in combination 
with the foreseen Planck sensitivity (full curve).}
\label{Figure12}
\end{figure}

In Fig. \ref{Figure12} the quantity defined in Eq. (\ref{est1}) is illustrated for different 
values of the magnetic field intensity and of the spectral index. Note that 
${\mathcal D}^{(\mathrm{TT})}_{\mathrm{B},0}$ estimates the difference induced by the 
presence of the magnetic field on the extrapolated three year WMAP best fit which can be used 
to deduce the nominal sensitivity of Planck for different regions in the multipole space.
Adopting the three year WMAP best fit as fiducial model, the  
1-$\sigma$ errors  can be inferred following the standard analysis also thoroughly 
reviewed in the Planck blue book  (see \cite{Planck})
\begin{eqnarray}
(\Delta C^{(\mathrm{TT})}_{\ell})^2 &=&\frac{2}{(2\ell+1) f_{\mathrm{sky}}}\left(C_{\ell}+w_{\mathrm{T}}^{-1} W_{\ell}^{-2}\right)^2,
\label{est2}\\
w_{\mathrm{T}} W_{\ell}^2 &=& \sum_{\mathrm{c}} w_{\mathrm{T}}^{(\mathrm{c})} e^{- \ell(\ell+1)/{\ell^{\mathrm{c}}_{\mathrm{beam}}}^2}.
\label{est3}
\end{eqnarray}

In Eqs. (\ref{est2})--(\ref{est3}) various assumptions should be made as far as the sky coverage and the 
relevant frequency channels are concerned. For this reason what we are presenting here 
are just preliminary indications of what could be the trend of the Planck accuracy on the basis of the 
figures customarily employed by the Planck team \cite{Planck}. Needless to say 
that the present estimate can be made more realistic once 
the effective Planck sensitivity will be available.  Thus, it will be assumed that $f_{\mathrm{sky}} \simeq 0.65$ 
corresponding to $\pm 20^{0}$ galactic cut. Furthermore the three lowest frequency channels of the 
high frequency instrument (i.e. $100$ GHz, $143$ GHz and $217$ GHz) are combined and the sum appearing in Eq. (\ref{est3}) then extends over these three channels. To make explicit the sum it should be noted that 
$w^{\mathrm{c}}_{\mathrm{T}} = (\sigma^{\mathrm{c}}_{\mathrm{pT}} \vartheta^{\mathrm{c}}_{\mathrm{FWHM}})^{-2}$ is the sensitivity 
per resolution element $\vartheta^{\mathrm{c}}_{\mathrm{FWHM}}\times\vartheta^{\mathrm{c}}_{\mathrm{FWHM}}$. The quantities $\sigma^{\mathrm{c}}_{\mathrm{pT}}$ and 
 $\vartheta^{\mathrm{c}}_{\mathrm{FWHM}}$ change for each  of the three aforementioned 
channels. For  $\sigma^{\mathrm{c}}_{\mathrm{pT}}$ and $\vartheta^{\mathrm{c}}_{\mathrm{FWHM}}$
the values reported in the Planck blue book \cite{Planck} have been selected. With these 
specifications in mind, the last quantity to be defined is 
$\ell^{\mathrm{c}}_{\mathrm{beam}} = \sqrt{8 \ln{2}}/ \theta_{\mathrm{FWHM}}^{\mathrm{c}}$ which measures 
the resolution of the Gaussian beam.

In Fig. \ref{Figure12} with the full curve the $1$-$\sigma$ error is reported. If the estimator 
leads to a value which is larger that the foreseen sensitivity it will be possible 
to make observational distinction between the magnetized model and the 
extrapolated three year best fit. In spite of the intrinsic uncertainty on the actual 
sensitivities of the instrument Fig. \ref{Figure12} is eloquent enough and then offers encouraging 
prospects for the region of high multipoles. 

\renewcommand{\theequation}{7.\arabic{equation}}
\setcounter{equation}{0}
\section{Concluding remarks}
\label{sec7}

Large-scale magnetic fields are an observed component of the 
present Universe. 
Why are they present at the $\mu$G level in galaxies with different morphologies 
and different evolutionary histories? Why are they present inside 
rich clusters? When did they originate? Are they a cosmic relic
 in the same way as the adiabatic mode of curvature perturbations has a primeval origin?

It is difficult to answer these rather motivated questions 
just by building new models or by emphasizing single (potentially 
interesting) effects. It is even more difficult 
to run complicated simulations trying to reproduce 
large-scale magnetic fields without knowing the initial conditions 
to be imposed at early times, i.e. after decoupling. 
What we need is a systematic 
scrutiny of pre-decoupling physics to answer more modest 
but necessary questions. The most urgent one concerns the effect 
of large-scale magnetic fields on CMB observables.
The combined numerical and analytical tools presented in this 
paper are a promising step along this direction.
Needless to say that we plan to scrutinize more 
deeply all the manifold and exciting implications of our approach.

\section*{Acknowledgments}

K. E. Kunze  acknowledges the support of the  ``Ram\'on y Cajal''  program as well as the grants FPA2005-04823 and FIS2006-05319 of the Spanish Science Ministry.

\newpage

\begin{appendix}
\renewcommand{\theequation}{A.\arabic{equation}}
\setcounter{equation}{0}
\section{From synchronous to longitudinal gauges}
\label{APPA}
The treatment employed in the present analysis is the 
one stemming from the synchronous gauge description.
We cross-checked all our results within the longitudinal 
approach. This cross-check is, under certain circumstances, mandatory.
Indeed, in the synchronous gauge the freedom of selecting the 
coordinate system is not completely fixed. It is therefore 
important to have at hand a gauge description where 
the gauge parameters are completely fixed.

The strategy we followed has been often to derive the same quantity 
in the two different gauges and then compare the 
results by transforming the obtained expressions from one gauge 
to the other. For the effectiveness of this approach the longitudinal gauge 
is not essential. Rather we should say that the only essential requirement is 
a gauge where the freedom of selecting the coordinate system 
is completely fixed. So, for instance, the uniform curvature gauge would 
work equally well for this purpose \cite{hwang1,hwang2}. 

In the longitudinal gauge the metric of Eq. (\ref{line})  is perturbed 
in such a way that non-vanishing entries 
of the first-order metric are
\begin{equation}
\delta_{\mathrm{s}} g_{00} = 2 a^2(\tau) \phi(k,\tau),\qquad  \delta_{\mathrm{s}} g_{ij} = 2 a^2(\tau)  \psi(k,\tau) \delta_{ij}.
\label{L1}
\end{equation}
The difference between the longitudinal and the synchronous 
coordinate systems is evident by comparing Eq. (\ref{L1}) 
with Eq. (\ref{pert1}). 
Following standard techniques we can find the precise 
relation between the longitudinal and the synchronous degrees 
of freedom:
\begin{eqnarray}
&& \phi(k,\tau) = - \frac{1}{2k^2} \{[h(k,\tau) + 6 \xi(k,\tau)]'' + {\mathcal H} [h(k,\tau) + 6 \xi(k,\tau)]'\},
\nonumber\\
&& \psi(k,\tau) = - \xi(k,\tau) + \frac{{\mathcal H}}{2 k^2} [h(k,\tau) + 6 \xi(k,\tau)]',
\nonumber\\
&& \overline{\delta}(k,\tau) = \delta(k,\tau) + \frac{3 {\mathcal H}(w + 1)}{2k^2}[h(k,\tau) + 6 \xi(k,\tau)]'
\nonumber\\
&& \overline{\theta}(k,\tau) = \theta(k,\tau) - \frac{1}{2}[h(k,\tau) + 6\xi(k,\tau)]'.
\label{L2}
\end{eqnarray}
The barred quantities (i.e. $\overline{\delta}$ and $\overline{\theta}$) 
are defined in the longitudinal gauge and $w$ is the barotropic 
index of the corresponding species.  Similarly 
the transformation for $\theta$ holds for a generic 
peculiar velocity. 
The inverse transformations can be also obtained and they are:
\begin{eqnarray}
&& \xi(k,\tau) =  -\psi(k,\tau) - \frac{{\mathcal H}}{a} \int^{\tau} a(\tau') \phi(k,\tau') d\tau',
\nonumber\\
&& h(k,\tau) = 6 \psi(k,\tau) + 6  \frac{{\mathcal H}}{a} \int^{\tau} a(\tau') \phi(k,\tau') d\tau' - 2 k^2 
\int^{\tau} \frac{d\tau'}{a(\tau')} \int^{\tau'} a(\tau'') \phi(k,\tau'') d\tau'',
\nonumber\\
&& \delta(k,\tau) =  \overline{\delta}(k,\tau) + \frac{3{\mathcal H}(w + 1)}{a} \int^{\tau} a(\tau') \phi(k,\tau') d\tau'
\nonumber\\
&& \theta(k,\tau) = \overline{\theta}(k,\tau) - \frac{k^2}{a} \int^{\tau} a(\tau') \phi(k,\tau') d\tau'.
\label{L3}
\end{eqnarray}
The integrals appearing in Eq. (\ref{L3}) for the expressions of $\theta$ and $h$ imply two integration constants which can be space dependent and which are fixed by demanding that $\theta_{\mathrm{c}}=0$ and that 
$h$ has no constant mode. 

The solution for the magnetized adiabatic mode will be, in the longitudinal gauge, 
\begin{eqnarray} 
&& \phi_{*}(k)= \frac{20 C(k)}{4 R_{\nu} + 15} - 2  \frac{R_{\gamma} [ 4 \sigma_{\mathrm{B}}(k) - R_{\nu} \Omega_{\mathrm{B}}(k)]}{ 4 R_{\nu} +15},
\nonumber\\
&&\psi_{*}(k) = \biggl(\frac{8 R_{\nu} + 20}{4R_{\nu} + 15}\biggr) C(k) + \frac{R_{\gamma} [4 \sigma_{\mathrm{B}}(k,\tau) - R_{\nu} \Omega_{\mathrm{B}}(k)]}{ 4 R_{\nu} +15}, 
\nonumber\\
&& \psi_{*}(k) = \biggl( 1 + \frac{2}{5} R_{\nu}\biggr) \phi_{*}(k)  + \frac{R_{\gamma}}{5} [ 4 \sigma_{\mathrm{B}}(k) - R_{\nu} \Omega_{\mathrm{B}}(k)],
\nonumber\\
&& \delta_{\gamma}(k,\tau) = -2 \phi_{*}(k) - R_{\gamma} \Omega_{\mathrm{B}}(k),
\nonumber\\
&& \delta_{\nu}(k) = -2 \phi_{*}(k) - R_{\gamma} \Omega_{\mathrm{B}}(k),
\nonumber\\
&& \delta_{\mathrm{c}}(k) = - \frac{3}{2} \phi_{*}(k) - \frac{3}{4}R_{\gamma} \Omega_{\mathrm{B}}(k),
\nonumber\\
&& \delta_{\mathrm{b}}(k) = - \frac{3}{2} \phi_{*}(k) - \frac{3}{4}R_{\gamma} \Omega_{\mathrm{B}}(k),
\nonumber\\
&& \sigma_{\nu}(k,\tau) = - \frac{R_{\gamma}}{R_{\nu}}  \sigma_{\mathrm{B}}(k) + \frac{k^2 \tau^2}{6 R_{\nu}} [ \psi_{*}(k) - \phi_{*}(k)],
\nonumber\\
&& \theta_{\gamma\mathrm{b}}(k,\tau) = \frac{k^2 \tau}{2} \biggl[ \phi_{*}(k) + \frac{R_{\nu} \Omega_{\mathrm{B}}(k)}{2} - 2 \sigma_{\mathrm{B}}(k)\biggr],
\nonumber\\
&& \theta_{\nu}(k,\tau)= \frac{k^2 \tau}{2}\biggl[ \phi_{*}(k) - \frac{R_{\gamma} \Omega_{\mathrm{B}}(k)}{2} + 2 \frac{R_{\gamma}}{R_{\nu}} \sigma_{\mathrm{B}}(k) \biggr],
\nonumber\\
&& \theta_{\mathrm{c}}(k,\tau) = \frac{k^2 \tau}{2} \phi_{*}(k).
\label{L4}
\end{eqnarray}
It can be easily checked that the solution (\ref{L5}) is a solution of the full system written in the longitudinal frame. 

There is an important point to be borne in mind when setting initial conditions. We set initial conditions deep in the radiation epoch. Now, the constant 
$C= C(k)$ that appears in the synchronous description can be actually related with curvature perturbations on comoving 
orthogonal hypersurfaces ${\mathcal R}$. In the longitudinal gauge we do know that, deep in the radiation epoch,
\begin{equation}
{\mathcal R}(k)=- \psi - \frac{{\mathcal H} \phi + \psi'}{{\mathcal H}^2 - {\mathcal H}'} \simeq  - \psi_{*}(k) - \frac{\phi_{*}(k)}{2} = - 2 C(k)
\label{L5}
\end{equation}
In the longitudinal gauge we can also easily express the variable
$\zeta$ introduced in Eq. (\ref{TDC6}). In terms of the longitudinal
degrees of freedom 
\begin{equation}
\zeta = - \psi - \frac{\delta^{\mathrm{L}}\rho_{\mathrm{t}} + \delta\rho_{\mathrm{B}}}{\rho_{\mathrm{t}}'},
\label{L6}
\end{equation}
where $\delta^{\mathrm{L}}\rho_{\mathrm{t}}$  is the total 
density fluctuation in the longitudinal gauge.
In the longitudinal gauge the Hamiltonian constraint 
reads
\begin{equation}
\nabla^2 \phi - 3 {\mathcal H} ({\mathcal H} \phi + \psi') = 
4\pi G a^2 (\delta^{\mathrm{L}}\rho_{\mathrm{t}} +\delta\rho_{\mathrm{B}}).
\label{L7}
\end{equation}
Using Eqs. (\ref{L5}) and (\ref{L6}) into Eq. (\ref{L7}) 
the Hamiltonian constraint can be written as
\begin{equation}
\zeta = {\mathcal R} + \frac{\nabla^2 \psi}{12\pi G a^2 (\rho_{\mathrm{t}} + p_{\mathrm{t}})}
\label{L8}
\end{equation}
where Eqs. (\ref{FL2}) and (\ref{FL3}) have been used. Equation (\ref{L8}) 
has been quoted and independently obtained in Eq. (\ref{TDC8})
\end{appendix}

\end{document}